\documentclass[conference, usenames, dvipsnames]{IEEEtran}
\IEEEoverridecommandlockouts

\usepackage[export]{adjustbox}      
\usepackage{amsmath,amssymb,amsfonts}   
\usepackage{amsthm}
\usepackage{algorithmic}    
\usepackage{booktabs}       
\usepackage{cite}           
\usepackage{graphics}
\usepackage{graphicx}       
\usepackage{mathtools}      
\usepackage{multirow}       
\usepackage{nicefrac}
\usepackage{subcaption}
\usepackage{tabularx}       
\usepackage{textcomp}       

\usepackage{thmtools, thm-restate}
\usepackage[dvipsnames, table]{xcolor}         
\usepackage{url}
\usepackage{xspace}
\usepackage{enumerate}
\usepackage[hidelinks]{hyperref}
\usepackage[normalem]{ulem} 

\makeatletter 
\newcommand{\linebreakand}{%
  \end{@IEEEauthorhalign}
  \hfill\mbox{}\par
  \mbox{}\hfill\begin{@IEEEauthorhalign}
}
\makeatother 

\newtheorem{theorem}{Theorem}

\newtheorem{lemma}[theorem]{Lemma}
\newtheorem{definition}[theorem]{Definition}

\newtheorem{example}[theorem]{Example}

\newcommand\Conc[1]	{\textnormal{\textsf{#1}}}
\newcommand\NF[2]       {\nicefrac{#1}{#2}}             

\newcommand{\calk}{\mathcal{K}}
\newcommand{\caln}{\mathcal{N}}
\newcommand{\calw}{\mathcal{W}}
\newcommand{\calx}{\mathcal{X}}
\newcommand{\caly}{\mathcal{Y}}

\newcommand{\ta}{\texttt{a}}
\newcommand{\tb}{\texttt{b}}
\newcommand{\tc}{\texttt{c}}

\newcommand\SV {V_{g_{\rm T}}}

\newcommand{\defeq}{\vcentcolon=}

\newcommand{\shrug}{\stackrel{?}{=}}
\newcommand{\qm}[1]{``#1''}

\allowdisplaybreaks

\newcommand\Tstrut{\rule{0pt}{2.6ex}}         
\newcommand\Bstrut{\rule[-0.9ex]{0pt}{0pt}}   

\newcommand{\colorcell}[2]{\cellcolor{#1}{#2}} 

\definecolor{c1}{HTML}{F2CF96} 
\definecolor{c2}{HTML}{F2CF96} 
\definecolor{c7}{HTML}{E1B0AB}
\definecolor{c9}{HTML}{B3A5B2}
\definecolor{c3}{HTML}{B299C1} 
\definecolor{c5}{HTML}{B299C1} 
\definecolor{c10}{HTML}{A498BE}
\definecolor{c8}{HTML}{8B8BAF}
\definecolor{c4}{HTML}{7E8CB1}

\definecolor{Nc1}{HTML}{E8FFE8} 
\definecolor{Nc5}{HTML}{D8F8D8} 
\definecolor{Nc8}{HTML}{C0EBC0} 
\definecolor{Nc4}{HTML}{A5D8A5} 

\definecolor{NSc1}{HTML}{FFEFEF} 
\definecolor{NSc9}{HTML}{FAE0E0} 
\definecolor{NSc5}{HTML}{F5D5D5} 
\definecolor{NSc10}{HTML}{EFC5C5} 
\definecolor{NSc8}{HTML}{EAB5B5} 
\definecolor{NSc4}{HTML}{D8A0A0} 

\definecolor{NSrc6}{HTML}{E8FFFF} 
\definecolor{NSrc1}{HTML}{CFF9F9} 
\definecolor{NSrc2}{HTML}{CFF9F9} 
\definecolor{NSrc7}{HTML}{BBEFEF} 
\definecolor{NSrc3}{HTML}{AAE3E3} 
\definecolor{NSrc5}{HTML}{AAE3E3} 
\definecolor{NSrc8}{HTML}{99DCDC} 
\definecolor{NSrc4}{HTML}{88D0D0} 

\newcommand{\prob}{\ensuremath{\pi}\xspace}     
\newcommand{\hyperDist}[1]{[\ensuremath{\prob \triangleright #1]}\xspace}

\newcommand{\channel}[1]{\Conc{#1}\xspace}

\newcommand{\gain}{\ensuremath{g}\xspace} 

\newcommand{\priorGVuln}[2]{\ensuremath{V_{#1}(#2)}\xspace} 
\newcommand{\postGVuln}[2]{\ensuremath{V_{#1}\hyperDist{#2}}\xspace} 
\newcommand{\multGLeakage}[3]{\ensuremath{\mathcal{L}^{\times}_{#1}(#2, #3)}\xspace} 

\newcounter{MJ}
\newcommand\MJ[1]{\textcolor{Peach}{$\langle\!\langle$MJ\arabic{MJ} #1 $\rangle\!\rangle$}\addtocounter{MJ}{1}}

\newcounter{CP}

\newcounter{MA}

\newcounter{RG}
\newcommand\RG[1]{\textcolor{RoyalBlue}{$\langle\!\langle$RG\arabic{RG} #1 $\rangle\!\rangle$}\addtocounter{RG}{1}}

\newcommand\edits[1]{#1}
\newcommand\review[1]{\textcolor{black}{ #1}} 
\newcommand\redundant[1]{\textcolor{Red}{ [#1]}}
 
\def\BibTeX{{\rm B\kern-.05em{\sc i\kern-.025em b}\kern-.08em
    T\kern-.1667em\lower.7ex\hbox{E}\kern-.125emX}}
\begin{document}

\title{Analyzing the Shuffle Model through the Lens of Quantitative Information Flow}

\date{January 2022}

\author{\IEEEauthorblockN{Mireya Jurado}
\IEEEauthorblockA{\textit{Florida International University} \\
Miami, USA \\
}
\and
\IEEEauthorblockN{Ramon G.\ Gonze}
\IEEEauthorblockA{\textit{Universidade Federal de Minas Gerais} \\
Belo Horizonte, Brazil\\
}
\IEEEauthorblockA{\textit{Inria Saclay, \'{E}cole Polytechnique} \\
Palaiseau, France \\
}
\and
\IEEEauthorblockN{M\'{a}rio S.\ Alvim}
\IEEEauthorblockA{\textit{Universidade Federal de Minas Gerais} \\
Belo Horizonte, Brazil \\
}
\and

\linebreakand 

\IEEEauthorblockN{Catuscia Palamidessi}
\IEEEauthorblockA{\textit{Inria Saclay and LIX, \'{E}cole Polytechnique} \\
Palaiseau, France \\
}
}

\maketitle
\thispagestyle{plain}
\pagestyle{plain}

\begin{abstract}
Local differential privacy (LDP) is a variant of differential privacy (DP) that avoids the necessity of a trusted central curator,
\review{at the expense of a worse}
trade-off between privacy and utility.
\review{The} shuffle model has emerged as a way to provide greater anonymity to users by randomly permuting \review{their} messages, so that the \edits{direct} link between users and their reported values is lost \review{to the data collector}.
By combining \review{an} LDP mechanism with a shuffler, privacy can be improved at \edits{no} cost for the accuracy \review{of operations  insensitive to permutations},
thereby improving \review{utility} in \review{many} analytic tasks.
However, the privacy implications \review{of shuffling} are not always \review{immediately evident}, and derivations of privacy bounds are made on a case-by-case basis.


\review{In this paper, we analyze the combination of LDP with
shuffling in the rigorous framework of quantitative information flow (QIF), and reason about the resulting resilience to inference attacks.
QIF naturally captures (combinations of) randomization mechanisms as information-theoretic channels, thus allowing for precise
modeling of a variety of inference attacks in a natural way
and for measuring
the leakage of private information under these attacks.
We exploit symmetries 
of the particular combination of $k$-RR mechanisms with the 
shuffle model to achieve closed formulas that express 
leakage exactly.
In particular, we provide formulas that show how shuffling improves  protection against leaks in the local model, and study how leakage behaves for various values of the privacy parameter of the LDP mechanism.}

\review{In contrast to the \emph{strong adversary} from differential privacy, who knows everyone's record in a dataset but the target's,
we focus on an \emph{uninformed adversary}, who does not
know the value of any individual in the dataset.
This adversary is often more realistic as a consumer of statistical datasets, and indeed we show that in some
situations mechanisms that are equivalent w.r.t.\ the strong adversary
can provide different privacy guarantees under the uninformed one.
Finally, we also illustrate the application of our model to the  typical strong adversary from DP.}

\end{abstract}

\begin{IEEEkeywords}
\review{quantitative information flow,
formal security models,
differential privacy,
shuffle model}
\end{IEEEkeywords}


\section{Introduction}  
\label{introduction}
Differential privacy (DP), introduced by Dwork and her colleagues~\cite{Dwork:06:EUROCRYPT,Dwork:06:TCC}, is one of the most successful frameworks for privacy protection. The original definition, 
which is now called \emph{central} DP, assumes the existence of a \emph{trusted curator}
who has access to the raw user data and is in charge of receiving queries to this data and reporting the corresponding answers, suitably obfuscated so to make them 
essentially
insensitive to any single data point. Differential privacy provides formal privacy guarantees and has various desirable properties, such as compositionality, which ensure its robustness to repeated queries. 
However, the fact that all user data is in the hands of one party means that there is a single point of failure: the privacy of all users depends on the integrity of the curator, and on his capability to protect them from security breaches.

As an alternative to the central model, researchers have proposed \emph{local} differential privacy (LDP)~\cite{Kasiviswanathan:11:SIAMJC,Duchi:13:FOCS}, which does not require
a trustworthy data curator. 
In this model, each user applies an LDP-compliant protocol to perturb her data on the client side and sends it to an aggregator. Then, an analyst can estimate the desired query based on the collected
noisy data from all users on the server side. The result is guaranteed to be private due to the postprocessing property.
LDP has become very popular, partially thanks to its adoption by tech giants such as Google~\cite{Erlingsson:14:CCS,Fanti:16:PETS}, Apple~\cite{AppleDPTeam:17:ML,Thakurta:17:Patent}, and Microsoft~\cite{Ding:17:NIPS}, that have deployed LDP-compliant algorithms into their products to collect users’ usage statistics.

In comparison with central DP, however, LDP suffers from a worse trade-off between privacy and utility. Namely, in order
to achieve the same accuracy as in the central model, 
we need either to lower the level of individual protection, or to provide more data samples.
Indeed, there are a number of lower bounds on the error of locally private protocols that strongly separate the local and the central model~\cite{Beimel:08:CRYPTO,Chan:12:ESA}. 

The respective drawbacks of the central and the local model have stimulated the search for 
different architectures, and in this context, the \emph{shuffle model}~\cite{Bittau:2017:SOSP} has emerged as an appealing alternative. This model, in its simplest form, assumes
a data collector who receives one message 
from each of the users, as in LDP.  
In contrast to the latter, however, it also assumes that 
a mechanism is in place to randomly permute the messages\review{, before they reach the data collector,} so that any 
direct
association between users  and their reported values is lost. 
In this way, shuffling 
provides privacy amplification at 
no
cost for the utility for the accuracy of commutative operations, i.e., operations that are insensitive to the order of data. These correspond to a large class of the most typical analytical tasks, \review{such as sum, average, and histogram queries}. In this way, the \review{trade-off} between 
privacy and utility is substantially improved for those queries, and in some cases, it even becomes very close to that of the central model~\cite{Cheu:19:EUROCRYPT}. 

Research on the shuffle model has focused on finding bounds for
the level of privacy obtained after shuffling, expressed in terms of the 
privacy parameter (or parameters, in case of approximate LDP) of the local obfuscation mechanism used in combination with the shuffler. These bounds have been improved, i.e., made stricter over the years,
\edits{but the proofs are quite involved and produced in a case-by-case basis, without a unifying framework to reason \review{about inference attacks} against the shuffle model altogether.}

In this paper, we take a different perspective by analyzing the shuffle model from the point of view of \emph{quantitative information flow} (QIF)~\cite{alvim2020QIF}, \edits{which is a rigorous framework grounded on sound \review{information--} and decision-theoretic principles to reason about the leakage caused by inference attacks.
QIF has been successfully applied to a variety of privacy and security analyses, including searchable encryption%
~\cite{Jurado:21:CSF}, 
intersection and linkage attacks against $k$-anonymity~\cite{Fernandes:18:FM}
and very large, longitudinal microdata collections~\cite{Alvim:22:PETSa}, and differential privacy~\cite{Alvim:15:JCS,Chatzikokolakis:19:CSF}.}

In the QIF framework, a system is modeled as 
\review{information-theoretic} channel taking in some secret input and producing some observable output. The information leakage \review{is defined} as the difference between the vulnerability of the secret 
\review{(i.e., the amount of useful information to perform an attack)}
before and after passing through the channel (prior and posterior vulnerabilities\review{, respectively}). 
The vulnerability, indeed, increases \review{because} the attacker can \edits{infer} information about the input by observing the output. 
\review{The vulnerability measure} also depends on the \edits{adversary's goals and capabilities}\review{, which are} captured in the state-of-the-art QIF framework of $g$-leakage~\cite{Alvim:12:CSF} by using suitable \emph{gain functions}. We model both the shuffler and the LDP mechanism as \edits{information-theoretic} channels, and their composition \review{with the operation of} \emph{channel cascading}, and we study their leakage.
In this paper, we understand vulnerability as the adversary's ability to infer personal data, and can thereby be considered the inverse of privacy; i.e., the higher the vulnerability, the lower the privacy, and \review{vice-versa}.

We examine the \emph{single-message shuffle model protocols}, in which each user sends \review{a} single data point to the collector, and also on the $k$-ary randomized response ($k$-RR) mechanism~\cite{Kairouz:16:JMLR} which is one of the most popular mechanisms for LDP, and the core of Google's RAPPOR~\cite{Erlingsson:14:CCS}. 
We focus on the case in which the goal of the adversary is to guess the secret value of a 
\emph{single} individual (target) from  the dataset, 
\edits{which corresponds to the typical scenario of differential privacy in which the reported answer should not allow an observer to distinguish with confidence between two adjacent datasets~\cite{Tschantz:20:SP,Cuff:16:CCS,Yang:15:SIGMOD,Li:13:CCS}.}\edits{~\footnote{\edits{Notice that depending on the variant of differential privacy and attack scenario considered, the adversary's prior on the secret value may vary. QIF, however, explicitly separates the adversary’s goals and capabilities, modeled as a gain function, from her prior knowledge on the secret, modeled as a prior distribution. Here our comparison is focused on the former.}}} 

\review{A key distinction of our work, however, is that we
focus on an \emph{uninformed adversary}, 
who does not know anyone's values before accessing a 
data release, and assumes
a uniform prior on datasets.
It is well known from the literature that the guarantees against inferences provided by differential privacy  hold only w.r.t.\  what is usually called the \emph{strong adversary}, who
knows everyone's data except those of the target individual.
However, adversaries with weaker, less informed priors, can 
often benefit \emph{more} from an inference attack than the strong adversary, simply because there is more to be learned by them%
~\cite{Tschantz:20:SP,Cuff:16:CCS,Yang:15:SIGMOD,Li:13:CCS}.
The fact that differential privacy's guarantees against a strong adversary do not necessarily carry over to less knowledgeable adversaries is commonly overlooked in some interpretations of the framework.\review{~\footnote{\review{For this reason, we believe that the term ``strong adversary'' is misleading, and that ``informed adversary'' would be preferable. However, in this paper we stick to the terminology from the literature.}}} 
We refer to recent work by Tschantz et al.~\cite{Tschantz:20:SP} for an excellent presentation of the issue.
Another important reason to study the uninformed adversary is that 
it is more realistic, as usually in DP  consumers cannot  access  the micro-data directly. This is important if we want to compare mechanisms for privacy: it could be the case that two mechanisms $M$ and  $M'$ are equivalent under the strong adversary, but they are not under the uninformed adversary. For example, assume for simplicity that each record contains one bit, and that $M$ outputs the last bit (i.e., the content of the last record) in the dataset, or its complement, with probabilities $\nicefrac{e^\epsilon}{1+e^\epsilon}$ and $\nicefrac{1}{1+e^\epsilon}$, respectively. 
In contrast, $M'$ computes the binary sum of the other bits, and 
behaves exactly as $M$ if this sum is $0$, otherwise, it outputs the same last bit or its complement, but with inverted probabilities w.r.t.\ what $M$ does. 
It is easy to see that both $M$ and $M'$ are exactly $\epsilon$-differentially private, so they are equivalent under the strong adversary (who knows all the records except the last one), but $M'$ is more private than $M$ under the uninformed adversary.
More specifically, in $M'$ the reported value does not give any information on the original value of the last record (the posterior probability of it being $0$ or $1$ are the same, as in the prior),
whereas in $M$ some information is gained.
}

\review{In any case,  once we have a QIF model for LDP and shuffle, the same
techniques can be applied to derive results about their information leakage properties of other variants of adversaries.}

We investigate extensively the binary case, i.e., 
when $k=2$.
Despite its simplicity, the binary case is quite ubiquitous, and deserves 
special attention. Databases often contain binary attributes, modeling the 
presence or absence of a given feature or storing ``yes'' or ``no'' 
answers to sensitive questions.
Indeed, randomized response was developed as a survey 
technique motivated by the need to collect binary responses~\cite{Warner1965RandomizedResponse}.
Successively, we extend our investigation to a generic $k$. 


\subsection{Contribution}
Our contributions are the following: 
\begin{itemize}

\item 
\review{We study the information leakage of different combinations of $k$-RR and of the shuffler (Section~\ref{sec:model}).
To the best our knowledge, this is the first formal QIF model for the combination of LDP and shuffle mechanisms.
In particular, we prove that 
they commute, i.e., that it is equivalent (w.r.t.\ leakage) to first apply $k$-RR and then the shuffler, or to do the opposite (Proposition~\ref{theorem:commutativity-full}).~\footnote{In Balle et al.  \cite{Balle:2019:AICC}  the LDP mechanism and the shuffler do not commute, but that is because  they use a notion of attacker 
\review{that, in addition to knowing the true values of all records but one, it can also observe whether users reports 
their true values or not (except for the user under attack).}}}

\item
\review{We investigate leakage under uninformed adversaries for $k{=}2$ (Section~\ref{sec:single}), and
derive the first exact, closed  formulas for  posterior vulnerability, 
and therefore for leakage \review{(Theorems~\ref{theorem:target_S2} and \ref{theorem:target_SN2_fast})}.} 


\item 
\review{We investigate leakage under uninformed adversaries for generic values $k{\geq}2$ (Section~\ref{sec:single_genk}).
Although we derive the first formulas for the vulnerabilities in this case (Propositions~\ref{proposition:target_Sk} and \ref{proposition:target_SNk}),} 
they are neither closed nor computationally efficient. Nevertheless, we are able to provide \review{novel} asymptotic bounds on 
leakage \review{(Theorem~\ref{proposition:target_Sk})} by uncovering a surprising connection between our
scenario of interest and a well-known combinatoric problem.

\item We use the above formulas to \review{study leakage} as the  size of the dataset increases, for various privacy parameters of $k$-RR \review{(Sections~\ref{sec:analyses-leakage-behavior-A}, \ref{sec:analyses-leakage-behavior-B} and \ref{sec:asymptotic})}.  

\item \review{We provide a brief discussion on how our QIF model,
being parametric on the adversary's prior knowledge, can be used to reason about the strong adversary from differential privacy  (Section~\ref{sec:abo}).}

\item We show that in the case of \review{uninformed adversaries,} shuffling is much more effective than noise for leakage reduction, hence the best trade-off between privacy and utility may be obtained by using the shuffling alone. In contrast, in the case of the \review{strong adversary,} noise obfuscation plays a crucial role in privacy protection, hence the best trade-off is achieved by combining noise and shuffling  \review{(Sections~\ref{sec:single} and \ref{sec:single_genk})}.   
\end{itemize}

\subsection{Related Work}
The shuffle model was the core idea in the Encode, Shuffle, Analyze (ESA) model introduced by Bittau et al. \cite{Bittau:2017:SOSP} (see also \cite{Erlingsson:20:ArXiv} for a revised version of that work). 
Cheu et al. ~\cite{Cheu:19:EUROCRYPT} formalized the definition of the shuffle model and also provided the first separation result showing
that the shuffle model is strictly between the central and the local models of DP. 
Characterizing the exact nature of this separation has been the aim of many subsequent works.
Erlingsson et al.~\cite{Erlingsson:2019:SDA} showed that the introduction of a trusted shuffler amplifies the privacy guarantees against an adversary who is not able to access the outputs from the local randomisers but only sees the shuffled output.
Balle et al.~\cite{Balle:2019:AICC} improved and generalized the results by Erlingsson et al. to a wider range of parameters, and provided a
family of methods to analyze privacy amplification in the shuffle model.
Feldman et al.~\cite{Feldman:21:FOCS} 
developed a similar concept of ``hiding in the crowd'', and suggested an asymptotically optimal dependence of the privacy amplification on the privacy parameter of the local randomizer.
\review{Koskela et al.~\cite{koskela2021tight, koskela2021tightdiscrete} proposed a numerical approach (i.e., not expressed by an analytical formula) to estimate tight bounds based on weak adversaries.}

Other directions of research related to shuffle models address summation queries~\cite{balle2019differentially,balle2020private,Cheu:19:EUROCRYPT, ishai2006cryptography,ghazi2019scalable,balle2019improved} and histogram queries~\cite{balcer2019separating, cheu2021differentially}. 
Balle et al.~\cite{Balle:2019:AICC} proposed a single-message protocol for messages in the interval $[0,1]$, and Cheu et al.~\cite{Cheu:19:EUROCRYPT} conducted a study on a bounded real-valued statistical queries using additional communication costs. 
\review{Ishai et al.}~\cite{ishai2006cryptography} analyzed a protocol that reduced the number of messages in the summation query under shuffle models, and Balcer et al. \cite{balcer2019separating} proposed a shuffle mechanism for histogram queries.

\review{Further relevant work involves robust shuffle differential privacy~\cite{balcer2021connecting, cheu2021differential, cheu2021limits}.} Shuffle models can provide a targeted level of privacy protection only with at least a specific number of users participating in the shuffle. 
If the number of data providers does not reach a certain quantity, the level of privacy protection degrades. 
Thus, studies on robust shuffle differential privacy
have gained recent attention in the community. Balcer et al. \cite{balcer2021connecting} explores robust shuffle private protocols and suggests a relationship between robust shuffle privacy and pan-privacy. 

\review{
Connections between QIF and differential privacy have been explored in the literature.
Barthe and K\"opf~\cite{Barthe:11:CSF} relate centralized differential privacy and information-theoretic notions of leakage, establishing bounds on the former in terms of the value of $\epsilon$.
In a similar line of work, Alvim et al.~\cite{Alvim:15:JCS} employ QIF to analyze leakage and utility in oblivious, centralized differentially-private mechanisms. They improve some of the bounds by Barthe and K\"opf, and provide a mechanism that, if some constraints are satisfied, maximizes utility for a given level of differential privacy.
Both of these works, however, focus on the centralized model of differential privacy (rather than on the local model, as we do), and do not consider shuffling at all.
Chatzikokolakis et al.~\cite{Chatzikokolakis:21:JCP} investigate 
leakage orderings induced by different differential privacy mechanisms, 
but, again, do not consider  shuffling. 
To the best of our knowledge, our work is the first to provide a QIF analysis of the combination of locally differentially-privacy and shuffle mechanisms. Moreover, we provide closed formulas (rather than bounds) to compute leakage in practical scenarios.
}

\subsection{Plan of the paper}
\review{In Section~\ref{prelim}, we review fundamentals from QIF, LDP, $k$-RR, and the shuffle model.
In Section~\ref{sec:model}, we provide a QIF model for $k$-RR and the shuffler as channels and investigate different combinations of such channels.
In Section~\ref{sec:single}, we study leakage under an adversary with
a uniform prior over datasets focusing on a single target, for $k{=}2$.
In Section~\ref{sec:abo}, we briefly consider the typical strong adversary from differential privacy also for $k{=}2$.
In Section~\ref{sec:single_genk}, we extend our investigation from Section~\ref{sec:single} to generic values of $k 
{\geq}2$.
In Section~\ref{sec:discussion}, we present consequences of our findings,
and in Section~\ref{conclusion}, we conclude.
}




\section{Preliminaries} \label{prelim}
Firstly, we review the analytical framework of \emph{quantitative information flow}.
While we introduce fundamental notation and vocabulary, the definitive resource on quantitative information flow with definitions, theorems, and proofs can be found in \cite{alvim2020QIF}.
Secondly, we review $k$-RR local differential privacy along with the shuffle model.
Readers familiar with both fields should feel free to move to the following section.

\subsection{QIF}\label{sec:qif_prelim}

\edits{The quantitative information flow (QIF) framework captures the adversary’s knowledge, goals, and capabilities, and from that, quantifies the leakage of information caused by a corresponding optimal inference attack.
The framework is grounded on sound information-- and decision-theoretic principles enabling the
rigorous assessment of how much information leakage a system allows
\emph{in principle}, and independently from the adversary's computational power~\cite{Clark:01:ENTCS,Smith:09:FOSSACS,McIver:10:ICALP,alvim2020QIF}.
Hence, QIF guarantees hold no matter the particular tactic
or algorithm the adversary employs to execute the attack, 
as what is measured is exactly how much
sensitive information is leaked by the best possible 
such tactic or algorithm. 
}

\edits{QIF separates
(1) the adversary's knowledge (modeled as a prior distribution on secret values) from
(2) her intentions and capabilities (modeled as a gain function), and that is separated
from 
(3) the description of the system being run (modeled as an information-theoretic channel).
We describe these components in more detail now.
}

QIF assumes the input has a probability distribution $\pi$ that is known to an adversary.
\review{We denote the random variables associated with the
channel's input and output as $X$ and $Y$, respectively.}
The system is modeled as a \review{channel} matrix \Conc{C}, where $\Conc{C}_{x, y}$ contains the conditional probability $Pr(y \mid x)$.
We assume the adversary knows how the channel works as well as the entries in $\Conc{C}_{x, y}$.
Knowing the distribution on secrets and the channel matrix, she can update her knowledge about $X$ to a posterior distribution $Pr(x \mid y)$.
Since each output $y$ also has a probability $Pr(y)$, the channel matrix \Conc{C} provides a mapping from any prior $\pi$ to \emph{distributions on posterior distributions}, which we call a
hyper-distribution and denote $\hyperDist{\channel{C}}$.

There is no one ``right'' way to measure how a system affects a secret since the adversary's probability of success depends on their goals and operational context.
To address this, QIF uses the $g$-leakage framework, introduced by \cite{Alvim:12:CSF}.
This framework defines the \emph{vulnerability} of secret $X$ with respect to specific operational scenarios.
An adversary is given a set $\calw$ of actions (or guesses) that she can make about the secret,
and given a \emph{gain function} $g(w,x)$ 
\edits{ranging over non-negative reals} 
which defines the gain of selecting the action $w$ when the real secret is $x$.\edits{~\footnote{In principle, the return value of a gain function may be negative, as long as the corresponding vulnerability for all priors is non-negative. This restriction is imposed so the ratio between posterior and prior vulnerabilities 
remains meaningful. However, the restriction \review{can be met by introducing} an action in the gain function having gain value of 0 for all secrets, representing, e.g., the action of not performing an attack. Another solution is to shift the gain function by adding a \review{suitable} constant positive real number to its return value.
}}
An optimal adversary will choose an action that maximizes her expected gain with respect to $\pi$.
The gain function then determines a secret's vulnerability.

Given a prior distribution on secrets $\pi$, prior $g$-vulnerability, denoted $\priorGVuln{\gain}{\pi}$, represents the adversary's \emph{expected gain} of her optimal action based only on $\pi$, i.e., before \review{observing the channel output}. 

\begin{restatable}[Prior vulnerability]{definition}{priorvuln}\label{def:prior_g_v}
Given a prior $\pi$ and a gain function $g$, the corresponding
\emph{prior vulnerability} is given by
\begin{equation}
	\mathit{V_g (\pi) \defeq \max_{w \in \mathcal{W}} \sum_{x\in \mathcal{X}}\pi_x \cdot g(w,x)}\, .
\end{equation}
\end{restatable}

In the posterior case, the adversary observes the output of the system which allows her to improve her action and consequent expected gain. 
\begin{restatable}[Posterior vulnerability]{definition}{postvuln}\label{def:posterior_g_v}
Given a prior $\pi$, a gain function $g$, and channel matrix \Conc{C} from $\calx$ to $\caly$, the corresponding \emph{posterior vulnerability} is given by
\begin{equation}
	\postGVuln{\gain}{\channel{C}} \defeq \sum_{y\in \mathcal{Y}}\max_{w \in \mathcal{W}} \sum_{x\in \mathcal{X}} \pi_x \cdot \channel{C}_{xy} \cdot g(w,x)\, .
\end{equation}
\end{restatable}

\review{The choice of the gain function $g$ covers a vast variety of adversarial 
scenarios~\cite{Alvim:12:CSF}.
Indeed, \emph{any} vulnerability
function satisfying a set of basic information-theoretic axioms can be expressed as $V_g$ for a properly constructed $g$~\cite{Alvim:16:CSF}.}

We measure the channel leakage by comparing the prior and posterior $g$-vulnerability, which quantifies how much a specific channel \Conc{C} \emph{increases} the
vulnerability of the system.
This comparison can be done additively or multiplicatively.
Channel matrices can be large and difficult to evaluate, necessitating simplification.
It is interesting, hence, to consider the concept of channel
equivalence w.r.t.\ information leakage properties.
Two channels $\Conc{C}$ and $\Conc{C}'$ defined on the same input set
(but possibly different output sets) are considered \emph{equivalent},
denoted by $\Conc{C} \equiv \Conc{C}'$, iff, for all priors
$\pi$ on their input set and gain function $g$, their posterior
vulnerabilities are the same, i.e., 
$\postGVuln{\gain}{\channel{C}} = \postGVuln{\gain}{\channel{C}'}$.
(Notice that if the posterior vulnerabilities for both
channels are same, then so will be their corresponding 
multiplicative and additive leakages, 
since the channels share the same prior vulnerability.)
One way to simplify a channel matrix into an equivalent one
is to adjust extraneous structure by deleting output labels and adding similar columns together (columns that are scalar multiples of each other)\review{, since output labels do not affect leakage.}

Importantly for this work, channel matrices can compose in \emph{cascades} such that the output of one channel becomes the input for another.
As defined in ~\cite{alvim2020QIF},
\begin{definition}[Channel cascade]\label{def:cascade}
    Given channel matrices $\Conc{C}: \mathcal{X} \rightarrow \mathcal{Y}$ and $\Conc{D}: \mathcal{Y} \rightarrow \mathcal{Z}$, the \emph{cascade} of \Conc{C} and \Conc{D} is the channel matrix \Conc{CD} of type $\mathcal{X} \rightarrow \mathcal{Z}$, where \Conc{CD} is given by ordinary matrix multiplication.
\end{definition}
Cascades model sequential operations, where the adversary observes the final output.
In the same way that $\Conc{C}_{x, y}$ specifies $Pr(y \mid x)$, a cascade $\Conc{CD}: \mathcal{X} \rightarrow \mathcal{Z}$ specifies $Pr(z \mid x)$.
From the perspective of information leakage, cascades have an important operational significance: \Conc{D} here acts as a sanitization policy, suppressing the release of $Y$.

In fact, by the data-processing inequality for the $g$-leakage framework (expressed in Theorem~\ref{theo:dpi} below),
for any \Conc{D} in cascade \Conc{CD},
we know that leakage can never be increased.
To understand this property, let us define the \textit{refinement relation} among channels of the same
input space as $\Conc{C} \sqsubseteq \Conc{C}'$, meaning
that $\Conc{C}$ is refined by $\Conc{C}'$ (or, equivalently, that $\Conc{C}'$ refines $\Conc{C}$) iff, for all priors
$\pi$ on their input set and gain function $g$, their posterior
vulnerabilities satisfy
$\postGVuln{\gain}{\channel{C}} \geq \postGVuln{\gain}{\channel{C}'}$.
Then the data-processing inequality for
the $g$-leakage framework is established as follows~\cite{alvim2020QIF}.

\begin{theorem}[Data-processing inequality]\label{theo:dpi}
Given channel matrices $\Conc{C}: \mathcal{X} \rightarrow \mathcal{Y}$ and $\Conc{C}': \mathcal{X} \rightarrow \mathcal{Z}$, we have that
$\Conc{C} \sqsubseteq \Conc{C}'$ iff there
exists a channel matrix 
$\Conc{D}: \mathcal{Y} \rightarrow \mathcal{Z}$ such that
$\Conc{C}\Conc{D} = \Conc{C}'$.
\end{theorem}

\subsection{Local Differential Privacy (LDP)}
Let $\mathcal{K}$ be the set of values \review{for a single individual's sensitive attribute,}
and $k$ be the size of $\mathcal{K}$. We say that a mechanism $\mathcal{R}$ is $\epsilon$-LDP if, for all $x,x',y\in \mathcal{K}$, we have 
\[
Pr( \mathcal{R}(x) = y) \leq e^\epsilon \; \edits{Pr( \mathcal{R}(x') = y)}\, , 
\]
where \edits{$Pr( a)$} represents the probability of the event $a$ and $e$ is the exponentiation operator. 
\subsection{$k$-ary Randomized Response ($k$-RR)}
The $k$-ary Randomized Response with privacy parameter $\epsilon$ is 
\review{a mechanism} $\mathcal{R}$ defined as:
\begin{equation}\label{eqn:kRR}
Pr(\mathcal{R}(x) = y)  = \left\{
\begin{array}{rl}
\review{\nicefrac{\displaystyle e^\epsilon}{\displaystyle (k-1+e^\epsilon)},} & \review{\text{if $y = x$,}}\\
\
\\
\review{\nicefrac{\displaystyle 1}{\displaystyle (k-1+e^\epsilon)},} & \review{\text{otherwise.}}
\end{array}
\right.
\end{equation}
It is easy to prove that such a mechanism satisfies $\epsilon$-DP \cite{wang2016using}. 
\subsection{Shuffler, simplified version}

We consider only the single-message shuffler model. 
We can define the shuffler as a mechanism $\mathcal{R}$ that takes as input a tuple of elements of $\mathcal{K}$ of some fixed length $n$, and produces a random permutation of it, with uniform probability.
Namely, for  all $(x_0,x_1,\ldots,x_{n-1})\in \mathcal{K}^{n}$ and all permutations 
$(y_0,y_1,\ldots,y_{n-1})$ of $(x_0,x_1,\ldots,x_{n-1})$, we have
\[
Pr(\mathcal{R}(x_0,x_1,\ldots,x_{n-1}) = (y_0,y_1,\ldots,y_{n-1})) \; = \; \frac{1}{n!}\;.
\]



\section{QIF model for local differential privacy and shuffling} \label{sec:model}

\review{In this section, we provide a formal model based on the QIF framework
for the application of LDP mechanisms 
and of shuffle mechanisms to datasets containing sensitive
values.
The model will be instrumental in the next sections, where
we investigate information leakage properties of 
these mechanisms and of their various compositions in different configurations of the adversary's prior knowledge.}

\subsection{Sensitive values, datasets, and the general scenario}
\label{sec:model-datasets}

We begin by formalizing the general scenario we consider.

\begin{definition}[Sensitive values and datasets]\label{def:datasets}
    Let $\caln = \{0,1,\ldots,n-1\}$ be a set of $n\geq1$ \emph{individuals of interest},
    and $\calk$ be a set of $k \geq 2$ possible values for each individual's \emph{sensitive attribute}.
    A \emph{dataset} $x$ is a tuple $(x_{0},x_{1},\ldots,x_{n-1})$ 
    in which each element $x_{i} \in \calk$ is the value of the sensitive attribute
    for individual $i\in\caln$.
    The domain of all possible datasets is, hence, $\calk^{n}$, and it has size $k^{n}$.
\end{definition}

Moreover, we adopt the following notation and terminology.
Given a set $\calk = \{\kappa_{0},\kappa_{1},\ldots,\kappa_{k-1}\}$ 
of values for the sensitive attribute, and a dataset $x \in \calk^{n}$, 
we let
\begin{itemize}
    \item the tuple $x=(x_{0},x_{1},\ldots,x_{n-1})$ be denoted by the sequence
    $x_{0}x_{1} \ldots x_{n-1}$ of its elements; 
    \item $n_{\kappa_{j}}(x)$ represent the the number of 
    individuals in the dataset $x$ having sensitive value $\kappa_{j} \in \calk$;
    \item \edits{$h(x) = (\kappa_{0}{:}n_{\kappa_{0}}(x), \,\, \kappa_{1}{:}n_{\kappa_{1}}(x), \,\,
    \ldots,
    \,\,
    \kappa_{k-1}{:}n_{\kappa_{k-1}}(x))$} be the
    the histogram of dataset $x$, containing a list of each possible value
    $\kappa_{j}\in\calk$ followed by the count $n_{\kappa_{j}}(x)$ of
    individuals in $x$ with that value; and 
    \item $\#h(x) = |\{ x' \in \calk^{n} \mid h(x') = h(x) \}|$ be the number
    of distinct datasets in $\calk^{n}$ having the same histogram as $x$.
\end{itemize}

We consider a 
dataset 
$x \in \calk^{n}$ 
containing the real value of 
the sensitive attribute for all individuals in $\caln$.
We also assume that a data analyst wants to infer 
some statistical information about the original dataset 
(e.g., 
an average or count), 
but the dataset's exact contents --in particular, the link
between individuals and their sensitive attribute-- are considered secret.

\begin{example}[Running example]\label{exa:running_initial}
Consider a scenario in which there are $n\,{=}\,3$ individuals 
and a sensitive binary attribute with possible values in $\calk = \{\ta,\tb\}$. 
The set of all possible datasets is then
$\{\ta,\tb\}^{3}= \{\ta\ta\ta, \ta\ta\tb, \ta\tb\ta, \ta\tb\tb, \tb\ta\ta, \tb\ta\tb, \tb\tb\ta, \tb\tb\tb\}.$
\end{example}

\subsection{Sanitization mechanisms as channels}
\label{sec:model-mechanisms-datasets}

We consider two sanitization mechanisms
which can be employed either individually or in combination.
The $k$-RR mechanism is applied to
each individual's sensitive value 
to introduce
uncertainty, while
the shuffle mechanism permutes dataset entries 
to obfuscate the link between each sensitive value 
and its owner.

\paragraph{The shuffle channel}
The full channels of these mechanisms are quite unwieldy.
For example, the shuffle mechanism, with the input dataset \ta\ta\tb, can output \ta\tb\tb, \ta\tb\ta, or \tb\ta\ta, each with probability $1/3$. 
Formally, the shuffle randomly permutes
the 
entries 
to obfuscate 
the link between each sensitive value and its owner.
Abstracting from implementation details and assuming perfect shuffling, 
each input dataset $x$ has an equal probability $\nicefrac{1}{\#h(x)}$ 
of mapping
to any of the $\#h(x)$ possible output
datasets with the same histogram.

\begin{definition}[Full shuffle channel]
\label{def:shuffling-channel-full}
Let $\caln = \{0,1,\ldots,n-1\}$ be a set of $n\geq1$ individuals
and $\calk$ be a set of $k \geq 2$ sensitive values.
A \emph{full shuffle channel} $\Conc{S}$ 
is a channel from $\calk^{n}$ to $\calk^{n}$ s.t.,
for all $x \in \calk^{n}$ and $y \in \calk^{n}$,
\begin{align*}
\Conc{S}_{x,y} =&\,\, 
\begin{cases}
        \nicefrac{1}{\#h(x)}, & \text{if $h(y)=h(x)$}, \\
        0, & \text{otherwise}.
        \end{cases}
\end{align*}
\end{definition}

Table~\ref{tab:full-channel-shuffling} contains the 
full channel $\Conc{S}$ representing the application of 
shuffling to the scenario of Example~\ref{exa:running_initial}.

Conveniently, there are symmetries here that we can exploit to
simplify the shuffle channel into something more tractable.
Notice that once an input dataset $x$ is 
shuffled into 
an output dataset $y$,
the position of each element in $y$ no longer identifies
its owner, and the only information preserved is $x$'s histogram.
For instance, in Table~\ref{tab:full-channel-shuffling},
the columns corresponding to datasets
$\ta\ta\tb$, 
$\ta\tb\ta$,
and 
$\tb\ta\ta$
are identical and,
from the point of view of the adversary, 
convey the same information.
Hence, these columns can be all merged into a new
column that represents only the histogram
\ta{:}2, \tb{:}1 that these datasets share.
This motivates the definition of
a reduced shuffle channel below, whose
output is simply the input dataset's histogram.

\begin{definition}[Reduced shuffle channel]
\label{def:shuffling-channel-reduced}
Let \Conc{S} be a full shuffle channel as per Def.~\ref{def:shuffling-channel-reduced}.
A \emph{reduced shuffle channel} $\Conc{S}^{r}$
corresponding to $\Conc{S}$ is a channel from $\calk^{n}$ to histograms on $\calk$ s.t.,
for all $x \in \calk^{n}$ and histogram $z$,
\begin{align*}
\Conc{S}^{r}_{x,z} \,\,\,=\,\,\,
\sum_{y \in \calk^{n}: h(y)=z} \Conc{S}_{x,y} \,\,\,=\,\,\,
\begin{cases}
        1, & \text{if $h(x)=z$}, \\
        0, & \text{otherwise}.
        \end{cases}
\end{align*}
\end{definition}

Table~\ref{tab:reduced-channel-shuffling} 
contains the channel $\Conc{S}^{r}$ representing the application of reduced 
shuffle to the scenario of Example~\ref{exa:running_initial}.

We can be confident that this simplification does not alter our analysis.
It is known from QIF literature that 
merging 
columns that are multiples of (or
identical to)
each other does not alter the leakage properties of a channel~\cite{alvim2020QIF}.
The original and resulting channels 
are
\emph{equivalent},
in the sense that, for every $g$-vulnerability 
measure and prior distribution on secret values, both yield the same quantification of information leakage.
Since $\Conc{S}^{r}$ is obtained from $\Conc{S}$ by merging
similar columns, these channels are equivalent.

\begin{restatable}[Equivalence between full and reduced shuffle]{proposition}{equivalentS}\label{prop:equivalence-S}
Let $\Conc{S}$ be a full shuffle channel as per Def.~\ref{def:shuffling-channel-full},
and $\Conc{S}^{r}$ be the reduced shuffle channel obtained from $\Conc{S}$ 
as per Def.~\ref{def:shuffling-channel-reduced}.
Then
\begin{align*}
  \Conc{S} \equiv \Conc{S}^{r}.
\end{align*}
\end{restatable}


\paragraph{The $k$-RR channel}

The $k$-RR mechanism independently randomizes 
the sensitive value of each entry of the dataset, 
and outputs the resulting dataset.
This process does not break the link between any 
individual and their sensitive value, but creates 
uncertainty about whether the reported
value for each individual is accurate.
To simplify notation, we will use $p$ to denote the probability 
representing that an individual's sensitive value
is reported accurately, i.e., $p= \nicefrac{e^\epsilon}{k-1+e^\epsilon}$ (cf. Equation~\ref{eqn:kRR}), and  $\overline{p}$ to represent $1{-}p$, i.e., the probability that the true value is swapped with some other one.

The $k$-RR mechanism can be modeled
as a channel that 
probabilistically maps each input dataset $x \in \calk^{n}$ to 
an output dataset $y\in\calk^{n}$ as follows.

\begin{definition}[$k$-RR channel]
\label{def:krr-channel}
Let $\caln = \{0,1,\ldots,n-1\}$ be a set of $n\geq1$ individuals
and $\calk$ be a set of $k \geq 2$ sensitive values.
A \emph{$k$-RR channel} $\Conc{N}$ (for \qm{\underline{n}oise}) with parameter
$p\in[\nicefrac{1}{k},1]$ is a channel from $\calk^{n}$ to $\calk^{n}$ s.t., 
for all $x \in \calk^{n}$ and $y \in \calk^{n}$,
\begin{align*}
\Conc{N}_{x,y} =&\,\, \prod_{i=0}^{n-1} Pr(y_{i} \mid x_{i}), 
\end{align*}
where
\begin{align*}
Pr(y_{i} \mid x_{i}) = 
\begin{cases}
p, & \text{if $y_{i}=x_{i}$}, \\
\nicefrac{\overline{p}}{(k-1)}, & \text{if $y_{i} \neq x_{i}$}
\end{cases}
\end{align*}

\end{definition}

Notice that 
$p$ is at least $\nicefrac{1}{k}$ 
so $p \geq \nicefrac{\overline{p}}{(k-1)}$, indicating that the real 
sensitive value never has a lower probability to be reported than any other value.
Table~\ref{tab:full-channel-krr} contains the 
full channel $\Conc{N}$ representing the application of a
$k$-RR mechanism to the scenario of Example~\ref{exa:running_initial}.

\begin{table}[tb]
\centering
\begin{subtable}{\linewidth}
    \centering
\begin{scriptsize}
    \begin{tabular}{c|c c c c c c c c}
		\Conc{N} & \ta\ta\ta & \ta\ta\tb & \ta\tb\ta & \tb\ta\ta &\ta\tb\tb &  \tb\ta\tb& \tb\tb\ta& \tb\tb\tb\\ 
		\ta\ta\ta & \colorcell{Nc1}{$p^3$} &	
        \colorcell{Nc5}{$p^2 \overline{p}$} &	
        \colorcell{Nc5}{$p^2 \overline{p}$} &	
        \colorcell{Nc5}{$p^2 \overline{p}$} & 
        \colorcell{Nc8}{$p \overline{p}^2$} &
         \colorcell{Nc8}{$p \overline{p}^2$} &	 
         \colorcell{Nc8}{$p \overline{p}^2$} &	
         \colorcell{Nc4}{$\overline{p}^3$} 
        \\ 
		\ta\ta\tb & \colorcell{Nc5}{$p^2 \overline{p}$} &	\colorcell{Nc1}{$p^3$} &	 \colorcell{Nc8}{$p \overline{p}^2$} &		 \colorcell{Nc8}{$p \overline{p}^2$} &\colorcell{Nc5}{$p^2 \overline{p}$} &	\colorcell{Nc5}{$p^2 \overline{p}$} &	\colorcell{Nc4}{$\overline{p}^3$} &	 \colorcell{Nc8}{$p \overline{p}^2$} \\
		\ta\tb\ta & \colorcell{Nc5}{$p^2 \overline{p}$} &	 \colorcell{Nc8}{$p \overline{p}^2$} &	\colorcell{Nc1}{$p^3$} &		 \colorcell{Nc8}{$p \overline{p}^2$} &\colorcell{Nc5}{$p^2 \overline{p}$} &	\colorcell{Nc4}{$\overline{p}^3$} &	\colorcell{Nc5}{$p^2 \overline{p}$} &	 \colorcell{Nc8}{$p \overline{p}^2$} \\
		\tb\ta\ta & \colorcell{Nc5}{$p^2 \overline{p}$} &	 \colorcell{Nc8}{$p \overline{p}^2$} &	 \colorcell{Nc8}{$p \overline{p}^2$} &		\colorcell{Nc1}{$p^3$} & \colorcell{Nc4}{$\overline{p}^3$} &	\colorcell{Nc5}{$p^2 \overline{p}$} &	\colorcell{Nc5}{$p^2 \overline{p}$} &	 \colorcell{Nc8}{$p \overline{p}^2$} \\
        \ta\tb\tb &  \colorcell{Nc8}{$p \overline{p}^2$} &	\colorcell{Nc5}{$p^2 \overline{p}$} &	\colorcell{Nc5}{$p^2 \overline{p}$} &		\colorcell{Nc4}{$\overline{p}^3$} &\colorcell{Nc1}{$p^3$} &	 \colorcell{Nc8}{$p \overline{p}^2$} &	 \colorcell{Nc8}{$p \overline{p}^2$} &	\colorcell{Nc5}{$p^2 \overline{p}$} \\
		\tb\ta\tb&  \colorcell{Nc8}{$p \overline{p}^2$} &	\colorcell{Nc5}{$p^2 \overline{p}$} &	\colorcell{Nc4}{$\overline{p}^3$} &	 	\colorcell{Nc5}{$p^2 \overline{p}$} &\colorcell{Nc8}{$p \overline{p}^2$} &	\colorcell{Nc1}{$p^3$} &	 \colorcell{Nc8}{$p \overline{p}^2$} &	\colorcell{Nc5}{$p^2 \overline{p}$} \\
        \tb\tb\ta&  \colorcell{Nc8}{$p \overline{p}^2$} &	\colorcell{Nc4}{$\overline{p}^3$} &	\colorcell{Nc5}{$p^2 \overline{p}$} &	 	\colorcell{Nc5}{$p^2 \overline{p}$} &	\colorcell{Nc8}{$p \overline{p}^2$} & \colorcell{Nc8}{$p \overline{p}^2$} &	\colorcell{Nc1}{$p^3$} &	\colorcell{Nc5}{$p^2 \overline{p}$} \\
		\tb\tb\tb & \colorcell{Nc4}{$\overline{p}^3$} &	 \colorcell{Nc8}{$p \overline{p}^2$} &	 \colorcell{Nc8}{$p \overline{p}^2$} &		 \colorcell{Nc8}{$p \overline{p}^2$} &	\colorcell{Nc5}{$p^2 \overline{p}$} &\colorcell{Nc5}{$p^2 \overline{p}$} &	\colorcell{Nc5}{$p^2 \overline{p}$} &	\colorcell{Nc1}{$p^3$} 
\end{tabular}
\end{scriptsize}
    \vspace{4mm}
    \caption{The $k$-RR channel \Conc{N} 
    where $p$ is the probability a user responds with their true value, and $\overline{p}=1-p$.
    Cells are shaded according to the entries' values when $p = 0.75$.}
    \label{tab:full-channel-krr}
\end{subtable}
\\ \vspace{2mm}
\begin{subtable}{\linewidth}
    \centering
    \begin{tabular}{c|c c c c c c c c}
        $\Conc{S}$ & \ta\ta\ta & \ta\ta\tb & \ta\tb\ta & \ta\tb\tb & \tb\ta\ta & \tb\ta\tb & \tb\tb\ta & \tb\tb\tb \\ 
        \hline
        \ta\ta\ta & 1 &	0 &	0 &	0 &	0 &	0 &	0 &	0 \\ 
        \ta\ta\tb & 0 &	\NF{1}{3} &	\NF{1}{3} &	0 &	\NF{1}{3} &	0 &	0 &	0 \\
        \ta\tb\ta & 0 &	\NF{1}{3} &	\NF{1}{3} &	0 &	\NF{1}{3} &	0 &	0 &	0 \\
        \ta\tb\tb & 0 &	0 &	0 &	\NF{1}{3} &	0 &	\NF{1}{3} &	\NF{1}{3} &	0 \\
        \tb\ta\ta & 0 &	\NF{1}{3} &	\NF{1}{3} &	0 &	\NF{1}{3} &	0 &	0 &	0 \\
        \tb\ta\tb & 0 &	0 &	0 &	\NF{1}{3} &	0 &	\NF{1}{3} &	\NF{1}{3} &	0 \\
        \tb\tb\ta & 0 &	0 &	0 &	\NF{1}{3} &	0 &	\NF{1}{3} &	\NF{1}{3} & 0 \\
        \tb\tb\tb & 0 &	0 &	0 &	0 &	0 &	0 &	0 &	1 
    \end{tabular}
    \caption{Full shuffle channel $\Conc{S}$ 
    which receives datasets as input and produces a datasets as output.}
    \label{tab:full-channel-shuffling}
\end{subtable}
\\ \vspace{2mm}
\begin{subtable}{\linewidth}
    \centering
    \begin{tabular}{c|c c c c}
	$\Conc{S}^{r}$ & \ta{:}3, \tb{:}0  & \ta{:}2, \tb{:}1 &	\ta{:}1, \tb{:}2 & \ta{:}0, \tb{:}3 \\
	\hline
        \ta\ta\ta &	1 &	0 &	0 &	0 \\ 
        \ta\ta\tb &	0 & 1 &	0 &	0\\
        \ta\tb\ta &	0 &	1 &	0 &	0\\
        \ta\tb\tb &	0 &	0 &	1 &	0\\
        \tb\ta\ta &	0 &	1 &	0 &	0\\
        \tb\ta\tb &	0 &	0 &	1 &	0\\
        \tb\tb\ta &	0 &	0 &	1 &	0\\
        \tb\tb\tb & 	0 &	0 &	0 &	1 
    \end{tabular}
    \caption{Reduced shuffle channel $\Conc{S}^{r}$ 
    which receives 
    datasets as inputs and produces histograms as outputs.}
    \label{tab:reduced-channel-shuffling}
\end{subtable}
    \caption{Examples of channels for the $k$-RR and shuffling
    mechanisms, with $k=2$ possible sensitive values and $n=3$ individuals.}
    \label{tab:full-channels}
\end{table}

\subsection{The combination of sanitization mechanisms}
\label{sec:model-sanitization-pipeline}

The sanitization mechanisms presented in the previous section can be
applied to an input dataset either in isolation or in combination.
In the QIF
framework, the 
effect of this combination is captured by 
channel cascading (Def.~\ref{def:cascade}).
We shall now cover the various possibilities.

We begin by analyzing the typical pipeline used in the literature, 
which
first applies the noisy $k$-RR mechanism to 
the original dataset, and then applies a shuffle mechanism 
to the result of the first sanitization~\cite{Bittau:2017:SOSP}.
The channel cascade $\Conc{N}\Conc{S}$ mirrors this process:
the datasets are processed by $\Conc{N}$, 
producing a randomized intermediate dataset, which are taken as input
into \Conc{S} which shuffles the entries and produces a final output dataset.
The matrix itself can be easily generated by matrix multiplication.
%

Interestingly, in our scenario the order of application of the full mechanisms
$\Conc{N}$ and $\Conc{S}$ is irrelevant, in the sense
that the resulting channels will be equivalent w.r.t.\ 
the information leakage they cause.
This commutativity property is 
formalized in the following result.

\begin{restatable}[Commutativity of $k$-RR and shuffle: Full case]{proposition}{commutativityFull}\label{theorem:commutativity-full}
Let $\caln = \{0,1,\ldots,n-1\}$ be a set of $n\geq1$ individuals
and $\calk$ be a set of $k \geq 2$ values for the sensitive attribute.
Let also $\Conc{N}$ be a $k$-RR channel as per Definition~\ref{def:krr-channel},
and $\Conc{S}$ be a full shuffle channel per Definition~\ref{def:shuffling-channel-full}.
Then 
\begin{align*}
    \Conc{N}\Conc{S} \equiv \Conc{S}\Conc{N}.
\end{align*}
\end{restatable}

Consider again the scenario of Example~\ref{exa:running_initial}.
The channel $\Conc{N}\Conc{S}$ and the channel 
$\Conc{S}\Conc{N}$ representing the application of $k$-RR followed
by full shuffle or vice-versa 
are identical, as
represented in Table~\ref{tab:NS-SN-full}.

\begin{table*}[tb]
\centering
    \begin{tabular}{@{}c|c c c c c c c c @{}}
        $\Conc{N}\Conc{S}$ / $\Conc{S}\Conc{N}$ & \ta\ta\ta & \ta\ta\tb & \ta\tb\ta &  \tb\ta\ta & \ta\tb\tb &\tb\ta\tb& \tb\tb\ta& \tb\tb\tb \\
        \ta\ta\ta &  \colorcell{NSc1}{$p^3$} &	 \colorcell{NSc5}{$p^2 \overline{p}$} &	 \colorcell{NSc5}{$p^2 \overline{p}$} &	 \colorcell{NSc5}{$p^2 \overline{p}$} &	 \colorcell{NSc8}{$p \overline{p}^2$} &   \colorcell{NSc8}{$p \overline{p}^2$} &	 \colorcell{NSc8}{$p \overline{p}^2$} &	 \colorcell{NSc4}{$\overline{p}^3$} \Tstrut \\
        \ta\ta\tb &  \colorcell{NSc5}{$p^2 \overline{p}$} &	\colorcell{NSc9}{$\NF{1}{3} (p^3 + 2p \overline{p}^2)  $} &	\colorcell{NSc9}{$\NF{1}{3} (p^3 + 2p \overline{p}^2)  $} &	\colorcell{NSc9}{$\NF{1}{3} (p^3 + 2p \overline{p}^2)  $} &	\colorcell{NSc10}{$\NF{1}{3} (2 p^2 \overline{p} + \overline{p}^3 )  $} & \colorcell{NSc10}{$\NF{1}{3} (2 p^2 \overline{p} + \overline{p}^3 )  $} &	\colorcell{NSc10}{$\NF{1}{3} (2 p^2 \overline{p} + \overline{p}^3 )  $} &	 \colorcell{NSc8}{$p \overline{p}^2$}\\
        \ta\tb\ta &  \colorcell{NSc5}{$p^2 \overline{p}$} &	\colorcell{NSc9}{$\NF{1}{3} (p^3 + 2p \overline{p}^2)  $} &	\colorcell{NSc9}{$\NF{1}{3} (p^3 + 2p \overline{p}^2)  $} &\colorcell{NSc9}{$\NF{1}{3} (p^3 + 2p \overline{p}^2)  $} &	\colorcell{NSc10}{$\NF{1}{3} (2 p^2 \overline{p} + \overline{p}^3 )  $} & \colorcell{NSc10}{$\NF{1}{3} (2 p^2 \overline{p} + \overline{p}^3 )  $} &	\colorcell{NSc10}{$\NF{1}{3} (2 p^2 \overline{p} + \overline{p}^3 )  $} &	 \colorcell{NSc8}{$p \overline{p}^2$}\\
        \tb\ta\ta &  \colorcell{NSc5}{$p^2 \overline{p}$} &	\colorcell{NSc9}{$\NF{1}{3} (p^3 + 2p \overline{p}^2)  $} &	\colorcell{NSc9}{$\NF{1}{3} (p^3 + 2p \overline{p}^2)  $} & \colorcell{NSc9}{$\NF{1}{3} (p^3 + 2p \overline{p}^2)  $} &	\colorcell{NSc10}{$\NF{1}{3} (2 p^2 \overline{p} + \overline{p}^3 )  $} & \colorcell{NSc10}{$\NF{1}{3} (2 p^2 \overline{p} + \overline{p}^3 )  $} &	\colorcell{NSc10}{$\NF{1}{3} (2 p^2 \overline{p} + \overline{p}^3 )  $} &	 \colorcell{NSc8}{$p \overline{p}^2$}\\
        \ta\tb\tb &  \colorcell{NSc8}{$p \overline{p}^2$} &	\colorcell{NSc10}{$\NF{1}{3} (2 p^2 \overline{p} + \overline{p}^3 )  $} &	\colorcell{NSc10}{$\NF{1}{3} (2 p^2 \overline{p} + \overline{p}^3 )  $} &	\colorcell{NSc10}{$\NF{1}{3} (2 p^2 \overline{p} + \overline{p}^3 )  $} &	\colorcell{NSc9}{$\NF{1}{3} (p^3 + 2p \overline{p}^2)  $} & \colorcell{NSc9}{$\NF{1}{3} (p^3 + 2p \overline{p}^2)  $} &	\colorcell{NSc9}{$\NF{1}{3} (p^3 + 2p \overline{p}^2)  $} &	 \colorcell{NSc5}{$p^2 \overline{p}$}\\
        \tb\ta\tb&  \colorcell{NSc8}{$p \overline{p}^2$} &	\colorcell{NSc10}{$\NF{1}{3} (2 p^2 \overline{p} + \overline{p}^3 )  $} &	\colorcell{NSc10}{$\NF{1}{3} (2 p^2 \overline{p} + \overline{p}^3 )  $} &	\colorcell{NSc10}{$\NF{1}{3} (2 p^2 \overline{p} + \overline{p}^3 )  $} &	\colorcell{NSc9}{$\NF{1}{3} (p^3 + 2p \overline{p}^2)  $} & \colorcell{NSc9}{$\NF{1}{3} (p^3 + 2p \overline{p}^2)  $} &	\colorcell{NSc9}{$\NF{1}{3} (p^3 + 2p \overline{p}^2)  $} &	\colorcell{NSc5}{$p^2 \overline{p}$}\\
        \tb\tb\ta&  \colorcell{NSc8}{$p \overline{p}^2$} &	\colorcell{NSc10}{$\NF{1}{3} (2 p^2 \overline{p} + \overline{p}^3 )  $} &	\colorcell{NSc10}{$\NF{1}{3} (2 p^2 \overline{p} + \overline{p}^3 )  $} &		\colorcell{NSc10}{$\NF{1}{3} (2 p^2 \overline{p} + \overline{p}^3 )  $} &	\colorcell{NSc9}{$\NF{1}{3} (p^3 + 2p \overline{p}^2)  $} & \colorcell{NSc9}{$\NF{1}{3} (p^3 + 2p \overline{p}^2)  $} &	\colorcell{NSc9}{$\NF{1}{3} (p^3 + 2p \overline{p}^2)  $} &	 \colorcell{NSc5}{$p^2 \overline{p}$}\\
        \tb\tb\tb &  \colorcell{NSc4}{$\overline{p}^3$} &	 \colorcell{NSc8}{$p \overline{p}^2$} &	 \colorcell{NSc8}{$p \overline{p}^2$} &	 	 \colorcell{NSc8}{$p \overline{p}^2$} & \colorcell{NSc5}{$p^2 \overline{p}$} &	 \colorcell{NSc5}{$p^2 \overline{p}$} &	 \colorcell{NSc5}{$p^2 \overline{p}$} &	 \colorcell{NSc1}{$p^3$} \Bstrut\\
    \end{tabular}
    \caption{Channel 
    representing both the cascade $\Conc{N}\Conc{S}$
    of $k$-RR followed by full shuffle and the cascade
    $\Conc{S}\Conc{N}$ of full shuffle followed by $k$-RR,
    with $k = 2$ possible sensitive values and $n = 3$ individuals.
    Here $p$ is the probability a user responds with their true value and $\overline{p}=1-p$.
    Cells are shaded according to the entries' values when $p = 0.75$ and $\overline{p} = 0.25$.}
    \label{tab:NS-SN-full}
\end{table*}

Finally, we notice that the equivalence between
the full $\Conc{S}$ and reduced $\Conc{S}^{r}$ versions
of shuffle channel (Prop.~\ref{prop:equivalence-S})
is carried over to the their composition with the 
$k$-RR mechanism.
This property will facilitate some proofs of our
technical results, and is formalized below.

\begin{restatable}[Equivalence of full and reduced compositions]{proposition}{equivalenceCompositionsA}\label{theorem:equivalenceCompositions-A}
Let $\caln = \{0,1,\ldots,n-1\}$ be a set of $n\geq1$ individuals
and $\calk$ be a set of $k \geq 2$ values for the sensitive attribute.
Let also $\Conc{N}$ be a full $k$-RR channel as per Definition~\ref{def:krr-channel},
$\Conc{S}$ be a full shuffle channel as per Definition~\ref{def:shuffling-channel-full},
and $\Conc{S}^{r}$ be a reduced shuffle channel obtained
from $\Conc{S}$ as per Definition~\ref{def:shuffling-channel-reduced}.
Then:
\begin{align}
    \Conc{N}\Conc{S} 
    \equiv&\,\, \Conc{N}\Conc{S}^{r} \label{eq:01a}
\end{align}
\end{restatable}
Continuing our example, the cascade $\Conc{N}\Conc{S}^r$ representing the application of $k$-RR followed
by the reduced shuffle is
represented in Table~\ref{tab:NS-SN-reducedTEXT}.

The relationship among channels discussed so far
is depicted in Fig.~\ref{fig:partial-diagram}.
Notice, however, that commutativity does not hold between
$\Conc{N}$ and $\Conc{S}^{r}$, as the composition
$\Conc{S}^{r}\Conc{N}$ is not even mathematically consistent
(the inner dimensions of the matrices do not match so multiplication
is impossible).
Commutativity can be recovered, however, if we define a
suitable reduced counterpart to the $k$-RR channel.
This possibility is explored in Appendix \ref{sec:model-reduced-krr}. 

\begin{table}
\centering
    \begin{tabular}{c| c c c c }
        $\Conc{N}\Conc{S}^{r}$ & (\ta{:}3,\tb{:}0) & (\ta{:}2,\tb{:}1) &	
        (\ta{:}1,\tb{:}2) & (\ta{:}0,\tb{:}3) \\
        \ta\ta\ta & \colorcell{NSrc1}{$p^3$}    &  \colorcell{NSrc2}{$3p^2 \overline{p}$}           & \colorcell{NSrc3}{$3p \overline{p}^2$}          &  \colorcell{NSrc4}{$\overline{p}^3$} \Tstrut\\
        \ta\ta\tb & \colorcell{NSrc5}{$p^2 \overline{p}$}  & \colorcell{NSrc6}{$p^3 + 2 p \overline{p}^2$}    & \colorcell{NSrc7}{$2p^2 \overline{p} + \overline{p}^3$}    & \colorcell{NSrc8}{$p \overline{p}^2$} \\
        \ta\tb\ta & \colorcell{NSrc5}{$p^2 \overline{p}$}  &  \colorcell{NSrc6}{$p^3 + 2 p \overline{p}^2$}    & \colorcell{NSrc7}{$2p^2 \overline{p} + \overline{p}^3$}    & \colorcell{NSrc8}{$p \overline{p}^2$} \\
        \tb\ta\ta & \colorcell{NSrc5}{$p^2 \overline{p}$}  &  \colorcell{NSrc6}{$p^3 + 2 p \overline{p}^2$}    & \colorcell{NSrc7}{$2p^2 \overline{p} + \overline{p}^3$}   & \colorcell{NSrc8}{$p \overline{p}^2$} \\
        \ta\tb\tb & \colorcell{NSrc8}{$p \overline{p}^2$}  &  \colorcell{NSrc7}{$2p^2 \overline{p} + \overline{p}^3$}    & \colorcell{NSrc6}{$p^3 + 2 p \overline{p}^2$}    &  \colorcell{NSrc5}{$p^2 \overline{p}$} \\
        \tb\ta\tb& \colorcell{NSrc8}{$p \overline{p}^2$}  &   \colorcell{NSrc7}{$2p^2 \overline{p} + \overline{p}^3$}    & \colorcell{NSrc6}{$p^3 + 2 p \overline{p}^2$}   & \colorcell{NSrc5}{$p^2 \overline{p}$} \\
        \tb\tb\ta& \colorcell{NSrc8}{$p \overline{p}^2$}  & \colorcell{NSrc7}{$2p^2 \overline{p} + \overline{p}^3$}    & \colorcell{NSrc6}{$p^3 + 2 p \overline{p}^2$}    & 
        \colorcell{NSrc5}{$p^2 \overline{p}$} \\
        \tb\tb\tb & \colorcell{NSrc4}{$\overline{p}^3$}    & \colorcell{NSrc3}{$3p \overline{p}^2$}           & \colorcell{NSrc2}{$3p^2 \overline{p}$}          & \colorcell{NSrc1}{$p^3$}
    \end{tabular}
    \caption{Channel 
    $\Conc{N}\Conc{S}^{r}$ representing $k$-RR followed by reduced shuffling, with $k = 2$ possible sensitive values and $n = 3$ individuals.
    Cells are shaded according to the entries' values when the probability $p$ a user responds with their true value is $0.75$ and $\overline{p} = 0.25$.
    Here $p^3 + 2 p \overline{p}^2$ represents the largest value at approx. 0.5156 while $\overline{p}^3$ is the smallest at approx. 0.0156.
    Interestingly, this ordering can change with $p$.}
    \label{tab:NS-SN-reducedTEXT}
\end{table}

\begin{figure}[tb]
\centering
\renewcommand{\arraystretch}{1.2}
\fbox{
$
\begin{array}{rcl}
     & \text{\small Prop.~\ref{prop:equivalence-S}} & \\
    \Conc{S} & \text{\LARGE $\equiv$} & \Conc{S}^{r} \\
\end{array}
$
}
\\
\vspace{2mm}
\fbox{
$
\begin{array}{ccccc}
    & \text{\small Prop.~\ref{theorem:commutativity-full}} & & \text{\small Prop.~\ref{theorem:equivalenceCompositions-A}} & \\
    \Conc{S}\Conc{N} & \text{\LARGE $\equiv$} & \Conc{N}\Conc{S} & \text{\LARGE $\equiv$} & \Conc{N}\Conc{S}^{r} \\
\end{array}
$
}
\caption{Relationship among compositions of full and reduced $k$-RR and shuffle. 
Here
$(\equiv)$ denotes
equivalence w.r.t.\ information leakage.}
\label{fig:partial-diagram}
\end{figure}


\review{\subsection{\review{On the priors on secrets, and on gain functions}}}
\label{sec:model-priors}

\review{
In this section we discuss the priors and gain functions employed in our analyses of the following sections.
}

\review{\textbf{On the choice of gain functions.}
In this work we focus on an adversary who has one try to correctly guess 
the value of a single individual chosen as a target,
and has maximum benefit from a correct guess, and no benefit from an incorrect one. 
The gain functions associated with this type of adversary yield what is usually called \emph{Bayes vulnerability}~\cite{alvim2020QIF}.
Given their intuitive operational interpretations and convenient mathematical properties, Bayes vulnerability and its variants have been used in many works on privacy and security~\cite{Braun:09:MPFS,Smith:09:FOSSACS,Alvim:12:CSF,Alvim:15:JCS,Jurado:21:CSF,Alvim:22:PETSa}, and our work is aligned with them.
}


\review{\textbf{On the choice of priors.}
In this work focus on what we call an \emph{uninformed adversary}. 
Besides the motivations already given in the introduction, we remark that in  QIF  uniform priors are closely linked to the maximum possible leakage that can be caused by a channel (over all priors and gain functions)~\cite{Alvim:14:CSF,alvim2020QIF}.
Hence the study of uniform priors is of particular relevance for  inference attacks. However, to illustrate the generality of our approach, 
we  also show how to apply it to the strong adversary of differential privacy and we derive numerically the leakage on some examples.}


\review{
\section{Single-target leakage of a binary value against an uninformed adversary} \label{sec:single}
}
\renewcommand{\arraystretch}{1}

In this section, we study the information leakage caused 
by different combinations 
of $k$-RR and shuffling.
We focus on an attack scenario with two central characteristics: 
(i) the adversary is attempting to guess the secret value of a 
single individual from $\caln$ chosen as the target in the dataset; and
(ii) the sensitive value can take binary values only, so
the size $\calk$ of the set of sensitive attributes is $k=2$.
\review{Moreover, we focus on an uninformed adversary,  with a uniform prior on datasets.}

We derive exact formulas to compute prior and posterior vulnerabilities, 
and therefore leakage, in this attack scenario.
Although these formulas are combinatorial in nature, 
they turn out to have computationally efficient 
equivalent formulations.
Finally, we analyze the behavior of leakage 
as the size of the dataset grows and other 
parameters vary. 
Sec.~\ref{sec:abo} discusses a variation to this single-target adversary, while
Sec.~\ref{sec:single_genk} expands our investigation to the case
of general size $k$ of the set of sensitive attributes.

\subsection{Attack scenario and leakage formulas}
\label{sec:single-scenario}

We start by providing an intuitive overview of the attack scenario.
Let $\caln = \{0,1,\ldots,n-1\}$ be the set of $n\geq1$ individuals of interest
and $\calk$ be the set of $k$ possible sensitive values.
The adversary starts off by knowing $\caln$ and $\calk$, 
but does not have 
access to the original dataset or to its sanitized and published version.
Moreover, we assume the adversary knows that all possible datasets
are equally likely a priori, so her prior knowledge is captured
by the uniform distribution defined,
for every possible dataset $x \in \calk^{n}$, as 
\begin{align}
    \label{eq:uniform-prior}
    \pi_{x} =  \nicefrac{1}{k^{n}}.
\end{align}

Without loss of generality, we assume the selected target to be
the first participant in the dataset, i.e., individual $0$.


The adversary's action consists in picking a value $\kappa \in \calk$
as her guess for the target's value $x_{0}$, 
obtaining some gain if her guess is correct, and no gain otherwise.
The adversary's prior $\pi$ on uniform datasets
implies that the adversary's a priori knowledge about
the target's secret value is a uniform distribution on 
$\kappa\in\calk$.
The goals and capabilities of this adversary are formalized 
as the following gain function.

\begin{restatable}[Single-target gain function]{definition}{targettwogf}
\label{def:target_k_gf}
Let $\calx = \calk^{n}$ be the set of all possible (secret) 
datasets, and $\calw = \calk$ be the set of actions available 
to the adversary, 
consisting in all possible guesses for the target individual's sensitive value.
The \emph{single target gain function} 
$g_{\rm T} : \calw \times \calx \rightarrow [0,1]$ 
is defined, for every action $w \in \calw$ and secret dataset $x \in \calx$,
as
\begin{align*}
    g_{\rm T}(w,x) :=&\,\, 
    \begin{cases}
    1, & \text{if $x_{0}=w$}, \\
    0, & \text{otherwise},
    \end{cases}
\end{align*}
where $x_{0}$ is the sensitive value of the first individual in $x$.
\end{restatable}

The prior vulnerability 
is 
exactly $\nicefrac{1}{k}$, representing exactly the adversary's
probability of correctly guessing the target individual's sensitive
value in one try given only her prior knowledge about the secret.

\begin{restatable}[Prior single-target vulnerability]{proposition}{singletargetprior}
\label{proposition:singletargetprior}
Given the uniform prior $\pi$ on the set $\calx = \calk^{n}$ of all possible datasets,
and the single-target gain function $g_{\rm T}$ per Def.~\ref{def:target_k_gf},
the corresponding prior vulnerability is given by
\begin{equation*}
    \SV(\pi) = \max_{w \in \calw} \sum_{x \in \calk^{n}} \pi_{x} g_{\rm T}(w,x) = \nicefrac{1}{k}.
\end{equation*}
\end{restatable}

We now turn our attention to the vulnerability of the secret after
a sanitized version of the original dataset is released and becomes
visible to the adversary.
We consider the sanitization process as consisting of the
application of 
the full $k$-RR channel \Conc{N}, 
the shuffle channel \Conc{S}, or the cascade \Conc{NS} 
representing the combination of both mechanisms.

First, let us consider posterior vulnerability when the 
$k$-RR mechanism is used in isolation.
Clearly, an adversary's optimal action is always to guess
the target's reported value 
since 
$p \geq \nicefrac{\overline{p}}{k-1}$, indicating it is always at least equally likely 
the reported value is the correct one.
Therefore, the adversary's probability of correctly guessing the target's value 
is $p$.

\begin{restatable}[Posterior single-target vulnerability under $k$-RR]{proposition}{targetnoise}\label{proposition:target_N}
Given the uniform prior $\pi$ on the set $\calk^{n}$ of all possible datasets, the single-target gain function $g_{\rm T}$ per Def.~\ref{def:target_k_gf}, and a full $k$-RR channel $\Conc{N}$ with parameter $p\in[\nicefrac{1}{k},1]$ per Def.~\ref{def:krr-channel},
the corresponding posterior vulnerability is given by
\begin{align*}
    \SV \hyperDist{\Conc{N}} = 
    \sum_{y \in \calk^{n}} \max_{w \in \calw} \sum_{x \in \calk^{n}} \pi_{x} \Conc{N}_{x,y} g_{\rm T}(w,x) = p.
\end{align*}
\end{restatable}

Now let us consider the posterior vulnerability corresponding to the 
shuffle mechanism alone.
Recall that, from Def.~\ref{def:shuffling-channel-reduced}, the reduced shuffle channel
\edits{$\Conc{S}^{r}$} simply maps each input dataset to its histogram. 
In this way, the correspondence between individuals and their sensitive value
is lost,
but the adversary can still observe the counts of people with each attribute.
Hence, intuitively, after observing any given histogram, an adversary's 
optimal action is always to guess the most common value in the histogram 
as the target individual's value.

\begin{restatable}[Posterior single-target vulnerability under shuffle and a binary sensitive attribute]{proposition}{targetStwo}\label{proposition:target_S2}
Given the uniform prior $\pi$ on the set $\calk^{n}$ of all possible datasets
over a binary attribute set $\calk$, the single-target gain function $g_{\rm T}$ 
per Def.~\ref{def:target_k_gf}, and
a full shuffle channel $\Conc{S}$ per Def.~\ref{def:shuffling-channel-full},
the corresponding posterior vulnerability is given by
\begin{align}\label{equ:target_S2}
    \SV \hyperDist{\Conc{S}} = \SV \hyperDist{\Conc{S}^{r}}
     = \frac{1}{2^n}\sum_{i=0}^{n} \binom{n}{i} \frac{\max (i, n-i)}{n},
\end{align}
\edits{where $\Conc{S}^{r}$ is the reduced channel equivalent to $\Conc{S}$, per Def.~\ref{def:shuffling-channel-reduced}.}
\end{restatable}

For intuition, 
consider that the binary set of sensitive values is $\calk=\{\ta,\tb\}$.
The formula for \eqref{equ:target_S2}
iterates through all histograms of size $n$ containing exactly $i$ 
individuals with value $\ta$ (and, hence, $n-i$ individuals with value $\tb$).
Clearly, there are $\binom{n}{i}$ of such histograms.
The adversary picks $\ta$ as her guess
if the number $i$ of \ta's in the histogram is at least as large as
the number $n-i$ of \tb's, and picks
$\tb$ as her guess if $i<n-i$.
Her probability of guessing correctly is the ratio
between the maximum among $i$ and $n-i$, normalized 
by the total $n$ of individuals in the histogram.
To obtain the final vulnerability, we take the expected success over
all possible datasets, and under a uniform prior each dataset has probability 
$\nicefrac{1}{2^n}$.

It is not trivial to immediately grasp the behavior of \eqref{equ:target_S2}
as the size $n$ of the dataset grows.
Moreover, the formula is not 
computationally efficient, 
as the number of binomials needed grows linearly with $n$.
While computational efficiency could be improved by taking advantage of the 
symmetry of binomial coefficients, a remarkably simple equivalent formulation 
for \eqref{equ:target_S2} exists, using only a single binomial coefficient.

\begin{restatable}[Posterior single-target vulnerability under shuffle and a binary sensitive attribute: Fast formula]{theorem}{targetSbinfast}\label{theorem:target_S2}
Given the same hypotheses as Prop.~\ref{proposition:target_S2},
\begin{align}~\label{equ:target_S2_fast}
    \SV \hyperDist{\Conc{S}} = \SV \hyperDist{\Conc{S}^{r}} =\frac{1}{2} + \frac{1}{2^n}\binom{n-1}{\lfloor\nicefrac{(n-1)}{2}\rfloor}
\end{align}
\end{restatable}

Notice that this formulation is not only faster to compute, but 
easier to analyze.
In particular, as $n$ grows, the second term in 
\eqref{equ:target_S2_fast} goes to 0, 
indicating the entire sum goes to $\nicefrac{1}{2}$.
This makes intuitive sense: given a large enough dataset, we expect that the 
frequency of each binary value in any histogram approaches the adversary's prior
on the target individual's sensitive value.
(Theorem~\ref{theorem:target_S2} is a case of
Theorem~\ref{theorem:target_SN2_fast} ahead.)

Finally, we consider the posterior vulnerability of the combined
use of $k$-RR and shuffling, modeled by the cascade \Conc{NS}.
At a first glance, this vulnerability may appear more challenging 
to compute than that of shuffle alone, but in fact it only requires 
a simple modification.

\begin{restatable}[Posterior single-target vulnerability under $k$-RR and shuffle, 
and a binary sensitive attribute]{proposition}{targetSNbin}\label{proposition:target_SN2}
Given the the uniform prior $\pi$ on the set $\calk^{n}$ of all possible datasets
over a binary attribute set $\calk$, the single-target gain function $g_{\rm T}$ 
per Def.~\ref{def:target_k_gf},
a $k$-RR channel $\Conc{N}$ per Def.~\ref{def:krr-channel}, 
and a reduced shuffle channel $\Conc{S}$ per Def.~\ref{def:shuffling-channel-reduced},
the corresponding posterior vulnerability is given by
\begin{align}\label{equ:target_NS2}
    \SV \hyperDist{\Conc{NS}} =&\,\, \SV \hyperDist{\Conc{NS}^{r}} = \nonumber \\
    \frac{1}{2^n} \sum_{i=0}^{n} \binom{n}{i} 
    &\times\frac{\max(i,n{-}i) p {+} \min(i, n{-}i)(1{-}p)}{n},
\end{align}
\edits{where $\Conc{S}^{r}$ is the reduced channel equivalent to $\Conc{S}$, per Def.~\ref{def:shuffling-channel-reduced}.}

\end{restatable}


Similarly to 
Prop.~\ref{proposition:target_S2}, 
the adversary's 
best guess is to pick the most represented value, either $i$ \ta's or $n-i$ \tb's.
With probability $p$, the target will have responded truthfully, 
and the adversary's guess will be correct proportionally to $n$.
However, with probability $1-p$, the target will have flipped their response
and ended up in the smaller group by chance, thereby still making the 
adversary's guess correct.
This sum is then weighed by the probability of each dataset, which, again under a uniform prior, 
is $\nicefrac{1}{2^n}$.

Here again, it may not be easy to immediately grasp 
from \eqref{equ:target_NS2} the behavior of vulnerability as $n$ grows, 
specially because the $k$-RR parameter $p$ influences the final result.
Fortunately, we discovered an equivalent formula.

\begin{restatable}[Posterior single-target vulnerability under $k$-RR and shuffle, 
and a binary sensitive attribute: Fast formula]{theorem}{targetSNBinfast}\label{theorem:target_SN2_fast}
Given the same hypotheses as those of Prop.~\ref{proposition:target_SN2},
\begin{align}~\label{equ:target_NS2FAST}
    \SV \hyperDist{\Conc{NS}} 
    =
    \SV \hyperDist{\Conc{NS}^{r}} 
    = \frac{1}{2} + \frac{1}{2^n}\binom{n-1}{\lfloor\nicefrac{(n-1)}{2}\rfloor} (2p-1).
\end{align}
\end{restatable}
Notice that the difference between
the formulation
\eqref{equ:target_S2_fast} 
for shuffle $\Conc{S}$ alone and
the formulation 
\eqref{equ:target_NS2FAST} 
for the cascade $\Conc{NS}$ is that, in the former, the second
term in the sum is scaled by a factor $2p-1$, which
is always in the range $[0,1]$ given that 
$p\in[\nicefrac{1}{2},1]$.

\subsection{Analyses of leakage behavior}
\label{sec:analyses-leakage-behavior-A}

We now turn our attention to how the leakage formulas 
behave as the size $n$ of the dataset grows, and the parameter $p$ controlling the noise
in the $k$-RR mechanism varies.

First, 
since
the cascade \Conc{NS} can be understood as the post-processing of 
\Conc{N} by \Conc{S},
by the data-processing-inequality (Sec.~\ref{sec:model}),
we know that \Conc{NS} can never cause more leakage than \Conc{N} alone.
Since \Conc{NS} is commutative with \Conc{SN}, per Prop.~\ref{theorem:commutativity-full}, 
we conclude that adding noise on top of shuffle, or shuffle on top of noise, can never increase
the leakage of information.
With our exact equations, we can isolate how the respective mechanisms affect posterior vulnerability.
\begin{figure}[tb]
    \centering
    \includegraphics[width=\linewidth]{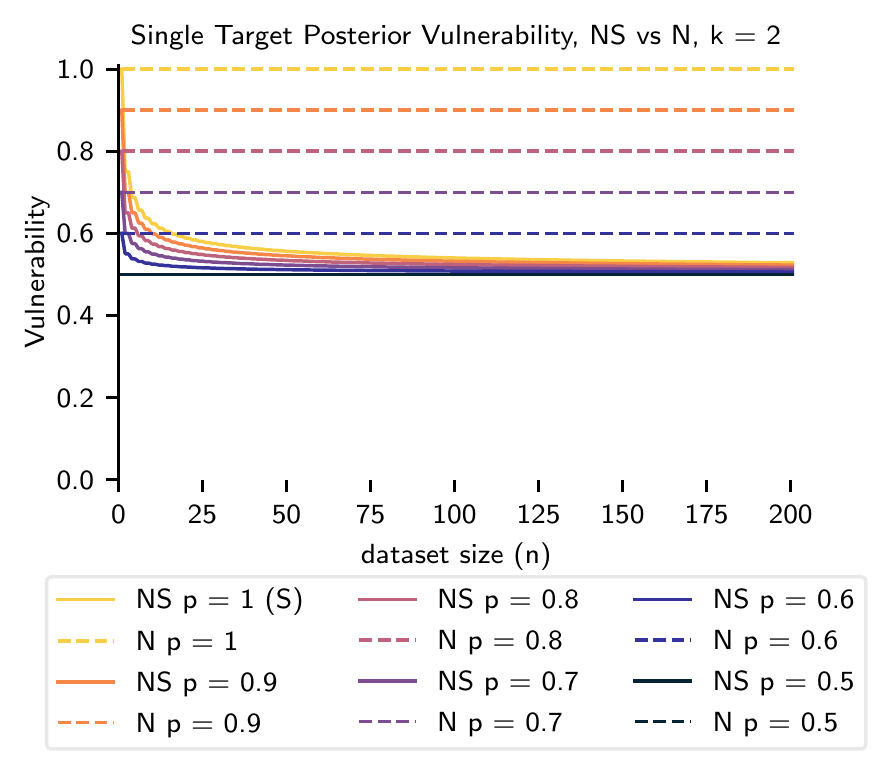}
    \caption{Posterior single-target vulnerability for binary attributes 
    under a uniform prior over various dataset sizes $n$.
    We consider the $k$-RR channel \Conc{N}, the composition of $k$-RR \& shuffle \Conc{NS},
    and the shuffle channel \Conc{S} (when $p = 1$).
    }
    \label{fig:t_NvsNSk2}
\end{figure}

Fig.~\ref{fig:t_NvsNSk2} compares 
posterior single-target vulnerability for binary attributes
under the shuffling channel $\Conc{S}$, 
under 
the $k$-RR mechanism $\Conc{N}$,
and under
the $k$-RR and shuffle cascade
$\Conc{NS}$, for various values
of the $k$-RR parameter $p$.

Notice that the dotted yellow line represents 
a baseline of no security measures:
when only  
$k$-RR $\Conc{N}$ is applied using parameter 
$p = 1$,
every participant reports their 
true value, and no shuffle is applied.
This line represents an upper bound on the posterior vulnerability,
since in the absence of any protective measure the adversary will know a 
target's value with certainty, and so the vulnerability is 1.
However, when a shuffle is performed, as represented by the solid yellow line, 
the vulnerability drops immediately and dramatically, even if no noise is added.
More precisely, since $p = 1$, this line represents exactly the vulnerability 
$\SV \hyperDist{\Conc{S}}$.

The dotted lines represent the posterior vulnerability under a $k$-RR channel 
\Conc{N} for different values of $p$.
We see that adding more noise improves the security of a single target's value.
As $p$ decreases, users are less likely to report their true values, thereby decreasing 
the adversary's probability of guessing correctly.
For example, if a user reports their true value with probability $p = 0.9$, the adversary will only guess the correct value $\nicefrac{9}{10}$ of the time, represented by the dotted orange line.
The vulnerability under $k$-RR is not affected by the dataset size, and thus remains constant for all $n$.

The remaining solid lines indicate the posterior vulnerability under the cascade \Conc{NS}.
We see here that 
shuffling noisy data exponentially decreases the vulnerability.
At $n = 1$, the adversary observes only one value, and will know with certainty that 
it belongs to the target.
When $p = 0.9$, she will guess the reported value and be correct $90\%$ of the time.
But when $n = 2$, she could observe any of the following histograms:
$\ta{:}2,\tb{:}0$, or $\ta{:}1, \tb{:}1$, or $\ta{:}0,\tb{:}2$.
If she observes  $\ta{:}2,\tb{:}0$, she will guess \ta and be correct $\nicefrac{9}{10}$ 
of the time; likewise if she observes $\ta{:}0,\tb{:}2$, she will guess \tb.
However, if she observes $\ta{:}1,\tb{:}1$ (which she will observe $\nicefrac{1}{2}$ the time), either guess has an equal probability of being correct, decreasing the total vulnerability from $\nicefrac{9}{10}$ to $\nicefrac{7}{10}$.
As $n$ grows, there are more possible outputs where the division of $n$ is less clear, bring the vulnerability closer to $\nicefrac{1}{2}$.
At $n = 200$, the posterior vulnerability of \Conc{NS} $p = 0.9$ is $0.5225$.

Note that the solid lines representing \Conc{NS} do not coincide at $n = 200$,
although they will converge as $n$ grows.
Concretely, given cascade $\Conc{NS}$ at $n = 200$, the vulnerability when $p = 1$ is approximately $0.5282$ and the vulnerability when $p = 0.6$ is approximately $0.5056$.\footnote{Also note that the step-like nature of the line representing \Conc{NS} is not a graphing artifact, but a
reflection of the equation itself\review{:
every two consecutive values of $n$ have the same vulnerability 
because of the floor function.}}

This graph shows that, for many values of $p$, shuffling alone is more effective in  reducing vulnerability than simply adding noise.
At $n = 200$, the posterior vulnerability of a shuffle with no noise represented by the solid yellow line is approximately $0.5286$, which means a shuffle alone has lower vulnerability than an application of noise where $p \geq 0.53$. This happens for most values of $p$.
The only choice of $p$ that improves upon the security of a shuffle alone is when $p$ approaches $\nicefrac{1}{2}$.

\edits{
\review{
\section{\review{Single-target leakage of a binary value against a strong adversary}}\label{sec:abo}
}
Thus far, our analysis has centered on specific adversary with limited prior knowledge\review{, since they can often
benefit the most from inference attacks.}
\review{However, since differential privacy guarantees are crafted for the
\emph{strong adversary}, who a priori knows \emph{everyone's} data but one,
in this section we introduce a preliminary discussion about how vulnerability changes under this strong, \emph{all-but-one (ABO)}, adversary.}

Returning to Example~\ref{exa:running_initial}, consider a dataset $n=3$ and assume the adversary knows a priori that the last two people have \ta's.
The set of secrets is reduced from $2^8$ to 2: \ta\ta\ta ~or \tb\ta\ta.
How does observing the channel output affect the adversary's ability to guess the target's value?

Under shuffling alone, the adversary could observe histogram $\ta:2, ~\tb: 1$ and know with certainty that $x_0 = \tb$, or she could observe $\ta:3$ and know with certainty that $x_0 = \ta$; the posterior ABO vulnerability is therefore 1.
Under the $k$-RR channel, she will guess what she sees and she can be confident about this guess with probability $p$.
Through QIF, we can illustrate how vulnerability can decrease when the two mechanisms are combined in $\Conc{NS}$.

From~\cite{alvim2020QIF}, posterior vulnerability can be calculated by summing column maximums and scaling by the prior probability.
\begin{align}
    V_{ABO} \hyperDist{\Conc{NS}^{r}} = \frac{1}{2} \sum_{y} \max_{\substack{x_{0}=\ta, \\ x_{0}=\tb}} \Conc{NS}^{r}_{x, y}
\end{align}

\begin{figure}
    \centering
    \includegraphics[width=.9\linewidth]{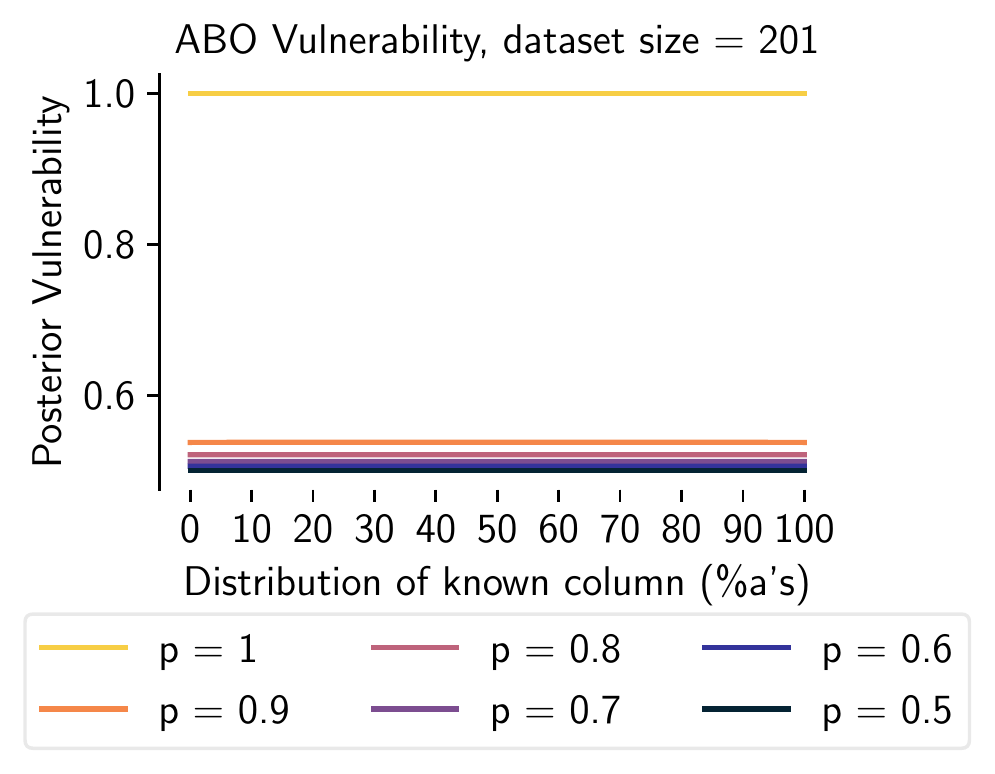}
    \caption{All-But-One posterior vulnerability when the dataset size $n = 201$ and the adversary knows 200 values a priori.}
    \label{fig:ABO_201}
\end{figure}

Figure~\ref{fig:ABO_201} shows the posterior ABO vulnerability when $n = 201$, over different sets of adversary prior knowledge.
The first $0\%$ tick represents an adversary who knows the last two hundred people have \tb's, and who is guessing the value of the first person.
Given this knowledge, when $p = 0.8$, the ABO vulnerability is approx. 0.52111.
Changing the prior distribution affects the ABO vulnerability, but only marginally. 
For instance, when the adversary knows that one hundred people have \ta's and the other hundred have \tb's (represented by the $50\%$ tick), the vulnerability increases a bit, but only to approx. 0.52116.

From the graph, we can confirm that the shuffle alone (represented by the yellow line when $p = 1$) does not affect vulnerability---the adversary can guess the last person with perfect confidence.
Recall that the opposite effect was observed in Figure~\ref{fig:t_NvsNSk2} where the shuffle alone dramatically reduced vulnerability; this speaks to how adversary choice influences our understanding of a given privacy mechanism.
However, when noise is introduced, the vulnerability decreases \emph{below} $p$, indicating that the combination of noise and shuffling provides the most security.
A formal investigation of this phenomena is set for future work.
}

\review{
\section{\review{Single-target leakage of a generic value $k{\geq}2$ against an uninformed adversary}}\label{sec:single_genk}
}

In this section, we extend our investigation of
Sec.~\ref{sec:single} to attack scenarios
in which:
(i) as before, the adversary is attempting to guess the secret value of a 
single individual from $\caln$ chosen as the target in the dataset; but
(ii) contrarily to before, the set $\calk$ of
sensitive values can have any size $k{\geq}2$.
\review{Here again, we focus on an uninformed adversary, with a uniform prior on datasets.}

For this general case, we develop exact equations with 
which to calculate single-target posterior vulnerability.
However, these equations are not as computationally efficient 
or as easy to analyze as 
the binary case.
Nevertheless, we are able to provide asymptotic bounds on 
leakage by uncovering a surprising connection between our
scenario 
and a well-known combinatorial problem.

\subsection{Leakage formulas}


The prior vulnerability remains $\nicefrac{1}{k}$, the
same as that of the binary case, per 
Prop.~\ref{proposition:singletargetprior}.
Indeed, given a uniform distribution, an adversary will 
guess any of the $k$ values and will have an equal probability 
of being correct.

For posterior vulnerability, we investigate the same
combinations of sanitization mechanisms as before:
the $k$-RR channel \Conc{N}, the full shuffle channel \Conc{S}, 
and the composition of $k$-RR with shuffle via the cascade \Conc{NS}. 
The posterior vulnerability of the $k$-RR channel $\Conc{N}$
has already been derived in Prop.~\ref{proposition:target_N} for
any $k\geq 2$, and it is simply $p$.
However, the vulnerability of the shuffle channel $\Conc{S}$ and
consequently of the cascade $\Conc{NS}$ depends on $k$, necessitating
a generalization of the earlier results.


Recall that under shuffling, the exact mapping from individuals to values is lost, and only the dataset's histogram is preserved.
In the binary case, the posterior vulnerability of shuffling
was computed by iterating through all histograms of size two, which 
could be counted using a summation of binomial coefficients.
In the general case, the adversary observes multi-sets, which 
must be counted using multinomial coefficients, as formalized below.

\begin{restatable}[Posterior single-target vulnerability under shuffle and $k\geq2$]{proposition}{targetSk}\label{proposition:target_Sk}
Given the same hypotheses as Prop.~\ref{proposition:target_S2}, except
that the sensitive set $\calk$ can have $k\geq2$ elements,
\begin{equation}\label{equ:target_Sk}
    \SV \hyperDist{\Conc{S}} =
    \SV \hyperDist{\Conc{S}^{r}} = \frac{1}{k^n} \sum_{\substack{n_1, \dots, n_k:\\n_1 + \dots + n_k =n}} \binom{n}{n_1, \dots, n_k} \frac{n^{*}}{n},
\end{equation}
where $n^{*} = \max(n_1, \dots, n_k)$.
\end{restatable}



Finally, the vulnerability of the combination of $k$-RR and shuffle
is an extension of the binary case from Prop.~\ref{theorem:target_SN2_fast}.

\begin{restatable}[Posterior single-target vulnerability under $k$-RR \& shuffle,  $k\geq2$]{proposition}{targetSNk}\label{proposition:target_SNk}
Given the same hypotheses as Prop.~\ref{proposition:target_SN2}, except
that the sensitive set $\calk$ can have $k\geq2$ elements,
\begin{align}~\label{equ:target_NSk}
    &\SV \hyperDist{\Conc{NS}} =
    \SV \hyperDist{\Conc{NS}^{r}} = \nonumber \\
    &\frac{1}{k^n}  \sum_{\substack{n_1, \dots, n_k:\\n_1 + \dots + n_k =n}} \binom{n}{n_1, \dots, n_k}
    \frac{ n^{*}p + (n{-}n^*)\nicefrac{(1{-}p)}{(k{-}1)}}{n}
\end{align}
where $n^{*} = \max(n_1, \dots, n_k)$.
\end{restatable}

Notice, however, that the formulas given in Prop.~\ref{proposition:target_Sk} and
Prop.~\ref{proposition:target_SNk}, although exact, are not easy to compute
or interpret.
Indeed, they depend on the computation of a number of binomial coefficients
that grows linearly with the dataset size $n$. 
In Sec.~\ref{sec:asymptotic}, we shall discuss asymptotic bounds for these
vulnerabilities, but first we analyze the exact 
behavior of leakage for tractable values of $n$, $k$, and $p$.

\subsection{Analyses of leakage behavior}
\label{sec:analyses-leakage-behavior-B}

\begin{figure}[tb]
    \centering
    \includegraphics[width=\linewidth]{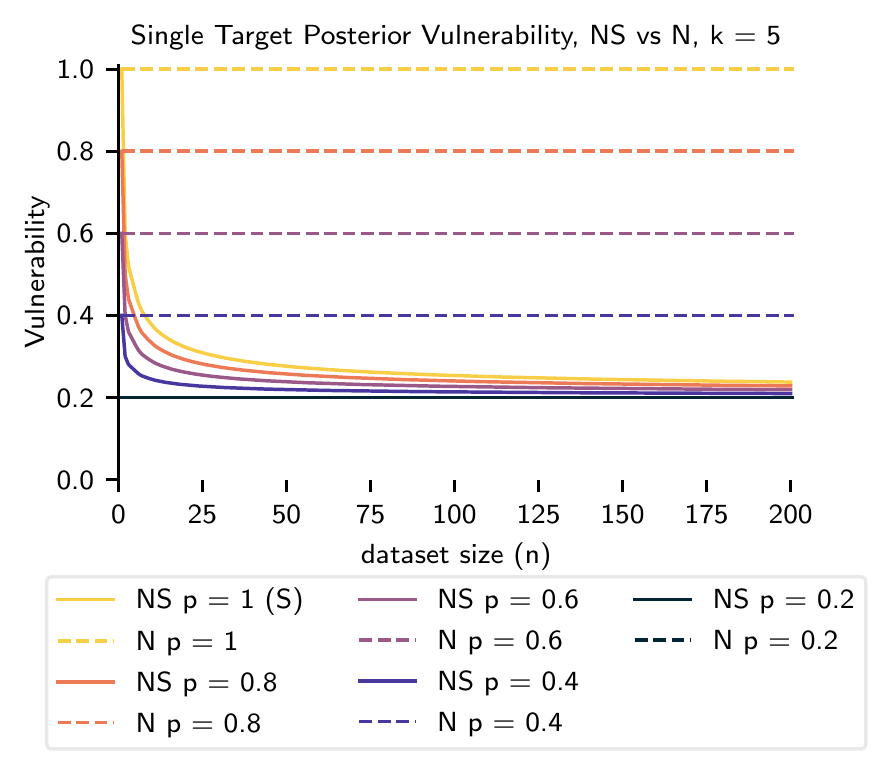}
    \caption{Posterior single-target vulnerability for 
    $k=5$ values,  
    under a uniform prior over various dataset sizes $n$. 
    We consider the $k$-RR channel \Conc{N}, the composition 
    of $k$-RR \& shuffle $\Conc{NS}$,
    and the shuffle channel $\Conc{S}$ ($p=1$).
    }
    \label{fig:t_NvsNSk5}
\end{figure}

Fig.~\ref{fig:t_NvsNSk5} compares the exact values for the
posterior vulnerabilities of the three channels $\Conc{N}$, $\Conc{S}$, and
$\Conc{NS}$ for a set $\calk$ of sensitive values with $k=5$ elements.

As before, the dotted yellow line represents the case of no noise nor shuffle; here, the adversary can observe and know the target's value exactly.
By observing the vulnerability under the shuffle channel \Conc{S}, represented by the solid yellow line, we can see that shuffle alone decreases the adversary's probability of success considerably.
The dotted lines represent the vulnerability under the $k$-RR channel \Conc{N} for different values of $p$, while the solid lines represent posterior vulnerability under the \Conc{NS} cascade.
For example, when $p = 0.8$, represented by the dotted orange line, the posterior vulnerability drops to $0.8$, regardless of the dataset size.
When shuffle is added, the vulnerability drops exponentially with $n$,
represented by the solid orange line, asymptotically approaching the lower
bound of $0.2=\nicefrac{1}{k}$ 
(the prior vulnerability).
The graph suggests that shuffle alone is a highly effective sanitization
mechanism, significantly outperforming $k$-RR's leakage guarantees in most cases.
Indeed, the only cases in which $k$-RR alone provides a sanitization
comparable to that provided by shuffle alone are when the parameter $p$ 
approaches $\nicefrac{1}{k}$.
However, these are exactly the cases in which the mechanisms are noisiest,
so their utility suffers the most.

\begin{figure}
    \centering
    \includegraphics{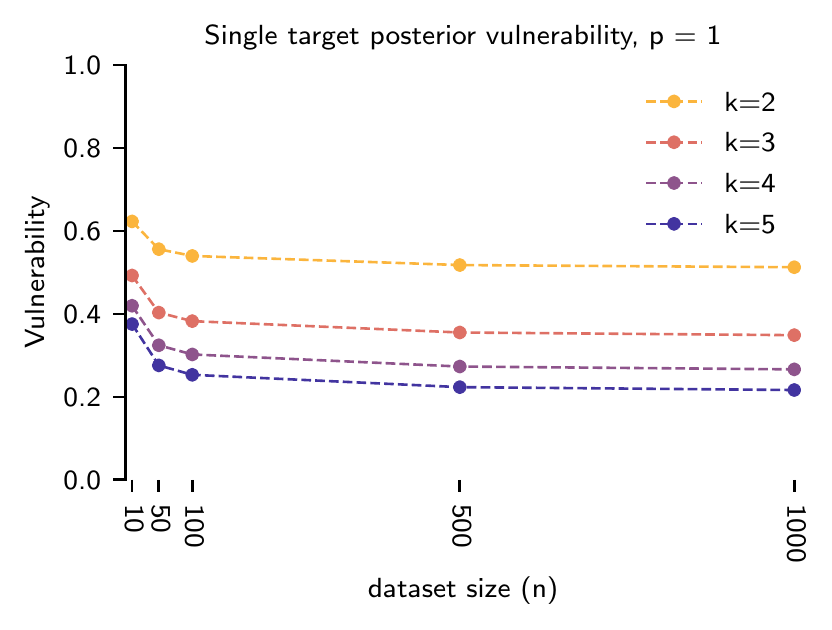}
    \caption{Posterior single-target vulnerability 
    of the shuffle channel $\Conc{S}$ under a uniform prior,
    for varying attribute set sizes $k$ and dataset
    sizes $n$.}
    \label{fig:t_range_k_S}
\end{figure}

Fig.~\ref{fig:t_range_k_S} provides more insight into how shuffle alone affects posterior vulnerability.
As exact values are computationally expensive to calculate as $n$ and $k$ increase, here we show the exact vulnerability for a subset of dataset sizes $n$ ranging from 10 to 1000.
We see that posterior vulnerability decreases with $n$, and approaches the prior vulnerability of $\nicefrac{1}{k}$.
For the sake of comparison, at $n = 100$ the posterior vulnerability when $k = 3$ is $0.3826$, and at $n = 1000$ it is $0.3488$.

\subsection{Asymptotic Bounds}
\label{sec:asymptotic}

Interestingly, \eqref{equ:target_Sk} has an important combinatorial interpretation when scaled up by $n$,
which can be written explicitly as
\begin{equation}\label{equ:max_load}
    \SV \hyperDist{\Conc{S}} \times n = \frac{1}{k^n} \sum_{\substack{n_1, n_2, \dots, n_k:\\n_1 + n_2 + \dots + n_k =n}} \binom{n}{n_1, n_2, \dots, n_k} n^{*}.
\end{equation}
Intuitively, when $n$ balls are thrown uniformly at random and independently into $k$ distinct bins, \eqref{equ:max_load} computes the exact number of expected balls in the fullest bin (also referred to as ``maximum load'').
This perspective provides an interpretation of \eqref{equ:target_Sk}: For every division of $n$ into $k$ parts, this equation counts the ways these parts can be labeled, chooses the largest part, then takes the average by dividing by the total number of the possibilities.
Brown's work on surmising remixed keys~\cite{Brown2015BoundsonRemixedKeys} posits an equivalent formula for Equ.~\ref{equ:max_load}, however he too was unable to find a faster way to compute these values exactly.
This connection is discussed in Appendix~\ref{sec:brown}. 

There are well-known asymptotic bounds for this combinatorial problem~\cite{Raab1998BallsIntoBins}.
In particular, when $n \geq k \ln k$,
\begin{equation}
    \SV \hyperDist{\Conc{S}} \times n = \frac{n}{k} + \Theta \left( \sqrt{\frac{n \ln k}{k}}\right) \quad \textnormal{(maximum load)} 
\end{equation}
This holds with high probability, meaning an event $M$ counting the maximum number of $n$ balls in $k$ bins occurs with probability Pr$(M) \geq 1 - k^{-c}$ for an arbitrarily chosen constant $c \geq 0$~\cite{Berenbrink2006BalancedAllocations}.
Dividing both sides by $n$, we get asymptotic bounds for posterior single-target vulnerability.
\begin{equation}\label{equ:Starget_asym}
    \SV \hyperDist{\Conc{S}} = \frac{1}{k} + \Theta \left(\sqrt{\frac{\ln k}{kn}}\right)
\end{equation}
These bounds show the maximum load grows sublinearly with $n$ while posterior vulnerability decreases monotonically.

\begin{figure}
    \centering
    \includegraphics[]{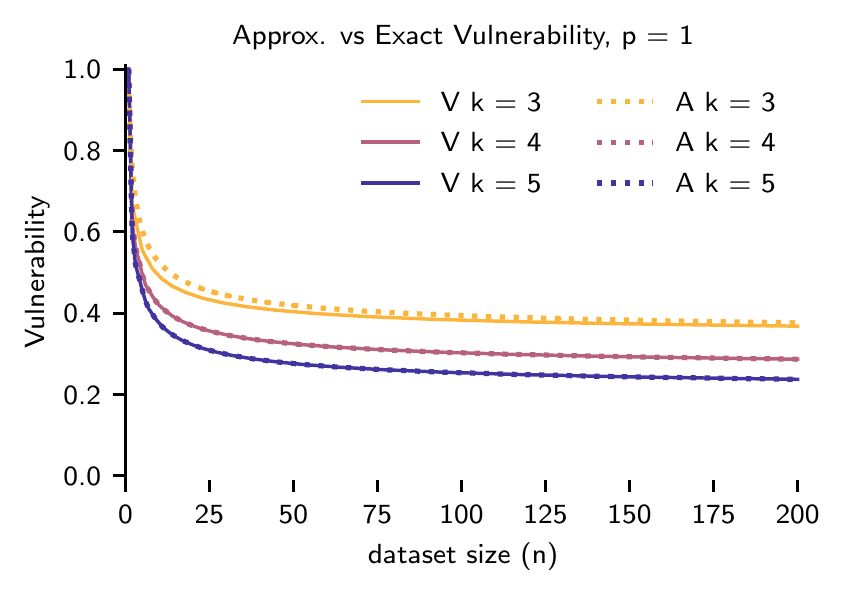}
    \caption{Comparison of the approximation with the true value of posterior single-target vulnerability through the shuffle channel for different values of $k$.}
    \label{fig:t_S_approx_vs_exact}
\end{figure}

While the big-theta implies the existence of a constant factor $f$, we can test how well this equation matches the true posterior vulnerability when $f = 1$. 
Fig.~\ref{fig:t_S_approx_vs_exact} shows how well the asymptotic bounds for posterior single-target vulnerability under the shuffle channel approximate the exact value.
The yellow line, representing the exact vulnerability when $k=3$, has a slight offset from the approximation, but the violet and blue lines, representing $k=4$ and $k=5$ respectively, are visually indistinguishable from the value provided by the asymptotic bounds.

Given an approximation for $\SV \hyperDist{\Conc{S}}$, we can derive an approximation for $\SV \hyperDist{\Conc{NS}}$.
\begin{restatable}{theorem}{NSgenkApprox}\label{theorem:NSgenkApprox}
Posterior single-target vulnerability for a general $k$ has the following asymptotic bounds.
\begin{equation}\label{equ:NStarget_asym}
    \SV \hyperDist{\Conc{NS}} = \frac{1}{k} + \Theta \left(\sqrt{\frac{\ln k}{kn}}\right) \left(\frac{kp-1}{k-1}\right) 
\end{equation}
\end{restatable}
The proof relies on the fact that the equation for posterior vulnerability under \Conc{NS} can be re-written as a function of the posterior vulnerability under \Conc{S}.
Explicitly, 
\begin{equation}
    \SV \hyperDist{\Conc{NS}} = \SV \hyperDist{\Conc{S}} \times \left( \frac{kp-1}{k-1} \right) + \frac{1-p}{k-1}.
\end{equation}

\begin{figure}
    \centering
    \includegraphics[]{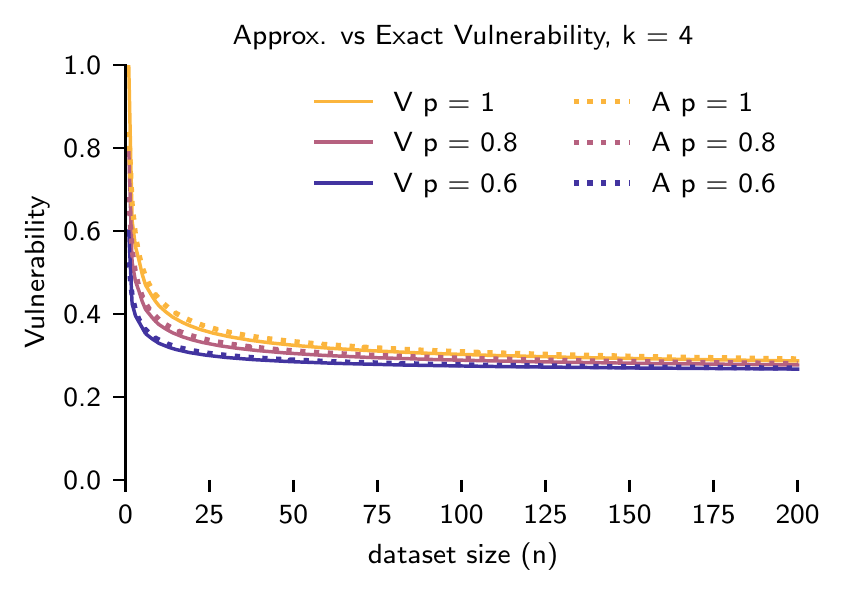}
    \caption{Comparison of the approximation with the true value of posterior single-target vulnerability through the shuffle and $k$-RR channels when $k = 4$ for different values of $p$.}
    \label{fig:t_NS_approx_vs_exact}
\end{figure}

Fig.~\ref{fig:t_NS_approx_vs_exact} sets $k = 4$ and compares the exact posterior single-target vulnerability with the asymptotic bounds assuming $f = 1$.
When $p = 1$, we see the dotted yellow line representing the bounds is slightly above the straight yellow line representing the exact value for small values of $n$.
As $n$ increases, the two lines converge.
This is seen again for $p = 0.8$ and $p = 0.6$.
From this anecdotal test, it seems the asymptotic bound closely approximates the true value when $f = 1$.

\review{
\section{Discussion}
\label{sec:discussion}
}
\review{The main lessons  learned from this work are the following:
\begin{itemize}
    \item The shuffle and the $k$-RR  mechanisms commute, in the sense that their composition as probabilistic functions   commute.
    We believe that this is the case for the composition of the shuffle with \emph{any} local obfuscation mechanism.  
    \item By using the formulas derived for vulnerabilities and leakage, we have shown that under the uninformed adversary the shuffling accounts for most of the privacy, as by itself it achieves almost the same level of resilience to inference attatcks as its composition with $k$-RR. (The only exception is when the probability of reporting the true value is $p=\nicefrac{1}{k}$, but this case has no utility.) We also noted  that the level of posterior vulnerability decreases very fast with the number of participants, and that it converges to the prior, which is $\nicefrac{1}{k}$, i.e., the probability of guessing the true value by making a random guess. 
    \item We have shown how to model the strong adversary in our framework, and how to calculate the posterior vulnerability. 
    Our experiments show that, under the strong adversary, shuffling alone is ineffective, but it substantially increases privacy when combined with $k$-RR. This reinforces the results from the shuffle model literature pointing out the merits of shuffling, like for example \cite{Balle:2019:AICC}, although in general those works consider adversaries even more powerful than the strong adversary of DP.
    \item As a consequence of the above findings, we point out that, for achieving the best trade-off between privacy and utility, in the strong model it is better to compose  $k$-RR and shuffle. In contrast, under the uninformed adversary, it may be better to use the shuffle alone, as $k$-RR reduces utility and does not have a significant impact on privacy. 
    \item Our results can help to choose, among two or more mechanisms (for instance, the shuffle model alone or the shuffle combined with $k$-RR), the one that gives the best trade-off between privacy and utility.   Indeed, by using the formulas derived for the posterior vulnerability, we can compute the level of privacy  (i.e., resilience to inference attacks) of the mechanisms and compare them. 
    This process may also require to  tune $\epsilon$ so to achieve the desired utility. 
    We remark that different adversary models may lead to different results in this comparison, as shown in  the introduction. We argue that, in case we don't know what kind of adversary we will have to face,  it is generally better to choose the more likely one. In the case of DP, a basic assumption is that the data consumer cannot access directly the database, it can only query it, via the interface (the database curator). Hence the uninformed adversary is more natural and likely than the strong one. 
\end{itemize}
}



\section{Conclusion}    \label{conclusion}

\review{In this work, we proved that $k$-RR and shuffling can commute without affecting information leakage, and
derived exact formulas for prior and posterior vulnerabilities for an uninformed adversary focusing on a single target.
For binary attributes, these formulas are computationally efficient, and for the general case we found 
asymptotic bounds. 
We studied how the leakage of shuffling and $k$-RR mechanisms behaves as the dataset size increases for different privacy parameters.
For the uninformed adversary, we found that shuffling alone may dramatically reduce leakage in many cases, outperforming the application of $k$-RR alone.}


\section*{Acknowledgements}
We are grateful to the anonymous reviewers for their helpful comments.
Mireya Jurado was supported by the U.S. Department of Homeland Security under Grant Award № 2017‐ST‐062‐000002.
Catuscia Palamidessi was supported by the European Research Council (ERC) under the Horizon 2020 research and innovation programme, grant agreement № 835294.
Ramon G.\ Gonze and M\'ario S.\ Alvim were supported by CNPq, CAPES, and FAPEMIG. 
The views and conclusions contained in this document are those of the authors and should not be interpreted as necessarily representing the official policies, either expressed or implied, of the U.S. Department of Homeland Security.

\bibliographystyle{IEEEtran}
\bibliography{bibs}

\begin{thebibliography}{10}
\providecommand{\url}[1]{#1}
\csname url@samestyle\endcsname
\providecommand{\newblock}{\relax}
\providecommand{\bibinfo}[2]{#2}
\providecommand{\BIBentrySTDinterwordspacing}{\spaceskip=0pt\relax}
\providecommand{\BIBentryALTinterwordstretchfactor}{4}
\providecommand{\BIBentryALTinterwordspacing}{\spaceskip=\fontdimen2\font plus
\BIBentryALTinterwordstretchfactor\fontdimen3\font minus
  \fontdimen4\font\relax}
\providecommand{\BIBforeignlanguage}[2]{{%
\expandafter\ifx\csname l@#1\endcsname\relax
\typeout{** WARNING: IEEEtran.bst: No hyphenation pattern has been}%
\typeout{** loaded for the language `#1'. Using the pattern for}%
\typeout{** the default language instead.}%
\else
\language=\csname l@#1\endcsname
\fi
#2}}
\providecommand{\BIBdecl}{\relax}
\BIBdecl

\bibitem{Dwork:06:EUROCRYPT}
C.~Dwork, K.~Kenthapadi, F.~McSherry, I.~Mironov, and M.~Naor, ``Our data,
  ourselves: Privacy via distributed noise generation,'' in \emph{Proc. of
  EUROCRYPT}, ser. LNCS, vol. 4004.\hskip 1em plus 0.5em minus 0.4em\relax
  Springer, 2006, pp. 486--503.

\bibitem{Dwork:06:TCC}
C.~Dwork, F.~Mcsherry, K.~Nissim, and A.~Smith, ``Calibrating noise to
  sensitivity in private data analysis,'' in \emph{In Proc. of {TCC}}, ser.
  LNCS, vol. 3876.\hskip 1em plus 0.5em minus 0.4em\relax Springer, 2006, pp.
  265--284.

\bibitem{Kasiviswanathan:11:SIAMJC}
S.~P. Kasiviswanathan, H.~K. Lee, K.~Nissim, S.~Raskhodnikova, and A.~D. Smith,
  ``What can we learn privately?'' \emph{{SIAM} Journal of Computing}, vol.~40,
  no.~3, pp. 793--826, 2011.

\bibitem{Duchi:13:FOCS}
J.~C. Duchi, M.~I. Jordan, and M.~J. Wainwright, ``Local privacy and
  statistical minimax rates,'' in \emph{Proc. of {FOCS}}.\hskip 1em plus 0.5em
  minus 0.4em\relax {IEEE} Computer Society, 2013, pp. 429--438.

\bibitem{Erlingsson:14:CCS}
{\'{U}}.~Erlingsson, V.~Pihur, and A.~Korolova, ``{RAPPOR:} randomized
  aggregatable privacy-preserving ordinal response,'' in \emph{Proc. of {ACM}
  {SIGSAC} {CCS}}.\hskip 1em plus 0.5em minus 0.4em\relax {ACM}, 2014, pp.
  1054--1067.

\bibitem{Fanti:16:PETS}
G.~Fanti, V.~Pihur, and {\'{U}}.~Erlingsson, ``Building a rappor with the
  unknown: Privacy-preserving learning of associations and data dictionaries,''
  \emph{PoPETS}, vol. 2016, no.~3, pp. 41--61, 2016.

\bibitem{AppleDPTeam:17:ML}
{Apple Differential Privacy Team}, ``Learning with privacy at scale,''
  \emph{Apple Machine Learning Journal}, vol.~1, no.~9, December 2017.

\bibitem{Thakurta:17:Patent}
A.~G. Thakurta, A.~H. Vyrros, U.~S. Vaishampayan, G.~Kapoor, J.~Freudinger,
  V.~V. Prakash, A.~Legendre, and S.~Duplinsky, ``Emoji frequency detection and
  deep link frequency,'' US Patent 9,705,908., July 11 2017.

\bibitem{Ding:17:NIPS}
B.~Ding, J.~Kulkarni, and S.~Yekhanin, ``Collecting telemetry data privately,''
  in \emph{Proc. of {NeurIPS}}, ser. NIPS'17.\hskip 1em plus 0.5em minus
  0.4em\relax Curran Associates Inc., 2017, pp. 3574--3583.

\bibitem{Beimel:08:CRYPTO}
A.~Beimel, K.~Nissim, and E.~Omri, ``Distributed private data analysis:
  Simultaneously solving how and what,'' in \emph{Advances in Cryptology --
  CRYPTO 2008}.\hskip 1em plus 0.5em minus 0.4em\relax Springer Berlin
  Heidelberg, 2008, pp. 451--468.

\bibitem{Chan:12:ESA}
T.-H.~H. Chan, E.~Shi, and D.~Song, ``Optimal lower bound for differentially
  private multi-party aggregation,'' in \emph{Proc. of the 20th ESA}, ser.
  ESA'12.\hskip 1em plus 0.5em minus 0.4em\relax Springer-Verlag, 2012, pp.
  277--288.

\bibitem{Bittau:2017:SOSP}
A.~Bittau, {\'U}.~Erlingsson, P.~Maniatis, I.~Mironov, A.~Raghunathan, D.~Lie,
  M.~Rudominer, U.~Kode, J.~Tinnes, and B.~Seefeld, ``Prochlo: Strong privacy
  for analytics in the crowd,'' in \emph{Proc. of SOSP}, 2017, pp. 441--459.

\bibitem{Cheu:19:EUROCRYPT}
A.~Cheu, A.~D. Smith, J.~R. Ullman, D.~Zeber, and M.~Zhilyaev, ``Distributed
  differential privacy via shuffling,'' in \emph{Proc. of EUROCRYPT}, ser.
  LNCS, vol. 11476.\hskip 1em plus 0.5em minus 0.4em\relax Springer, 2019, pp.
  375--403.

\bibitem{alvim2020QIF}
M.~S. Alvim, K.~Chatzikokolakis, A.~McIver, C.~Morgan, C.~Palamidessi, and
  G.~Smith, \emph{The Science of Quantitative Information Flow}, ser.
  Information Security and Cryptography.\hskip 1em plus 0.5em minus 0.4em\relax
  Springer Int. Publishing, 2020.

\bibitem{Jurado:21:CSF}
M.~Jurado, C.~Palamidessi, and G.~Smith, ``A formal information-theoretic
  leakage analysis of order-revealing encryption,'' in \emph{CSF}, 2021, pp.
  1--16.

\bibitem{Fernandes:18:FM}
N.~Fernandes, M.~Dras, and A.~McIver, ``Processing text for privacy: an
  information flow perspective,'' in \emph{FM}, 2018, pp. 3--21.

\bibitem{Alvim:22:PETSa}
M.~S. Alvim, N.~Fernandes, A.~McIver, C.~Morgan, and G.~H. Nunes, ``Flexible
  and scalable privacy assessment for very large datasets, with an application
  to official governmental microdata,'' \emph{PoPETS}, vol. 2022, pp. 378--399,
  2022.

\bibitem{Alvim:15:JCS}
M.~S. Alvim, M.~E. Andr{\'{e}}s, K.~Chatzikokolakis, P.~Degano, and
  C.~Palamidessi, ``On the information leakage of differentially-private
  mechanisms,'' \emph{J. of Comp. Security}, vol.~23, no.~4, pp. 427--469,
  2015.

\bibitem{Chatzikokolakis:19:CSF}
K.~Chatzikokolakis, N.~Fernandes, and C.~Palamidessi, ``Comparing systems:
  Max-case refinement orders and application to differential privacy,'' in
  \emph{CSF}.\hskip 1em plus 0.5em minus 0.4em\relax {IEEE}, 2019, pp.
  442--457.

\bibitem{Alvim:12:CSF}
M.~S. {Alvim}, K.~{Chatzikokolakis}, C.~{Palamidessi}, and G.~{Smith},
  ``Measuring information leakage using generalized gain functions,'' in
  \emph{Proc. of CSF}, 2012, pp. 265--279.

\bibitem{Kairouz:16:JMLR}
P.~Kairouz, S.~Oh, and P.~Viswanath, ``Extremal mechanisms for local
  differential privacy,'' \emph{The Journal of Machine Learning Research},
  vol.~17, no.~1, pp. 492--542, Jan. 2016.

\bibitem{Tschantz:20:SP}
M.~C. Tschantz, S.~Sen, and A.~Datta, ``Sok: Differential privacy as a causal
  property,'' in \emph{2020 IEEE S\&P}, 2020, pp. 354--371.

\bibitem{Cuff:16:CCS}
P.~Cuff and L.~Yu, ``Differential privacy as a mutual information constraint,''
  in \emph{Proc. of {ACM} {SIGSAC} {CCS}}.\hskip 1em plus 0.5em minus
  0.4em\relax ACM, 2016, p. 43–54.

\bibitem{Yang:15:SIGMOD}
B.~Yang, I.~Sato, and H.~Nakagawa, ``Bayesian differential privacy on
  correlated data,'' in \emph{Proc. of the 2015 ACM SIGMOD Int. Conf. on
  Management of Data}, ser. SIGMOD '15.\hskip 1em plus 0.5em minus 0.4em\relax
  ACM, 2015, p. 747–762.

\bibitem{Li:13:CCS}
N.~Li, W.~Qardaji, D.~Su, Y.~Wu, and W.~Yang, ``Membership privacy: A unifying
  framework for privacy definitions,'' in \emph{Proc. of {ACM} {SIGSAC} {CCS}},
  ser. CCS '13.\hskip 1em plus 0.5em minus 0.4em\relax ACM, 2013, p. 889–900.

\bibitem{Warner1965RandomizedResponse}
S.~L. Warner, ``Randomized response: A survey technique for eliminating evasive
  answer bias,'' \emph{Journal of the American Statistical Association},
  vol.~60, no. 309, pp. 63--69, 1965.

\bibitem{Balle:2019:AICC}
B.~Balle, J.~Bell, A.~Gasc{\'o}n, and K.~Nissim, ``The privacy blanket of the
  shuffle model,'' in \emph{CRYPTO}, 2019, pp. 638--667.

\bibitem{Erlingsson:20:ArXiv}
{\'U}.~Erlingsson, V.~Feldman, I.~Mironov, A.~Raghunathan, S.~Song, K.~Talwar,
  and A.~Thakurta, ``Encode, shuffle, analyze privacy revisited: Formalizations
  and empirical evaluation,'' \emph{arXiv:2001.03618}, 2020.

\bibitem{Erlingsson:2019:SDA}
{\'U}.~Erlingsson, V.~Feldman, I.~Mironov, A.~Raghunathan, K.~Talwar, and
  A.~Thakurta, ``Amplification by shuffling: From local to central differential
  privacy via anonymity,'' in \emph{Proc. of SODA}, 2019, pp. 2468--2479.

\bibitem{Feldman:21:FOCS}
V.~Feldman, A.~McMillan, and K.~Talwar, ``Hiding among the clones: {A} simple
  and nearly optimal analysis of privacy amplification by shuffling,'' in
  \emph{Proc. of {FOCS}}.\hskip 1em plus 0.5em minus 0.4em\relax {IEEE}, 2021,
  pp. 954--964.

\bibitem{koskela2021tight}
A.~Koskela, M.~A. Heikkil{\"a}, and A.~Honkela, ``Tight accounting in the
  shuffle model of differential privacy,'' \emph{arXiv:2106.00477}, 2021.

\bibitem{koskela2021tightdiscrete}
A.~Koskela, J.~J{\"a}lk{\"o}, L.~Prediger, and A.~Honkela, ``Tight differential
  privacy for discrete-valued mechanisms and for the subsampled gaussian
  mechanism using fft,'' in \emph{Proc. of AISTATS}, 2021, pp. 3358--3366.

\bibitem{balle2019differentially}
B.~Balle, J.~Bell, A.~Gascon, and K.~Nissim, ``Differentially private summation
  with multi-message shuffling,'' \emph{arXiv:1906.09116}, 2019.

\bibitem{balle2020private}
B.~Balle, J.~Bell, A.~Gasc{\'o}n, and K.~Nissim, ``Private summation in the
  multi-message shuffle model,'' in \emph{Proc. of {ACM} {SIGSAC} {CCS}}, 2020,
  pp. 657--676.

\bibitem{ishai2006cryptography}
Y.~Ishai, E.~Kushilevitz, R.~Ostrovsky, and A.~Sahai, ``Cryptography from
  anonymity,'' in \emph{Proc. of {FOCS}}, 2006, pp. 239--248.

\bibitem{ghazi2019scalable}
B.~Ghazi, R.~Pagh, and A.~Velingker, ``Scalable and differentially private
  distributed aggregation in the shuffled model,'' \emph{arXiv:1906.08320},
  2019.

\bibitem{balle2019improved}
B.~Balle, J.~Bell, A.~Gasc{\'o}n, and K.~Nissim, ``Improved summation from
  shuffling,'' \emph{arXiv:1909.11225}, 2019.

\bibitem{balcer2019separating}
V.~Balcer and A.~Cheu, ``Separating local \& shuffled differential privacy via
  histograms,'' \emph{arXiv:1911.06879}, 2019.

\bibitem{cheu2021differentially}
A.~Cheu and M.~Zhilyaev, ``Differentially private histograms in the shuffle
  model from fake users,'' \emph{arXiv:2104.02739}, 2021.

\bibitem{balcer2021connecting}
V.~Balcer, A.~Cheu, M.~Joseph, and J.~Mao, ``Connecting robust shuffle privacy
  and pan-privacy,'' in \emph{Proc. of {SODA}}, 2021, pp. 2384--2403.

\bibitem{cheu2021differential}
A.~Cheu, ``Differential privacy in the shuffle model: A survey of
  separations,'' \emph{arXiv:2107.11839}, 2021.

\bibitem{cheu2021limits}
A.~Cheu and J.~Ullman, ``The limits of pan privacy and shuffle privacy for
  learning and estimation,'' in \emph{Proc. of the 53rd Annual ACM SIGACT
  Symposium on Theory of Computing}, 2021, pp. 1081--1094.

\bibitem{Barthe:11:CSF}
G.~Barthe and B.~K{\"{o}}pf, ``Information-theoretic bounds for differentially
  private mechanisms,'' in \emph{Proc. of CSF}.\hskip 1em plus 0.5em minus
  0.4em\relax {IEEE} Computer Society, 2011, pp. 191--204.

\bibitem{Chatzikokolakis:21:JCP}
K.~Chatzikokolakis, N.~Fernandes, and C.~Palamidessi, ``Refinement orders for
  quantitative information flow and differential privacy,'' \emph{Journal of
  Cybersecurity and Privacy}, vol.~1, no.~1, pp. 40--77, 2021.

\bibitem{Clark:01:ENTCS}
D.~Clark, S.~Hunt, and P.~Malacaria, ``Quantitative analysis of the leakage of
  confidential data,'' \emph{Electron. Notes Theor. Comput. Sci.}, vol.~59,
  no.~3, pp. 238--251, 2001.

\bibitem{Smith:09:FOSSACS}
G.~Smith, ``On the {F}oundations of {Q}uantitative {I}nformation {F}low,'' in
  \emph{FOSSACS}, ser. LNCS, vol. 5504.\hskip 1em plus 0.5em minus 0.4em\relax
  Springer, 2009, pp. 288--302.

\bibitem{McIver:10:ICALP}
A.~McIver, L.~Meinicke, and C.~Morgan, ``Compositional closure for {Bayes Risk}
  in probabilistic noninterference,'' in \emph{ICALP}, vol. 6199.\hskip 1em
  plus 0.5em minus 0.4em\relax Springer Verlag, 2010, pp. 223--235.

\bibitem{Alvim:16:CSF}
M.~S. {Alvim}, K.~{Chatzikokolakis}, A.~{McIver}, C.~{Morgan},
  C.~{Palamidessi}, and G.~{Smith}, ``Axioms for information leakage,'' in
  \emph{Proc. of CSF}, 2016, pp. 77--92.

\bibitem{wang2016using}
Y.~Wang, X.~Wu, and D.~Hu, ``Using randomized response for differential privacy
  preserving data collection.'' in \emph{EDBT/ICDT Workshops}, vol. 1558, 2016,
  pp. 0090--6778.

\bibitem{Braun:09:MPFS}
C.~Braun, K.~Chatzikokolakis, and C.~Palamidessi, ``Quantitative notions of
  leakage for one-try attacks,'' in \emph{Proc. {MFPS}}, ser. ENTCS, vol.
  249.\hskip 1em plus 0.5em minus 0.4em\relax Elsevier, 2009, pp. 75--91.

\bibitem{Alvim:14:CSF}
M.~S. Alvim, K.~Chatzikokolakis, A.~McIver, C.~Morgan, C.~Palamidessi, and
  G.~Smith, ``Additive and multiplicative notions of leakage, and their
  capacities,'' in \emph{{IEEE} CSF}.\hskip 1em plus 0.5em minus 0.4em\relax
  {IEEE} Computer Society, 2014, pp. 308--322.

\bibitem{Brown2015BoundsonRemixedKeys}
D.~R.~L. Brown, ``Bounds on surmising remixed keys,'' \emph{{IACR} Cryptol.
  ePrint Arch.}, p. 375, 2015.

\bibitem{Raab1998BallsIntoBins}
M.~Raab and A.~Steger, ````balls into bins'' --- a simple and tight analysis,''
  in \emph{Randomization and Approximation Techniques in Computer
  Science}.\hskip 1em plus 0.5em minus 0.4em\relax Springer Berlin Heidelberg,
  1998, pp. 159--170.

\bibitem{Berenbrink2006BalancedAllocations}
P.~Berenbrink, A.~Czumaj, A.~Steger, and B.~V\"{o}cking, ``Balanced
  allocations: The heavily loaded case,'' \emph{SIAM J. Comput.}, vol.~35,
  no.~6, p. 1350–1385, jun 2006.

\bibitem{juradoDE19}
M.~Jurado and G.~Smith, ``Quantifying information leakage of deterministic
  encryption,'' in \emph{Proc. of the 2019 ACM SIGSAC Conf. on Cloud Computing
  Security Workshop}, ser. CCSW'19.\hskip 1em plus 0.5em minus 0.4em\relax ACM,
  2019, p. 129–139.

\end{thebibliography}

\appendix


\subsection{\review{Reduced} $k$-RR mechanism and its compositional properties}
\label{sec:model-reduced-krr}

Here we complement the discussion on reduced mechanisms,
presented in Sec.~\ref{sec:model-mechanisms-datasets},
by introduced a reduced version of the $k$-RR mechanism
and proving its compositional properties with the
reduced shuffling mechanism.

\paragraph{Reduced channel for the $k$-RR mechanism}
For completeness, we can also consider a reduced version of the
$k$-RR mechanism that operates by taking a histogram 
as input and producing a histogram over randomized values
as output.

We want to define the reduced channel $k$-RR operating directly over histograms
\review{so that} its final effect is the same as the expected effect of a 
full $k$-RR channel operating over all concrete datasets with the same histogram as 
the input histogram, weighted \review{by} uniform prior distribution on \review{datasets}.

Formally, \review{let} $Pr(x,y,z_{1},z_{2})$ \review{denote} the joint probability
over datasets $x,y\in\calk^{n}$ and histograms $z_{1},z_{2}$ over $\calk$.
We will derive the reduced $k$-RR channel $\Conc{N}^{r}$ mapping histograms
to histograms by imposing restrictions on this joint probability.
We start by deriving, for all histograms $z_{1},z_{2}$ over $\calk$:
\begin{align}
\label{prop:reducedN-1}
&\Conc{N}^{r}_{z_{1},z_{2}} = \text{(def.\ of channel)} \nonumber \\ 
&Pr(z_{2} \mid z_{1}) = \text{(marginalization)} \nonumber \\
&\textstyle \sum_{\substack{x \in \calk^n \\ y \in \calk^{n}} } \,\, Pr(x,y,z_{2} \mid z_{1}) = \text{(chain rule)} \nonumber \\ 
&\textstyle \sum_{\substack{x \in \calk^n \\ y \in \calk^{n}} } \,\, Pr(x \mid z_{1}) Pr(y \mid x,z_{1}) Pr(z_{2} \mid x,y,z_{1})
\end{align}

Now, we impose the restriction that $Pr(y \mid x, z_{1})$ must be
exactly the probability $\Conc{N}_{x,y}$ that the corresponding
full channel $\Conc{N}$ operating on a dataset $x$ would produce
dataset $y$ as output.
We also impose that $z_{2}$ must be the histogram of dataset 
$y$, so $Pr(z_{2} \mid x,y,z_{1})$ must be 1 exactly when
$h(y)=z_{2}$, and 0 otherwise.
By applying that to \eqref{prop:reducedN-1}, we get:
\begin{align}
\label{prop:reducedN-2}
\Conc{N}^{r}_{z_{1},z_{2}} =&\,\, \textstyle \sum_{\substack{x \in \calk^n \\ y \in \calk^{n}: h(y)=z_{2}}} Pr(x \mid z_{1}) \Conc{N}_{x,y} & \text{}  
\end{align}

Now let us denote by $\pi_{x \mid z_{1}}$ the conditional probability $Pr(x \mid z_{1})$
that the input dataset is $x$ given that its histogram must be $z_{1}$.
Considering that the distribution on input datasets is uniform,
as per \eqref{eq:uniform-prior}, we then derive:
\begin{align}
\label{prop:reducedN-3}
\pi_{x \mid z_{1}}
=
Pr(x \mid z_{1}) 
= 
\begin{cases}
 \nicefrac{1}{\#h(x)}, & \text{if $h(x)=z_{1}$}, \\
        0,  & \text{otherwise}.
\end{cases}
\end{align}

Finally, by substituting \eqref{prop:reducedN-3} in \eqref{prop:reducedN-2},
we reach the definition of a reduced shuffling channel $\Conc{N}^{r}$ below.

\begin{definition}[Reduced $k$-RR channel]
\label{def:krr-reduced}
Given a uniform prior $\pi$ on all possible original datasets $x \in \calk^{n}$,
a \emph{reduced $k$-RR channel} $\Conc{N}^{r}$ (again, for \qm{\underline{n}oise})
is a channel from histograms on $\calk$ to histograms on $\calk$ s.t.,
for every input histogram $z_{1}$ and output histogram $z_{2}$,
\begin{align*}
\Conc{N}^{r}_{z_{1},z_{2}} 
=
\textstyle 
\sum_{\substack{x \in \calk^n: h(x)=z_{1} \\ y \in \calk^{n}: h(y)=z_{2}}} \frac{1}{\#h(x)} \Conc{N}_{x,y}. 
\end{align*}
\end{definition}

\begin{example}[Reduced $k$-RR channel]
\label{exa:running-krr-channel-reduced}
Table~\ref{tab:reduced-channel-krr} 
contains the channel $\Conc{N}^{r}$ representing the application of a reduced 
$k$-RR mechanism to the scenario of Example~\ref{exa:running_initial}.
\begin{table}[bt]
\centering
    \begin{tabular}{c|c c c c}
	$\Conc{N}^{r}$ & \ta{:}3, \tb{:}0  & \ta{:}2, \tb{:}1 &	\ta{:}1, \tb{:}2 & \ta{:}0, \tb{:}3 \\
	\hline
        \ta{:}3, \tb{:}0 & $p^{3}$ & $3p^{2}\overline{p}$ & $3p\overline{p}^{2}$ & $\overline{p}^{3}$  \\
        \ta{:}2, \tb{:}1 & $p^{2}\overline{p}$ & $\nicefrac{(p^{3} + 2p\overline{p}^{2})}{3}$ & $\nicefrac{(2p^{2}\overline{p} + \overline{p}^{3})}{3}$ & $p\overline{p}^{2}$ \\
        \ta{:}1, \tb{:}2 & $p\overline{p}^{2}$ & $\nicefrac{(2p^{2}\overline{p} + \overline{p}^{3})}{3}$ & $\nicefrac{(p^{3} + 2p\overline{p}^{2})}{3}$ & $p^{2}\overline{p}$ \\
        \ta{:}0, \tb{:}3 & $\overline{p}^{3}$ & $3p\overline{p}^{2}$ & $3p^{2}\overline{p}$ & $p^{3}$ 
    \end{tabular}
    \caption{Reduced $k$-RR channel $\Conc{N}^{r}$ for Example~\ref{exa:running-krr-channel-reduced}, 
    with $k{=}2$ possible sensitive values and $n{=}3$ individuals.
    The channel receives histograms as inputs and produces histograms as outputs.}
    \label{tab:reduced-channel-krr}
\end{table}
\end{example}

In contrast to the case regarding shuffling channels, 
the full $k$-RR channel $\Conc{N}$ and its reduced counterpart $\Conc{N}^{r}$
are not equivalent.
Indeed, they are not even comparable w.r.t, since
they do not have the same input type.

\begin{restatable}[Non-equivalence between full and reduced $k$-RR]{proposition}{nonequivalentN}\label{prop:nonequivalence-N}
Let $\Conc{N}$ be a full $k$-RR channel as per Def.~\ref{def:krr-channel},
and $\Conc{N}^{r}$ be the reduced $k$-RR channel obtained from $\Conc{N}$ 
as per Def.~\ref{def:krr-reduced}.
Then
\begin{align*}
  \Conc{N} \not\equiv \Conc{N}^{r}.
\end{align*}
\end{restatable}

\paragraph{Compositions of reduced mechanisms}
As with their full counterparts (\review{mapping datasets to datasets}),
the reduced sanitization mechanisms (\review{operating over histograms}) 
can be applied to an input dataset either in isolation or in combination.

\begin{figure}[tb]
    \centering
    \includegraphics[width=0.8\linewidth]{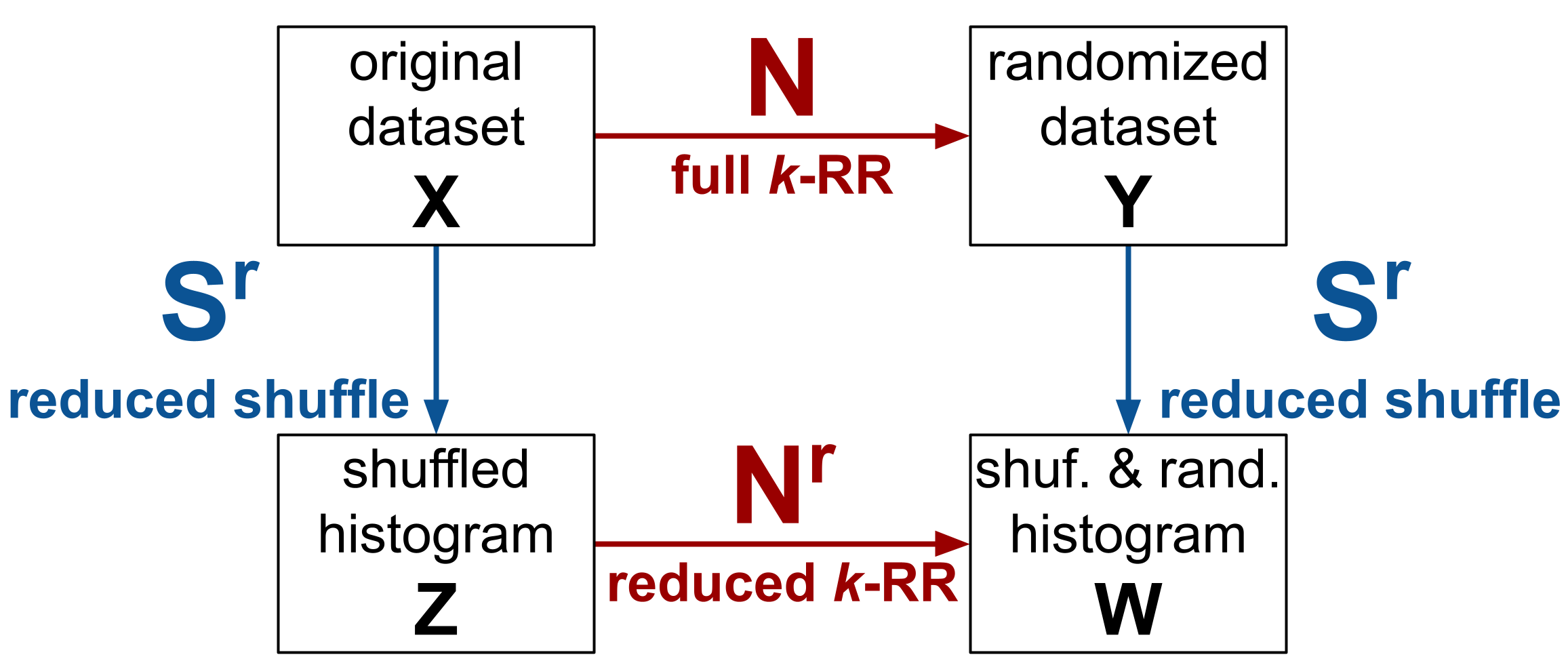}
    \caption{Commutativity in the reduced case:
    $\Conc{N}\Conc{S}^{r} = \Conc{S}^{r}\Conc{N}^{r}$.
    Channel $\Conc{N}$ maps datasets to datasets,
    channel $\Conc{S}^{r}$ maps datasets to histograms, and
    channel $\Conc{N}^{r}$ maps histograms to histograms.}
    \label{fig:commutativity-diagrams-reduced}
\end{figure}

The situation regarding reduced mechanisms, however, is more subtle.
Notice that we are allowed to cascade a reduced shuffling channel $\Conc{S}^{r}$
with a reduced $k$-RR channel $\Conc{N}^{r}$, since the output of the former and the
input of the latter are both histograms, and the first channel operates on datasets.
But we cannot start our sanitization process by applying a reduced $k$-RR channel
$\Conc{N}^{r}$ to a dataset, since this channel receives as input only histograms.
It turns out, however, that a meaningful form of commutativity between shuffling 
and $k$-RR holds if we are careful with the types of the channels in the cascade,
as depicted in Fig.~\ref{fig:commutativity-diagrams-reduced}.
That is formalized in the following result.

\begin{restatable}[Commutativity of $k$-RR and shuffling: Reduced case]{proposition}{commutativityReduced}\label{theorem:commutativity-reduced}
Let $\caln = \{0,1,\ldots,n-1\}$ be a set of $n\geq1$ individuals
and $\calk$ be a set of $k \geq 2$ values for the sensitive attribute.
Let also $\Conc{N}$ be a full $k$-RR channel as per Definition~\ref{def:krr-channel},
$\Conc{N}^{r}$ be a reduced $k$-RR channel as per Definition~\ref{def:krr-reduced},
and $\Conc{S}^{r}$ be a reduced shuffling channel per Definition~\ref{def:shuffling-channel-reduced}.
Then 
\begin{align*}
    \Conc{N}\Conc{S}^{r} = \Conc{S}^{r}\Conc{N}^{r}.
\end{align*}
\end{restatable}

\begin{example}[Cascading of $k$-RR and shuffling in the reduced case]
\label{exa:cascading-reduced}
Consider again the scenario of Example~\ref{exa:running_initial}. 
The channel $\Conc{N}\Conc{S}^{r}$ representing the application 
of a full $k$-RR mechanism followed by reduced shuffling and
the channel $\Conc{S}^{r}\Conc{N}^{r}$ representing the application
of reduced shuffling followed by a reduced $k$-RR mechanism are identical, as
represented in Table~\ref{tab:NS-SN-reduced}.

\begin{table}[tb]
\centering
    \begin{tabular}{c| c c c c c}
        $\Conc{N}\Conc{S}^{r}$/$\Conc{S}^{r}\Conc{N}^{r}$ & (\ta{:}3,\tb{:}0) &	(\ta{:}2,\tb{:}1) &	(\ta{:}1,\tb{:}2) &	(\ta{:}0,\tb{:}3) \\
        \ta\ta\ta & \colorcell{NSrc1}{$p^3$}    &  \colorcell{NSrc2}{$3p^2 \overline{p}$}           & \colorcell{NSrc3}{$3p \overline{p}^2$}          & \colorcell{NSrc4}{$\overline{p}^3$} \Tstrut\\
        \ta\ta\tb & \colorcell{NSrc5}{$p^2 \overline{p}$}  & \colorcell{NSrc6}{$p^3 + 2 p \overline{p}^2$}    & \colorcell{NSrc7}{$2p^2 \overline{p} + \overline{p}^3$}    & \colorcell{NSrc8}{$p \overline{p}^2$} \\
        \ta\tb\ta & \colorcell{NSrc5}{$p^2 \overline{p}$}  & \colorcell{NSrc6}{$p^3 + 2 p \overline{p}^2$}    & \colorcell{NSrc7}{$2p^2 \overline{p} + \overline{p}^3$}    & \colorcell{NSrc8}{$p \overline{p}^2$} \\
        \tb\ta\ta & \colorcell{NSrc5}{$p^2 \overline{p}$}  & \colorcell{NSrc6}{$p^3 + 2 p \overline{p}^2$}    & \colorcell{NSrc7}{$2p^2 \overline{p} + \overline{p}^3$}   & \colorcell{NSrc8}{$p \overline{p}^2$} \\
        \ta\tb\tb & \colorcell{NSrc8}{$p \overline{p}^2$}  & \colorcell{NSrc7}{$2p^2 \overline{p} + \overline{p}^3$}    & \colorcell{NSrc6}{$p^3 + 2 p \overline{p}^2$}    & \colorcell{NSrc5}{$p^2 \overline{p}$} \\
        \tb\ta\tb& \colorcell{NSrc8}{$p \overline{p}^2$}  & \colorcell{NSrc7}{$2p^2 \overline{p} + \overline{p}^3$}    & \colorcell{NSrc6}{$p^3 + 2 p \overline{p}^2$}   & \colorcell{NSrc5}{$p^2 \overline{p}$} \\
        \tb\tb\ta& \colorcell{NSrc8}{$p \overline{p}^2$}  & \colorcell{NSrc7}{$2p^2 \overline{p} + \overline{p}^3$}    & \colorcell{NSrc6}{$p^3 + 2 p \overline{p}^2$}    & \colorcell{NSrc5}{$p^2 \overline{p}$} \\
        \tb\tb\tb & \colorcell{NSrc4}{$\overline{p}^3$}    & \colorcell{NSrc3}{$3p \overline{p}^2$}           & \colorcell{NSrc2}{$3p^2 \overline{p}$}          & \colorcell{NSrc1}{$p^3$}
    \end{tabular}
    \caption{\review{Channel for Example~\ref{exa:cascading-reduced}, 
    representing both the cascading $\Conc{N}\Conc{S}^{r}$
    of full $k$-RR followed by reduced shuffling and the cascading
    $\Conc{S}^{r}\Conc{N}^{r}$ of reduced shuffling followed by reduced $k$-RR,
    with $k{=}2$ possible sensitive values and $n{=}3$ individuals.
    Here $p$ is the probability a user responds with their true value, and $\overline{p}{=}1{-}p$.
    The cells are colored according to the entries' values when $p{=}0.75$ and $\overline{p}{=}0.25$.
    Here $p^3 + 2 p \overline{p}^2$ represents the largest value at approximately 0.5156 while $\overline{p}^3$ is the smallest at approximately 0.0156.
    }}
    \label{tab:NS-SN-reduced}
\end{table}
\end{example}

Finally, we provide the following result, stating
that the composed mechanisms of shuffling and $k$-RR in their full and
reduced forms are equivalent, in the sense that, for every $g$-vulnerability 
measure and prior distribution on secret values, both yield the same quantification of information leakage.

\begin{restatable}[Equivalence of full and reduced compositions]{proposition}{equivalenceCompositionsB}\label{theorem:equivalenceCompositions-B}
Let $\caln = \{0,1,\ldots,n-1\}$ be a set of $n\geq1$ individuals
and $\calk$ be a set of $k \geq 2$ values for the sensitive attribute.
Let also $\Conc{N}$ be a full $k$-RR channel as per Definition~\ref{def:krr-channel},
$\Conc{N}^{r}$ be a reduced $k$-RR channel as per Definition~\ref{def:krr-reduced},
and $\Conc{S}^{r}$ be a reduced shuffling channel per Definition~\ref{def:shuffling-channel-reduced}.
Then:
\begin{align}
    \Conc{S}\Conc{N} \equiv&\,\, \Conc{S}^{r}\Conc{N}^{r} \label{eq:01b}
\end{align}
\end{restatable}

\begin{figure}[tb]
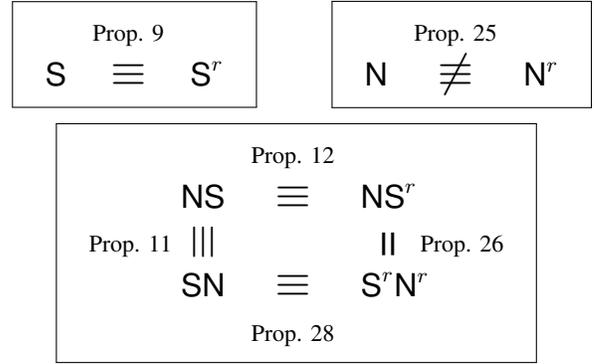

\centering
\begin{large}
\renewcommand{\arraystretch}{1.2}
\fbox{
$
\begin{array}{rcl}
     & \text{\small Prop.~\ref{prop:equivalence-S}} & \\
    \Conc{S} & \text{\LARGE $\equiv$} & \Conc{S}^{r} \\
\end{array}
$
}
\qquad
\fbox{
$
\begin{array}{rcl}
     & \text{\small Prop.~\ref{prop:nonequivalence-N}} & \\
    \Conc{N} & \text{\LARGE $\not\equiv$} & \Conc{N}^{r} \\
\end{array}
$
}
\\
\vspace{2mm}
\fbox{
$
\begin{array}{rcl}
    & \text{\small Prop.~\ref{theorem:equivalenceCompositions-A}} & \\
    \Conc{N}\Conc{S} & \text{\LARGE $\equiv$} & \Conc{N}\Conc{S}^{r} \\
    \text{\small Prop.~\ref{theorem:commutativity-full}} \,\,\,\, \rotatebox[origin=c]{90}{\LARGE $\equiv$} \,\,\, &  & \,\,\,\, \rotatebox[origin=c]{90}{\LARGE =} \,\,\,\, \text{\small Prop.~\ref{theorem:commutativity-reduced}}\\
     \Conc{S}\Conc{N} &  \text{\LARGE $\equiv$} & \Conc{S}^{r}\Conc{N}^{r} \\
     & \text{\small Prop.~\ref{theorem:equivalenceCompositions-B}} & \\
\end{array}
$
}
\end{large}
\caption{All relationships among compositions of full and reduced $k$-RR and shuffling, extending the results from Fig.~\ref{fig:partial-diagram}.
Here $(=)$ denotes syntactic equality, whereas $(\equiv)$ denotes
equivalence w.r.t.\ information leakage.}
\label{fig:overall-diagram}
\end{figure}

Fig.~\ref{fig:overall-diagram} 
extends Fig.~\ref{fig:partial-diagram} and
summarizes all relationships among 
compositions of the full and reduced versions of both the $k$-RR 
mechanism and of shuffling.

\subsection{Using Integer Partitions}
\label{sec:brown}

\review{Although Prop.~\ref{proposition:target_Sk} and Prop.~\ref{proposition:target_SNk} 
provide exact values for posterior vulnerability under 
the shuffle and $k$-RR channels, 
the formulas are computationally expensive.}

\review{
We can 
improve these formulas
by eliminating computational redundancies in \eqref{equ:target_Sk}. 
E.g., when $n{=}6$ and $k{=}3$, an adversary could observe \ta:2, \tb:3, \tc:1 (so $n_1{=}2$, $n_2{=}3$, $n_3{=}1$), or they could observe \ta:3, \tb:1, \tc:2 (so $n_1{=}3$, $n_2{=}1$, $n_3{=}2$). 
Both cases have the same multinomial coefficient and maximum $n^{*}$.
Notice that the exact assignment of value to $n^{*}$ does not matter, as long as they are all accounted for.
With this insight, we follow~\cite{juradoDE19} and use \emph{integer partitions}, i.e.,
partitions of $n$ into exactly $k$ parts such that their elements sum to $n$.
E.g., when $n{=}6$ and $k{=}3$, all relevant integer partitions are $[6, 0, 0], [5, 1, 0], [4, 2, 0], [4, 1, 1], [3, 3, 0], [3, 2, 1]$, and $[2, 2, 2]$.
Instead of iterating through 28 multi-sets, we only need to consider 7 cases.
Hence, we can rewrite \eqref{equ:target_Sk} as
\begin{equation}\label{equ:v_intpart}
    \SV \hyperDist{\Conc{S}} =
    \SV \hyperDist{\Conc{S}^{r}} = \frac{1}{k^n} \sum_{\lambda_{n, k}} \binom{n}{\lambda} \binom{k}{\bar{\lambda} \quad k - \ell(\lambda)}\frac{\lambda^{*}}{n},
\end{equation}
where $\lambda$ is the integer partition of $n$, 
$\bar{\lambda}$ is the multiplicity of each value in the partition, 
$l$ is the length of $\lambda$, and 
$\lambda^{*}{=}\max_{i}\lambda_{i}$ is the maximum value in $\lambda$.
E.g., given $n{=}20$ and $k{=}4$,  when $\lambda = (8, 8, 1, 1, 1, 1)$, then $\ell{=}6$, and $\bar{\lambda}{=}(4, 2)$.
Note that both coefficients here are multinomials, as $\lambda$ and often $\bar{\lambda}$ represent sets of numbers, and they always sum to $n$ and $k$ respectively.
We can simplify this equation further if we adjust \eqref{equ:v_intpart} by removing 
the prior distribution of the secrets, and multiplying the summation by $n$, therefore rewriting it as
}
\begin{equation}\label{equ:ADJv_intpart}
    \textstyle \sum_{\lambda_{n, k}} \binom{n}{\lambda} \binom{k}{\bar{\lambda} \quad k - \ell(\lambda)}\lambda^{*}~.
\end{equation}
\review{It turns out that Brown has investigated this problem on surmising remixed keys~\cite{Brown2015BoundsonRemixedKeys}, where he uses the formulation}
\begin{equation}\label{equ:brown}
    B_{n,k} = 
    \textstyle
    \sum_{|\lambda| = n} \lambda^{*} \frac{n!}{\lambda! \bar{\lambda}!} \ell! \binom{k}{\ell},
\end{equation}
where $\lambda^{*}$ is the maximum value in the integer partition.
It can be shown that our formulation and Brown's are equivalent.

\begin{restatable}{proposition}{equivBrown}\label{prop:equivalenceBrown}
Equations~\ref{equ:ADJv_intpart} and  ~\ref{equ:brown} are equivalent.
\end{restatable}

Brown, however, states that he was unable to find a faster way to compute these values exactly.

\onecolumn

\newpage


\subsection{Proofs}
\label{sec:proofs}

In this section we provide proofs absent from the main body of the paper.


\subsubsection{Auxiliary technical results}
\label{sec:proofs-aux}

\begin{lemma}[Probability that full $k$-RR produces an output with a specific histogram for a given input: Binary case]
\label{lemma:aux}
Let $\caln = \{0,1,\ldots,n-1\}$ be a set of $n\geq1$ distinct individuals of interest, $\calk = \{\ta,\tb\}$ be binary set (i.e., $k=2$) for each individual's possible sensitive attribute, and
$\Conc{N}$ be a full $k$-RR channel as per Def.~\ref{def:krr-channel}.
Then, the probability that an input dataset $x\in\calk^{n}$
is mapped through $\Conc{N}$ to an output dataset 
with histogram $h(y)= \ta{:}n_{\ta}(y), \tb{:}n_{\tb}(y)$, for some $y\in\calk^{n}$, is given by
\begin{align}
\label{eq:lemmaaux}
\sum_{w\in\calk^{n}:h(w)=h(y)}
\Conc{N}_{x,w} = &\,\,
\sum_{m_{\ta} = \max(n_{\ta}(x)-n_{\tb}(y),0)}^{\min(n_{\ta}(x),n_{\ta}(y))} \binom{n_{\ta}(x)}{m_{\ta}} \binom{n_{\tb}(x)}{n_{\tb}(x)-n_{\ta}(y)+m_{\ta}} p^{2m_{\ta}+n_{\tb}(x)-n_{\ta}(y)} \overline{p}^{n-2m_{\ta}-n_{\tb}(x)+n_{\ta}(y)}~.
\end{align}
\end{lemma}


\begin{proof}
The proof can be constructed from the following observations:
\begin{enumerate}[(i)]
    \item From Def.~\ref{def:krr-channel}, the mapping from an input dataset $x$ to an output dataset $w$ via the $k$-RR channel $\Conc{N}$ can be seen as a Bernoulli process with  $n$ independent tries
    (one for each individual in $x$), 
    each having probability $p$ of $w_{i}$ matching the original 
    value of $x_{i}$ (a \qm{match}), and probability $\overline{p}=1-p$ 
    of $w_{i}$ not matching the original value of $x_{i}$ (a \qm{mismatch}).
    \item Now, fix the dataset $x$, so all its elements, and, consequently, $n_{\ta}(x)$ and $n_{\tb}(x)$, are fixed.
    Fix also the histogram  $h(w)$ of dataset $w$ to be the same 
    histogram $h(y)$ of dataset $y$, so $n_{\ta}(w)=n_{\ta}(y)$ and $n_{\tb}(w)=n_{\tb}(y)$ are fixed, but not necessarily each entry $w_{i} = y_{i}$.
    Let $m(x,w)$ be the total number of matches in the mapping from a particular 
    $x$ to a particular $w$ (i.e., the count of positions $i$ s.t.\ $x_{i} = w_{i}$).
    We can split $m$ into $m(x,w) = m_{\ta}(x,w) + m_{\tb}(x,w)$, where 
    $m_{\ta}(x,w)$ and $m_{\tb}(x,w)$ are, respectively, the total
    number of matching $\ta$'s between $x$ and $w$ 
    (i.e., the count of positions $i$ s.t.\ $x_{i} = w_{i} = \ta$) and
    the total number of matching $\tb$'s 
    (i.e., the count of positions $i$ s.t.\ $x_{i} = w_{i} = \tb$).
    \item The number $m_{\ta}(x,w)$ of matching $\ta$'s between $x$ and $w$ must respect the following inequalities.
    \begin{enumerate}
        \item Its lower limit is
        \begin{align}
        \label{eq:lemmaauxlower}
            m_{\ta}(x,w) \geq \max(n_{\ta}(x)-n_{\tb}(y),0)
        \end{align}
    since for each mismatch of an $\ta$ it is necessary to exist a $\tb$ in $w$
    paired up with an $\ta$ in $x$. 
    Hence, the number of mismatching $\ta$'s,
    given by $n_{\ta}(x) - m_{\ta}(x,w)$, can be at most the number $n_{\tb}(y)$ of
    $\tb$'s in $w$.
    This gives us $n_{\ta}(x) - m_{\ta}(x,w) \leq n_{\tb}(y)$, or, equivalently,
    $m_{\ta}(x,w) \geq n_{\ta}(x)-n_{\tb}(y)$.
    When $n_{\tb}(y) > n_{\ta}(x)$, this only tells us that $m_{\ta}(x,w)$ can be 0,
    since it can never be negative.
    \item Its upper limit is
    \begin{align}
        \label{eq:lemmaauxupper}
        m_{\ta}(x,w) \leq \min(n_{\ta}(x),n_{\ta}(y)),
    \end{align}
    since the number of matching $\ta$'s cannot be greater than the total number of $\ta$'s in $x$ or in $w$.
    \end{enumerate}
    
    \item The number $m_{\tb}(x,w)$ of matching $\tb$'s is the following function of the numbers $n_{\ta}(x),n_{\tb}(x),n_{\ta}(y),n_{\tb}(y)$ of $\ta$'s and $\tb$'s in $x$ and $w$, and
    the number $m_{\ta}\edits{(x,w)}$ of matching $\ta$'s between $x$ and $w$:
    $$
        m_{\tb}(x,w) = n_{\tb}(x) -  n_{\ta}(y) + m_{\ta}(x,w),
    $$
    since once $m_{\ta}(x,w)$ values $\ta$ are matched between $x$ and $w$, the other $n_{\ta}(x)-m_{\ta}(x,w)$ values $\ta$ in $x$ must be paired up with that same number of $\tb$'s from $w$.
    So, from the total number $n_{\tb}(y)$ of $\tb$'s in $w$, only the
    remaining $n_{\tb}(y) - (n_{\ta}(x)-m_{\ta}\edits{(x,w)})$ will produce a match with $\tb$'s from $x$.
    By substituting $n_{\tb}(y) = n - n_{\ta}(y)$ and
    $n_{\ta}(x) = n - n_{\tb}(x)$, we get to the equality above.
    \item Now recall that we have already fixed $x$ (i.e., the value of each entry of $x_i$ of $x$ is fixed) and the histogram of $h(w)=h(y)$ (i.e.,  $n_{\ta}(w) = n_{\ta}(y)$ and $n_{\tb}(w) = n_{\tb}(y)$), but \edits{not}
    necessarily the positions in which each $\ta$ and $\tb$ are in $w$.
    Let us go further and also fix the number $m_{\ta}(x,w)$ 
    of matching $\ta$'s between $x$ and $w$ as well.
    From that, we have the following:
    \begin{enumerate}
        \item To get exactly $m_{\ta}(x,w)$ matching $\ta$'s between $x$ and $w$, we need to pick from the $n_{\ta}(x)$ positions in $x$ with $\ta$'s the $m_{\ta}(x,w)$ ones which will be paired up with one of the $n_{\ta}(y)$ $\ta$'s from $w$. 
        There are $\binom{n_{\ta}(x)}{m_{\ta}}$ ways of doing that.
        \item To get exactly $m_{\tb}(x,\edits{w})$ matching $\tb$'s between $x$ and $w$, we need to pick from the $n_{\tb}(x)$ positions in $x$ with $\tb$'s the $m_{\tb}(x,w) = n_{\tb}(x)-n_{\ta}(y) + m_{\ta}(x,w)$ ones which will be paired up with one of the
        $n_{\tb}(y)$ $\tb$'s from $w$. 
        There are $\binom{n_{\tb}(x)}{n_{\tb}(x)-n_{\ta}(y) + m_{\ta}(x,w)}$ ways of doing that.
        \item The total number of matches between $x$ and $w$ is, then, $m_{\ta}(x,w)+m_{\tb}(x,w) = 2m_{\ta}(x,w)+n_{\tb}(x)-n_{\ta}(y)$, and the total number of failures is $n-2m_{\ta}(x,w)-n_{\tb}(x)+n_{\ta}(y)$.
    \end{enumerate}
    Hence, the probability that $\Conc{N}$ maps a fixed input dataset $x$
    to an output dataset $w$ having histogram $h(y)$ and a fixed number $m_{\ta}(x,w)=m_{\ta}$ of matching $\ta$'s between input and output is
    \begin{align}
        \label{eq:lemmaux01}
        \binom{n_{\ta}(x)}{m_{\ta}} \binom{n_{\tb}(x)}{n_{\tb}(x)-n_{\ta}(y)+m_{\ta}} p^{2m_{\ta}+n_{\tb}(x)-n_{\ta}(y)} \overline{p}^{n - 2m_a - n_{\tb}(x) + n_{\ta}(y)}
    \end{align}
\item Finally, we obtain the desired result in \eqref{eq:lemmaaux} by summing over all possible values $m_{\ta}$ of matching $\ta$'s given
by \eqref{eq:lemmaauxlower} and \eqref{eq:lemmaauxupper}.
\end{enumerate}
\end{proof}
\subsubsection{Proofs of Sec.~\ref{sec:model}}
\label{sec:proofs-model}

\equivalentS*

\begin{proof}
This follows from the fact that $\Conc{S}^{r}$ is obtained from
$\Conc{S}$ only by the operation of merging columns that 
are multiples of each other, which does not alter vulnerability
measures~\cite{alvim2020QIF}.
\end{proof}

\commutativityFull*

\begin{proof}
For simplicity, we provide a proof for the case $\calk$ has $k=2$ values;
the case when $k\geq2$ can be obtained as a direct generalization.

Notice that to achieve the desired result, it is sufficient to show the followin three points.
\begin{enumerate}[(i)]
    \item All columns of $(\Conc{N}\Conc{S})$ that represent output datasets with the same histogram are identical, i.e.,
    for all datasets $x,y,y'$ such that $h(y)=h(y')$,
    \begin{align}
    \label{eq:comm01}
        (\Conc{N}\Conc{S})_{xy} = (\Conc{N}\Conc{S})_{xy'}.
    \end{align}
    \item All columns of $(\Conc{S}\Conc{N})$ that represent output datasets with the same histogram are identical, i.e.,
    for all datasets $x,y,y'$ such that $h(y)=h(y')$,
    \begin{align}
    \label{eq:comm02}
        (\Conc{S}\Conc{N})_{xy} = (\Conc{S}\Conc{N})_{xy'}.
    \end{align}
    \item For every $x \in \calk^{n}$ and histogram $z$ on $\calk$, 
    \begin{align}
    \label{eq:comm03}
    \sum_{\substack{y \in \calk^{n}: \\ h(y)=z}} (\Conc{N}\Conc{S})_{xy}
    =
    \sum_{\substack{y \in \calk^{n}: \\ h(y)=z}} (\Conc{S}\Conc{N})_{xy}.
\end{align}
\end{enumerate}

Note that \eqref{eq:comm01} and \eqref{eq:comm02} tell us that in each channel the columns that represent the same histogram can be merged, since they are identical and that does not change leakage properties.
Then, \eqref{eq:comm03} tells us that the resulting channels are identical, since they both are defined as mappings from datasets to histograms, and all entries are identical.

To derive \eqref{eq:comm01}, notice that for all datasets $x,y,y'$ s.t.\ 
$h(y)=h(y')$:
\begin{align*}
     (\Conc{N}\Conc{S})_{xy} =&\,\, \sum_{w \in \calk^{n}} \Conc{N}_{x,w} \Conc{S}_{w,y} & \text{(matrix multiplication)} \\
    =&\,\, \sum_{\substack{w \in \calk^{n}: \\ h(w) = h(y)}} \frac{1}{\#h(y)}  \Conc{N}_{x,w}  & \text{(Definition of $\Conc{S}$)}\\
    =&\,\, \sum_{\substack{w \in \calk^{n}: \\ h(w) = h(y')}} \frac{1}{\#h(y')}  \Conc{N}_{x,w}  & \text{($h(y)=h(y')$)}\\
    =&\,\,  \sum_{w \in \calk^{n}} \Conc{N}_{x,w} \Conc{S}_{w,y'}   & \text{(matrix multiplication)}\\
    =&\,\, (\Conc{N}\Conc{S})_{xy'} 
\end{align*}

To derive \eqref{eq:comm02}, notice that for all datasets $x,y,y'$ s.t.\ 
$h(y)=h(y')$:
\begin{align}
     (\Conc{S}\Conc{N})_{xy} =&\,\, \sum_{w \in \calk^{n}} \Conc{S}_{x,w} \Conc{N}_{w,y} & \text{(matrix multiplication)} \nonumber \\
    =&\,\, \sum_{\substack{w \in \calk^{n}: \\ h(w) = h(x)}} \frac{1}{\#h(x)}  \Conc{N}_{w,y}  & \text{(Definition of $\Conc{S}$)} \nonumber \\
    =&\,\, \frac{1}{\#h(x)}  \sum_{\substack{w \in \calk^{n}: \\ h(w) = h(x)}}  \Conc{N}_{w,y} & \text{(rearranging)} \nonumber \\
     =&\,\, \frac{1}{\#h(x)}  \sum_{\substack{w \in \calk^{n}: \\ h(w) = h(x)}}  \Conc{N}_{y,w} & \text{($\Conc{N}_{w,y}=\Conc{N}_{y,w}$)} \nonumber \\
     =&\,\, \frac{1}{\#h(x)}  \sum_{\substack{w \in \calk^{n}: \\ h(w) = h(x)}}  \Conc{N}_{y',w} & \text{(see below)} \label{eq:commX}\\
     =&\,\, \frac{1}{\#h(x)}  \sum_{\substack{w \in \calk^{n}: \\ h(w) = h(x)}}  \Conc{N}_{w,y'} & \text{($\Conc{N}_{y',w}=\Conc{N}_{w,y'}$)} \nonumber \\
     =&\,\, \sum_{\substack{w \in \calk^{n}: \\ h(w) = h(x)}} \frac{1}{\#h(x)}  \Conc{N}_{w,y'}  & \text{(rearranging)} \nonumber \\
     =&\,\, \sum_{w \in \calk^{n}} \Conc{S}_{x,w} \Conc{N}_{w,y'} & \text{(Definition of $\Conc{S}$)} \nonumber \\
     =& (\Conc{S}\Conc{N})_{xy'} & \text{(matrix multiplication)} \nonumber
\end{align}
Note that step \eqref{eq:commX} follows because, from the proof of Lemma~\ref{lemma:aux}, $\textstyle \sum_{\substack{w \in \calk^{n}: \\ h(w) = h(x)}}  \Conc{N}_{y,w}$ depends only on the histogram of $y$, and not on its exact contents, so we can replace $y$ with $y'$ as they have the same histogram.

Finally, to derive  \eqref{eq:comm03} we can reason:
\begin{align}
     &  \sum_{\substack{y \in \calk^{n}: \\ h(y)=z}} (\Conc{N}\Conc{S})_{xy} \nonumber \\
    =&\quad \text{(matrix multiplication)} \nonumber \\
     & \sum_{\substack{y \in \calk^{n}: \\ h(y)=z}} \sum_{w \in \calk^{n}} \Conc{N}_{x,w} \Conc{S}_{w,y} \nonumber \\
    =& \quad \text{(Definition of $\Conc{S}$)} \nonumber \\
     & \sum_{\substack{y \in \calk^{n}: \\ h(y)=z}} \sum_{\substack{w \in \calk^{n}: \\ h(w) = h(y)}} \frac{1}{\#h(y)}  \Conc{N}_{x,w} \nonumber \\
    =&\quad \text{(rearranging)} \nonumber \\
     & \sum_{\substack{y \in \calk^{n}: \\ h(y)=z}} \frac{1}{\#h(y)} \sum_{\substack{w \in \calk^{n}: \\ h(w) = h(y)}}  \Conc{N}_{x,w}  \nonumber \\
    =&\quad \text{(Lemma~\ref{lemma:aux})} \nonumber \\ 
    & \sum_{\substack{y \in \calk^{n}: \\ h(y)=z}} \frac{1}{\#h(y)}  \sum_{m_{\ta} = \max(n_{\ta}(x)-n_{\tb}(y),0)}^{\min(n_{\ta}(x),n_{\ta}(y))} \binom{n_{\ta}(x)}{m_{\ta}} \binom{n_{\tb}(x)}{n_{\tb}(x)-n_{\ta}(y)+m_{\ta}} p^{2m_{\ta}+n_{\tb}(x)-n_{\ta}(y)} \overline{p}^{n-2m_{\ta}-n_{\tb}(x)+n_{\ta}(y)} \nonumber \\
    =& \quad \text{($h(y)=h(z) \rightarrow n_{\ta}(y)=n_{\ta}(z)$ and $n_{\tb}(y)=n_{\tb}(z)$)} \nonumber \\
    & \sum_{\substack{y \in \calk^{n}: \\ h(y)=z}} \frac{1}{\#h(y)}  \sum_{m_{\ta} = \max(n_{\ta}(x)-n_{\tb}(z),0)}^{\min(n_{\ta}(x),n_{\ta}(z))} \binom{n_{\ta}(x)}{m_{\ta}} \binom{n_{\tb}(x)}{n_{\tb}(x)-n_{\ta}(z)+m_{\ta}} p^{2m_{\ta}+n_{\tb}(x)-n_{\ta}(z)} \overline{p}^{n-2m_{\ta}-n_{\tb}(x)+n_{\ta}(z)} \nonumber \\
    =& \quad \text{(rearranging the summation)} \nonumber \\
    & \sum_{m_{\ta} = \max(n_{\ta}(x)-n_{\tb}(z),0)}^{\min(n_{\ta}(x),n_{\ta}(z))} \binom{n_{\ta}(x)}{m_{\ta}} \binom{n_{\tb}(x)}{n_{\tb}(x)-n_{\ta}(z)+m_{\ta}} p^{2m_{\ta}+n_{\tb}(x)-n_{\ta}(z)} \overline{p}^{n-2m_{\ta}-n_{\tb}(x)+n_{\ta}(z)} \sum_{\substack{y \in \calk^{n}: \\ h(y)=z}} \frac{1}{\#h(y)} \nonumber \\
    =& \quad \text{(there are exactly $\#h(y)$ datasets with histogram $h(y)$)} \nonumber \\
    & \sum_{m_{\ta} = \max(n_{\ta}(x)-n_{\tb}(z),0)}^{\min(n_{\ta}(x),n_{\ta}(z))} \binom{n_{\ta}(x)}{m_{\ta}} \binom{n_{\tb}(x)}{n_{\tb}(x)-n_{\ta}(z)+m_{\ta}} p^{2m_{\ta}+n_{\tb}(x)-n_{\ta}(z)} \overline{p}^{n-2m_{\ta}-n_{\tb}(x)+n_{\ta}(z)} \frac{1}{\#h(y)} \#h(y) \nonumber \\
    =& \quad \text{(simplification)} \nonumber \\
    & \sum_{m_{\ta} = \max(n_{\ta}(x)-n_{\tb}(z),0)}^{\min(n_{\ta}(x),n_{\ta}(z))} \binom{n_{\ta}(x)}{m_{\ta}} \binom{n_{\tb}(x)}{n_{\tb}(x)-n_{\ta}(z)+m_{\ta}} p^{2m_{\ta}+n_{\tb}(x)-n_{\ta}(z)} \overline{p}^{n-2m_{\ta}-n_{\tb}(x)+n_{\ta}(z)} \label{eq:commutativity-ns}
\end{align}

Analogously, we can derive:
\begin{align}
     &  \sum_{\substack{y \in \calk^{n}: \\ h(y)=z}} (\Conc{S}\Conc{N})_{xy} \nonumber \\
    =&\quad \text{(matrix multiplication)} \nonumber \\
     & \sum_{\substack{y \in \calk^{n}: \\ h(y)=z}} \sum_{w \in \calk^{n}} \Conc{S}_{x,w} \Conc{N}_{w,y} \nonumber \\
    =& \quad \text{(Definition of $\Conc{S}$)} \nonumber \\
     & \sum_{\substack{y \in \calk^{n}: \\ h(y)=z}} \sum_{\substack{w \in \calk^{n}: \\ h(w) = h(x)}} \frac{1}{\#h(x)}  \Conc{N}_{w,y} \nonumber \\
    =&\quad \text{(rearranging)} \nonumber \\
     & \sum_{\substack{w \in \calk^{n}: \\ h(w) = h(x)}} \frac{1}{\#h(x)} 
     \sum_{\substack{y \in \calk^{n}: \\ h(y)=z}}  \Conc{N}_{w,y}  \nonumber \\
    =&\quad \text{(Lemma~\ref{lemma:aux})} \nonumber \\ 
    & \sum_{\substack{y \in \calk^{n}: \\ h(y)=z}} \frac{1}{\#h(x)}  \sum_{m_{\ta} = \max(n_{\ta}(w)-n_{\tb}(z),0)}^{\min(n_{\ta}(w),n_{\ta}(z))} \binom{n_{\ta}(w)}{m_{\ta}} \binom{n_{\tb}(w)}{n_{\tb}(w)-n_{\ta}(z)+m_{\ta}} p^{2m_{\ta}+n_{\tb}(w)-n_{\ta}(z)} \overline{p}^{n-2m_{\ta}-n_{\tb}(w)+n_{\ta}(z)} \nonumber \\
    =& \quad \text{($h(w)=h(x) \rightarrow n_{\ta}(w)=n_{\ta}(x)$ and $n_{\tb}(w)=n_{\tb}(x)$)} \nonumber \\
    & \sum_{\substack{y \in \calk^{n}: \\ h(y)=z}} \frac{1}{\#h(x)}  \sum_{m_{\ta} = \max(n_{\ta}(x)-n_{\tb}(z),0)}^{\min(n_{\ta}(x),n_{\ta}(z))} \binom{n_{\ta}(x)}{m_{\ta}} \binom{n_{\tb}(x)}{n_{\tb}(x)-n_{\ta}(z)+m_{\ta}} p^{2m_{\ta}+n_{\tb}(x)-n_{\ta}(z)} \overline{p}^{n-2m_{\ta}-n_{\tb}(x)+n_{\ta}(z)} \nonumber \\
    =& \quad \text{(rearranging the summation)} \nonumber \\
    & \sum_{m_{\ta} = \max(n_{\ta}(x)-n_{\tb}(z),0)}^{\min(n_{\ta}(x),n_{\ta}(z))} \binom{n_{\ta}(x)}{m_{\ta}} \binom{n_{\tb}(x)}{n_{\tb}(x)-n_{\ta}(z)+m_{\ta}} p^{2m_{\ta}+n_{\tb}(x)-n_{\ta}(z)} \overline{p}^{n-2m_{\ta}-n_{\tb}(x)+n_{\ta}(z)} \sum_{\substack{y \in \calk^{n}: \\ h(y)=z}} \frac{1}{\#h(x)}  \nonumber \\
    =& \quad \text{(there are exactly $\#h(x)$ datasets with histogram $h(x)$)} \nonumber \\
    & \sum_{m_{\ta} = \max(n_{\ta}(x)-n_{\tb}(z),0)}^{\min(n_{\ta}(x),n_{\ta}(z))} \binom{n_{\ta}(x)}{m_{\ta}} \binom{n_{\tb}(x)}{n_{\tb}(x)-n_{\ta}(z)+m_{\ta}} p^{2m_{\ta}+n_{\tb}(x)-n_{\ta}(z)} \overline{p}^{n-2m_{\ta}-n_{\tb}(x)+n_{\ta}(z)} \frac{1}{\#h(x)} \#h(x) \nonumber \\
    =& \quad \text{(simplification)} \nonumber \\
    & \sum_{m_{\ta} = \max(n_{\ta}(x)-n_{\tb}(z),0)}^{\min(n_{\ta}(x),n_{\ta}(z))} \binom{n_{\ta}(x)}{m_{\ta}} \binom{n_{\tb}(x)}{n_{\tb}(x)-n_{\ta}(z)+m_{\ta}} p^{2m_{\ta}+n_{\tb}(x)-n_{\ta}(z)} \overline{p}^{n-2m_{\ta}-n_{\tb}(x)+n_{\ta}(z)} \label{eq:commutativity-sn}
\end{align}
To conclude the proof, notice that \eqref{eq:commutativity-ns} and \eqref{eq:commutativity-sn} are equal.
\end{proof}


\equivalenceCompositionsA*

\begin{proof}
We need to show that $\Conc{N}\Conc{S} \sqsubseteq \Conc{N}\Conc{S}^{r}$ and
that $\Conc{N}\Conc{S}^{r} \sqsubseteq \Conc{N}\Conc{S}$.
But for that, we first notice that, by Prop.~\ref{prop:equivalence-S}, 
$\Conc{S} \equiv \Conc{S}^{r}$, so both $\Conc{S} \sqsubseteq \Conc{S}^{r}$ and
$\Conc{S}^{r} \sqsubseteq \Conc{S}$ hold. 
Then we use the fact that cascading is monotonic on when pre-processed by a same channel~\cite{alvim2020QIF}.
\end{proof}


\subsubsection{Proofs of Sec.~\ref{sec:single}}
\label{sec:proofs-single}

\singletargetprior*

\begin{proof}
\begin{align*}
    \SV(\pi) &= \max_{w \in \mathcal{W}} \sum_{x\in \mathcal{X}}\pi_x \cdot g_{\rm T}(w,x) & \text{(Def.~\ref{def:prior_g_v})} \\
    &= \max_{w \in \mathcal{W}} \sum_{x\in \mathcal{X}} \frac{1}{k^n} \cdot g_{\rm T}(w,x) & \text{(since the prior is uniform)} \\
    &= \max_{w \in \mathcal{W}} \sum_{\substack{x\in \mathcal{X}: \\ x_0 = w}} \frac{1}{k^n} & \text{(Def.~\ref{def:target_k_gf})}\\
    &= \frac{1}{k^n} \max_{w \in \mathcal{W}} \sum_{\substack{x\in \mathcal{X}: \\ x_0 = w}} 1 & \text{(rearrange)}\\
    &= \frac{1}{k^n} k^{n-1} & \text{(each of the remaining $n-1$ can take $k$ possible values)}\\
    &= \frac{1}{k} \text{ .}
\end{align*}
\end{proof}

\targetnoise*

\begin{proof}
As stated in Lemma~\ref{lemma:aux}, channel \Conc{N} can be understood as $n$ independent Bernoulli trials.
The single target gain function determines the expected value of only the first try, which is known to be $p$. 
We can also derive it from the matrix definition as follows:
\begin{align*}
    \SV \hyperDist{\Conc{N}} =& \sum_{y \in \calk^{n}} \max_{w \in \mathcal{W}} \sum_{x \in \calk^{n}} \pi_x \cdot \Conc{N}_{x,y} \cdot g_{\rm T}(w,x) & (\text{Def.~\ref{def:posterior_g_v}}) \\
    =& \sum_{y \in \calk^{n}} \max_{w \in \mathcal{W}} \sum_{\substack{x \in \calk^{n} \\ x_0 = w}} \frac{1}{k^n} \cdot \Conc{N}_{x,y} & (\text{Def.~\ref{def:target_k_gf} and Equ.~\eqref{eq:uniform-prior}}) \\
    =& \frac{1}{k^n} \sum_{y \in \calk^{n}} \max_{w \in \mathcal{W}} \left\{ \sum_{\substack{x \in \calk^{n} \\ x_0 = w}} \text{Pr}(y_0 \mid x_0{=}w) \cdot \prod_{i = 1}^{n-1} \text{Pr}(y_i \mid x_i)\right\}\\
    =& \frac{1}{k^n} \sum_{\substack{y \in \calk^{n}:\\y_0{=}\kappa_0}} \max_{w \in \mathcal{W}}\left\{ \sum_{\substack{x \in \calk^{n} \\ x_0 = w}} \text{Pr}(y_0{=}\kappa_0 \mid x_0{=}w) \cdot \prod_{i = 1}^{n-1} \text{Pr}(y_i \mid x_i)\right\} + \ldots + \\
    & \frac{1}{k^n} \sum_{\substack{y \in \calk^{n}:\\y_0{=}\kappa_{k-1}}} \max_{w \in \mathcal{W}}\left\{ \sum_{\substack{x \in \calk^{n} \\ x_0 = w}} \text{Pr}(y_0{=}\kappa_{k-1} \mid x_0{=}w) \cdot \prod_{i = 1}^{n-1} \text{Pr}(y_i \mid x_i)\right\} & (\text{Split the sum over all $y$ by the value of $y_0$})\\
    =& \frac{1}{k^n} \sum_{\substack{y \in \calk^{n}:\\y_0{=}\kappa_0}} \sum_{\substack{x \in \calk^{n} \\ x_0 = \kappa_0}} \text{Pr}(y_0{=}\kappa_0 \mid x_0{=}\kappa_0) \cdot \prod_{i = 1}^{n-1} \text{Pr}(y_i \mid x_i) + \ldots + \\
    & \frac{1}{k^n} \sum_{\substack{y \in \calk^{n}:\\y_0{=}\kappa_{k-1}}}  \sum_{\substack{x \in \calk^{n} \\ x_0 = \kappa_{k-1}}} \text{Pr}(y_0{=}\kappa_{k-1} \mid x_0{=}\kappa_{k-1}) \cdot \prod_{i = 1}^{n-1} \text{Pr}(y_i \mid x_i) & \text{\parbox{15em}{(As $p\geq\nicefrac{1}{k}$, then Pr$\edits{(}y_0{=}\kappa_i|x_i{=}w\edits{)}$ is maximized when $w=\kappa_i$)}}\\
    =& \frac{p}{k^n} \sum_{\substack{y \in \calk^{n}:\\y_0{=}\kappa_0}} \sum_{\substack{x \in \calk^{n} \\ x_0 = \kappa_0}} \prod_{i = 1}^{n-1} \text{Pr}(y_i \mid x_i) + \ldots + \\
    & \frac{p}{k^n} \sum_{\substack{y \in \calk^{n}:\\y_0{=}\kappa_{k-1}}}  \sum_{\substack{x \in \calk^{n} \\ x_0 = \kappa_{k-1}}} \prod_{i = 1}^{n-1} \text{Pr}(y_i \mid x_i) & \text{(Lemma~\ref{lemma:aux})}\\
    =& \frac{p}{k^n} \sum_{\substack{y \in \calk^{n}:\\y_0{=}\kappa_0}} \sum_{i=0}^{n-1} {n-1\choose i}p^i\left(\frac{1-p}{k-1}\right)^{n-1-i} + \ldots + \\
    & \frac{p}{k^n} \sum_{\substack{y \in \calk^{n}:\\y_0{=}\kappa_{k-1}}}  \sum_{i=0}^{n-1} {n-1\choose i}p^i \left(\frac{1-p}{k-1}\right)^{n-1-i} & \text{\parbox{15em}{(Setting the first index ($x_0$), we count the ways $n-1$ indices could stay or flip.)} }\\
    =& \frac{p}{k^n} \sum_{\substack{y \in \calk^{n}:\\y_0{=}\kappa_0}} 1 + \ldots + \frac{p}{k^n} \sum_{\substack{y \in \calk^{n}:\\y{=}\kappa_{k-1}}} 1 & \text{}\\
    =& \frac{p}{k^n} k^{n-1} + \ldots + \frac{p}{k^n} k^{n-1} & \text{}\\
    =& \frac{p}{k} + \ldots + \frac{p}{k} & \text{}\\
    =& p
\end{align*}
\end{proof}

\targetStwo*

\begin{proof}
    \begin{align}
        \SV \hyperDist{\Conc{S}^r} =& \sum\limits_{y \in \mathcal{Y}} \max\limits_{w \in \mathcal{W}} \sum\limits_{x \in \mathcal{X}} \pi_x \cdot \Conc{S}^r_{x,y} \cdot g_{\rm T}(w,x) & (\text{Def.~\ref{def:posterior_g_v}}) \nonumber \\
        =& \sum\limits_{i=0}^{n} \max\limits_{w \in \mathcal{W}} \sum\limits_{\substack{x \in \mathcal{X}:\\n_a(x) = i}} \frac{1}{2^n} \cdot g_{\rm T}(w,x) & (\text{Def.~\ref{eq:uniform-prior} and \ref{def:shuffling-channel-reduced}}) \nonumber \\
        =& \frac{1}{2^n} \sum\limits_{i=0}^{n} \max\left(\sum\limits_{\substack{x \in \mathcal{X}:\\n_a(x) = i}} g_{\rm T}(\ta,x), \sum\limits_{\substack{x \in \mathcal{X}:\\n_a(x) = i}} g_{\rm T}(\tb,x) \right) &  \nonumber \\
        =& \frac{1}{2^n} \sum\limits_{i=0}^{n} \max\left(\sum\limits_{\substack{x \in \mathcal{X}:\\n_a(x) = i\\x_0=\ta}} 1, \sum\limits_{\substack{x \in \mathcal{X}:\\n_a(x) = i\\x_0=\tb}} 1\right) &  (\text{Def.~\ref{def:target_k_gf}}) \nonumber \\
        =& \frac{1}{2^n} \sum\limits_{i=0}^{n} \max \left({n-1\choose i-1}, {n-1\choose i} \right) \nonumber \\
        =& \frac{1}{2^n} \sum\limits_{i=0}^{n} \max \left({n\choose i} \cdot \frac{i}{n}, {n\choose i} \cdot \frac{n-i}{n} \right) \nonumber \\
        =& \frac{1}{2^n} \sum\limits_{i=0}^{n} {n\choose i} \frac{\max (i,n-i)}{n}. & \nonumber
    \end{align}
    
\noindent Finally we have that $\SV \hyperDist{\Conc{S}^r} = \SV \hyperDist{\Conc{S}}$ by Prop.~\ref{prop:equivalence-S}.
\end{proof}

\targetSbinfast*

\begin{proof}
The proof follows immediately from  Theorem~\ref{theorem:target_SN2_fast}, since making $p=1$ in the $k$-RR mechanism is equivalent to not adding any noise to the dataset.
\end{proof}

\targetSNbin*

\begin{proof}
The proof follows immediately from Proposition~\ref{proposition:target_SNk} when $k = 2$.
\end{proof}

\targetSNBinfast*

\begin{proof}
\begin{align*}
    &\frac{1}{2} + \frac{1}{2^n}(2p-1)\binom{n-1}{\lfloor\frac{n-1}{2}\rfloor} 
    \shrug 
    \frac{1}{2^n} \sum_{i=0}^{n} \binom{n}{i} \times\frac{\max (i, n-i) p + \min(i, n-i)(1-p)}{n} 
    \intertext{Multiply both sides by $2^n$}
     &2^{n-1} + (2p-1)\binom{n-1}{\lfloor\frac{n-1}{2}\rfloor} 
     \shrug
     \sum_{i=0}^{n} \left(\binom{n}{i} \times\frac{\max (i, n-i) p + \min(i, n-i)(1-p)}{n} \right)
    \intertext{Split the right side into two cases.}
    \shrug&
    \sum_{i=0}^{\lfloor n/2 \rfloor} \left(\binom{n}{i} \times \left(\frac{(n-i)p}{n} + \frac{i(1-p)}{n}\right) \right) 
    + \sum_{i=\lfloor n/2 \rfloor +1}^{n} \left(\binom{n}{i} \times \left(\frac{i(p)}{n} + \frac{(n-i)(1-p)}{n} \right) \right)
    \intertext{Remove the cases where $i = 0$ and $i = n$.}
    \shrug&
    \sum_{i=1}^{\lfloor n/2 \rfloor} \left(\binom{n}{i} \times \left(\frac{(n-i)p}{n} + \frac{i(1-p)}{n}\right) \right)
    + \sum_{i=\lfloor n/2 \rfloor +1}^{n-1} \left(\binom{n}{i} \times \left(\frac{i(p)}{n} + \frac{(n-i)(1-p)}{n} \right) \right)
    +2p
    \intertext{Convert binomial to factorial.}
    \shrug&
    \sum_{i=1}^{\lfloor n/2 \rfloor} \left( \frac{n!}{i!(n-i)!} \times \left(\frac{(n-i)p}{n} 
    + \frac{i(1-p)}{n}\right) \right)
    + \sum_{i=\lfloor n/2 \rfloor +1}^{n-1} \left( \frac{n!}{i!(n-i)!} \times \left(\frac{i(p)}{n} 
    + \frac{(n-i)(1-p)}{n} \right) \right)
    +2p 
    \intertext{Distribute.}
    \shrug&
    \sum_{i=1}^{\lfloor n/2 \rfloor} \left( \frac{n!(n-i)p}{i!(n-i)!n} 
    + \frac{n!i(1-p)}{i!(n-i)!n} \right) 
    + \sum_{i=\lfloor n/2 \rfloor +1}^{n-1} \left( \frac{n!ip}{i!(n-i)!n} 
    + \frac{n!(n-i)(1-p)}{i!(n-i)!n} \right)
    +2p 
    \intertext{Cancel.}
    \shrug&
    \sum_{i=1}^{\lfloor n/2 \rfloor} 
    \left(\frac{(n-1)!p}{i!(n-i-1)!} 
    + \frac{(n-1)!(1-p)}{(i-1)!(n-i)!} \right)
    + \sum_{i=\lfloor n/2 \rfloor +1}^{n-1} 
    \left( \frac{(n-1)!p}{(i-1)!(n-i)!} 
    + \frac{(n-1)!(1-p)}{i!(n-i-1)!} \right)
    +2p 
    \intertext{Simplify and factor out $p$ and $(1-p)$.}
    \begin{split}
        \shrug&
        \sum_{i=1}^{\lfloor n/2 \rfloor}
        \Bigg( p\left(\frac{(n-1)!}{i!(n-i-1)!}\right) + (1-p)\left(\frac{(n-1)!}{(i-1)!(n-i)!}\right) \Bigg)
    \end{split}\\
    \begin{split}
        &+ \sum_{i=\lfloor n/2 \rfloor +1}^{n-1} 
        \Bigg( p\left(\frac{(n-1)!}{(i-1)!(n-i)!}\right) + (1-p)\left(\frac{(n-1)!}{i!(n-i-1)!}\right) \Bigg)
        +2p
    \end{split}
    \intertext{Represent as binomial coefficients}
    \shrug&
    p\sum_{i=1}^{\lfloor n/2 \rfloor}\binom{n-1}{i} 
    + (1-p)\sum_{i=1}^{\lfloor n/2 \rfloor} \binom{n-1}{i-1} 
    + p \sum_{i=\lfloor n/2 \rfloor +1}^{n-1} \binom{n-1}{i-1} 
    + (1-p)\sum_{i=\lfloor n/2 \rfloor +1}^{n-1} \binom{n-1}{i} 
    +2p
\intertext{Note}
    &\sum_{i=1}^{\lfloor n/2 \rfloor - 1}\binom{n-1}{i} \equiv \sum_{i=2}^{\lfloor n/2 \rfloor}\binom{n-1}{i-1}
    \text{ and }
    \sum_{i=\lfloor n/2 \rfloor +2}^{n-1} \binom{n-1}{i-1} \equiv
    \sum_{i=\lfloor n/2 \rfloor +1}^{n-2} \binom{n-1}{i}
\intertext{because the index shifts are counterbalanced in the binomial coefficients.
Therefore we can remove the cases when $i = \lfloor \frac{n}{2} \rfloor, 1, \lfloor \frac{n}{2} \rfloor +1$ and $n-1$ from the four respective summations.}
\intertext{Remove the four cases from their respective summations}
    \begin{split}
        \shrug&
        p \sum_{i=1}^{\lfloor n/2 \rfloor -1}\binom{n-1}{i} 
        + p \binom{n-1}{\lfloor \frac{n}{2} \rfloor} 
        + (1-p)\sum_{i=2}^{\lfloor n/2 \rfloor} \binom{n-1}{i-1}
        + (1-p)\binom{n-1}{1-1}
    \end{split}\\
    \begin{split}
        + p \sum_{i=\lfloor n/2 \rfloor +2}^{n-1} \binom{n-1}{i-1}
        + p \binom{n-1}{\lfloor \frac{n}{2} \rfloor +1 -1} 
        + (1-p) \sum_{i=\lfloor n/2 \rfloor +1}^{n-2} \binom{n-1}{i}
        + (1-p) \binom{n-1}{n-1} + 2p
    \end{split}\\
    \intertext{Rearrange and simplify.}
    \begin{split}
        \shrug&
        p \sum_{i=1}^{\lfloor n/2 \rfloor -1}\binom{n-1}{i} 
        + (1-p)\sum_{i=2}^{\lfloor n/2 \rfloor} \binom{n-1}{i-1}
        + p \sum_{i=\lfloor n/2 \rfloor +2}^{n-1} \binom{n-1}{i-1}
        + (1-p) \sum_{i=\lfloor n/2 \rfloor +1}^{n-2} \binom{n-1}{i}
            \end{split}\\
    \begin{split}
        + 2p \binom{n-1}{\lfloor \frac{n}{2} \rfloor} 
        + 2(1-p) + 2p
    \end{split}\\
    \intertext{Now the two left sums are equal and the two right sums are equal. Both $p$ and $1-p$ cancel out for both sets of equal terms since $p(x) + (1-p)x = x$.}
    \shrug& 
    \sum_{i=1}^{\lfloor n/2 \rfloor -1}\binom{n-1}{i}
    + \sum_{i=\lfloor n/2 \rfloor +1}^{n-2} \binom{n-1}{i}
    + 2p \binom{n-1}{\lfloor \frac{n}{2} \rfloor} +2(1-p) + 2p\\
    \intertext{Simplify.}
    \shrug& 
    \sum_{i=1}^{\lfloor n/2 \rfloor -1}\binom{n-1}{i}
    + \sum_{i=\lfloor n/2 \rfloor +1}^{n-2} \binom{n-1}{i}
    + 2p \binom{n-1}{\lfloor \frac{n}{2} \rfloor} +2\\ %
    \intertext{Note the two sums almost go from 0 to n-1, minus the cases where $i = 0, \lfloor \frac{n}{2} \rfloor$, and $n-1$. We can represent the entire sum, minus these cases.}
    \shrug& 
    \sum_{i = 0}^{n-1} \binom{n-1}{i} 
    - \binom{n-1}{0} 
    - \binom{n-1}{\lfloor \frac{n}{2} \rfloor}
    - \binom{n-1}{n-1}
    + 2p \binom{n-1}{\lfloor \frac{n}{2} \rfloor} +2\\
    \intertext{Evaluate binomials.}
    \shrug& 
    2^{n-1} 
    - 1 
    - \binom{n-1}{\lfloor \frac{n}{2} \rfloor}
    - 1
    + 2p \binom{n-1}{\lfloor \frac{n}{2} \rfloor} +2\\ 
    \intertext{Subtract terms.}
    \shrug& 
    2^{n-1} 
    - \binom{n-1}{\lfloor \frac{n}{2} \rfloor}
    + 2p \binom{n-1}{\lfloor \frac{n}{2} \rfloor}\\
    \intertext{Rearrange.}
    \shrug& 
    2^{n-1} 
    +(2p-1)\binom{n-1}{\lfloor \frac{n}{2} \rfloor}\\
    \intertext{Both sides are equal since the binomial coefficients are equivalent.}
    &2^{n-1} + (2p-1)\binom{n-1}{\lfloor\frac{n-1}{2}\rfloor} 
    = 2^{n-1} 
    +(2p-1)\binom{n-1}{\lfloor \frac{n}{2} \rfloor}
\end{align*}
\end{proof}


\subsubsection{Proofs of Sec.~\ref{sec:single_genk}}
\label{sec:proofs-single_genk}

\targetSk*

\begin{proof}
Let $n_j$ indicate $n_{\kappa_{j}}(x)$ such that $n_{\kappa_{0}} + n_{\kappa_{\edits{1}}} + \dots + n_{\kappa_{k-1}} = n$, per Definition~\ref{def:datasets}.
    \begin{align}
    \intertext{(Def.~\ref{def:posterior_g_v})}
        \SV \hyperDist{\Conc{S}^r} =& \sum\limits_{y{\in}\caly} \max\limits_{w{\in}\calw} \sum\limits_{x{\in}\calx} \pi_x \cdot \Conc{S}^r_{x,y} \cdot g_T(w,x)  \nonumber \\
    \intertext{(Def. \ref{def:datasets} and \ref{def:shuffling-channel-reduced})}
        =& \sum_{n_0, n_{\edits{1}}, \dots, n_{k-1}} \max\limits_{w{\in}\calw} \sum\limits_{\substack{x{\in}\calx:\\h(x)=n_{\edits{0}}\ldots n_{\edits{k-1}}}} \frac{1}{k^n} \cdot g_T(w,x)  \nonumber \\
    \intertext{(Rearranging and Def.~\ref{def:target_k_gf})}
        =& \frac{1}{k^n} \sum_{n_0, n_{\edits{1}}, \dots, n_{k-1}} \max\limits_{w{\in}\calw} \sum\limits_{\substack{x{\in}\calx:\\h(x)=n_{\edits{0}}\ldots n_{\edits{k-1}},\\x_0=w}} 1 \nonumber \\
    \intertext{When we fix $x_0=n_i$, the other $n{-}1$ individuals must be distributed in bins of size $n_1,\ldots,n_i{-}1,\ldots,n_k$}
        =& \frac{1}{k^n} \sum_{n_0, n_{\edits{1}}, \dots, n_{k-1}} \max \left\{{n{-}1\choose n_0{-}1,n_1,\ldots,n_{k-1}},{n{-}1\choose n_0,n_1{-}1,\ldots,n_{k-1}},\ldots,{n{-}1\choose n_0,n_1,\ldots,n_{k{-}2},n_{k{-}1}{-}1}\right\}  \nonumber \\
        =& \frac{1}{k^n} \sum_{n_0, n_{\edits{1}}, \dots, n_{k-1}} \max \left\{{n\choose n_0,n_1,\ldots,n_{k{-}1}}\frac{n_1}{n},\ldots,{n\choose n_0,n_1,\ldots,n_{k{-}1}}\frac{n_k}{n}\right\}  \nonumber \\
        =& \frac{1}{k^n} \sum_{n_0, n_{\edits{1}}, \dots, n_{k-1}} {n\choose n_0,n_1,\ldots,n_{k{-}1}} \frac{\max (n_1,n_2,\ldots,n_k)}{n} \nonumber
    \end{align}
\noindent Finally we have that $\SV \hyperDist{\Conc{S}^r} = \SV \hyperDist{\Conc{S}}$ by Prop.~\ref{prop:equivalence-S}.
\end{proof}

\targetSNk*

\begin{proof}
(Note we shift the indices to begin at $n_0$).
Let $n_j$ indicate $n_{\kappa_{j}}(x)$ such that $n_{\kappa_{0}} + n_{\kappa_{\edits{1}}} + \dots + n_{\kappa_{k-1}} = n$, per Definition~\ref{def:datasets}. \edits{Let also $\mathcal{Z} = \{n_0,n_1,\ldots,n_{k-1}~|~n_0+n_1+\ldots+n_{k+1}=n\}$ be the set of all possible histograms.}
\begin{align*}
\intertext{(Def.~\ref{def:posterior_g_v})}
     &\SV \hyperDist{\Conc{NS}} = \sum_{z \in \mathcal{Z}} \max_{w \in \mathcal{W}} \sum_{x \in \mathcal{X}} \pi_x \cdot \Conc{NS}^{r}_{x,z} \cdot g_T(w,x)  \\
\intertext{(Def.~\ref{def:target_k_gf} and Equ.~\eqref{eq:uniform-prior})}
    =& \sum_{\edits{z \in \mathcal{Z}}} \max_{w \in \mathcal{W}} \sum_{\substack{x \in \mathcal{X} \\ x_0 = w}} \frac{1}{k^n} \cdot \Conc{NS}^{r}_{x,z}  \\
    =& \frac{1}{k^n} \sum_{\edits{z \in \mathcal{Z}}} \max_{w \in \mathcal{W}} \sum_{\substack{x \in \mathcal{X} \\ x_0 = w}} \sum_{\substack{y \in \edits{\calk^n} }} \Conc{N}_{x,y}\Conc{S}^{r}_{y, z} \\
    =& \frac{1}{k^n} \sum_{\edits{z \in \mathcal{Z}}} \max_{w \in \mathcal{W}} \sum_{\substack{x \in \mathcal{X} \\ x_0 = w}} \sum_{\substack{y \in \edits{\calk^n} \\ h(y) = z }} \Conc{N}_{x,y}  \\
    =& \frac{1}{k^n} \sum_{\edits{z \in \mathcal{Z}}} \max_{w \in \mathcal{W}} \sum_{\substack{x \in \mathcal{X} \\ x_0 = w}} \sum_{\substack{y \in \edits{\calk^n} \\ h(y) = z }} \prod_{i = 0}^{n} \text{Pr}(y_i \mid x_i) \\
    =& \frac{1}{k^n} \sum_{\edits{z \in \mathcal{Z}}} \max_{w \in \mathcal{W}} \sum_{\substack{x \in \mathcal{X} \\ x_0 = w}} \sum_{\substack{y \in \edits{\calk^n} \\ h(y) = z }} \text{Pr}(y_0  \mid x_0 ) \cdot \prod_{i = 1}^{n} \text{Pr}(y_i \mid x_i)  \\
    =& \frac{1}{k^n} \sum_{\edits{z \in \mathcal{Z}}} \max_{w \in \mathcal{W}} \sum_{\substack{x \in \mathcal{X} \\ x_0 = w}} \sum_{\substack{y \in \edits{\calk^n} \\ h(y) = z \\y_0 = (x_0 = w) }} \text{Pr}(y_0  \mid x_0 ) \cdot \prod_{i = 1}^{n} \text{Pr}(y_i \mid x_i) + \sum_{\substack{x \in \mathcal{X} \\ x_0 = w}} \sum_{\substack{y \in \edits{\calk^n} \\ h(y) = z \\y_0 \neq (x_0 = w)}} \text{Pr}(y_0  \mid x_0 ) \cdot \prod_{i = 1}^{n} \text{Pr}(y_i \mid x_i)  \\ 
    =& \frac{1}{k^n} \sum_{\edits{z \in \mathcal{Z}}} \max_{w \in \mathcal{W}}
        \sum_{\substack{x \in \mathcal{X} \\ x_0 = w}} \sum_{\substack{y \in \edits{\calk^n} \\ h(y) = z \\y_0 = (x_0 = w) }} p \cdot \prod_{i     = 1}^{n} \text{Pr}(y_i \mid x_i) 
            + \sum_{\substack{x \in \mathcal{X} \\ x_0 = w}}\sum_{\substack{y \in \edits{\calk^n} \\ h(y) = z \\y_0 \neq (x_0 = w)}} \left(\frac{1-p}{k-1}\right) \cdot \prod_{i = 1}^{n} \text{Pr}(y_i \mid x_i) \\
    =& \frac{1}{k^n} \sum_{\edits{z \in \mathcal{Z}}} \max_{w \in \mathcal{W}} 
        p \sum_{\substack{x \in \mathcal{X} \\ x_0 = w}} \sum_{\substack{y \in \edits{\calk^n} \\ h(y) = z \\y_0 = (x_0 = w) }}  \prod_{i     = 1}^{n} \text{Pr}(y_i \mid x_i) 
            + \left(\frac{1-p}{k-1}\right) \sum_{\substack{x \in \mathcal{X} \\ x_0 = w}}\sum_{\substack{y \in \edits{\calk^n} \\ h(y) = z \\y_0 \neq (x_0 = w)}} \prod_{i = 1}^{n} \text{Pr}(y_i \mid x_i) \\
    =& \frac{1}{k^n} \sum_{\edits{z \in \mathcal{Z}}} \max_{w \in \mathcal{W}}
        p  \sum_{\substack{y \in \edits{\calk^n} \\ h(y) = z \\y_0 = w }} \sum_{\substack{x \in \mathcal{X} \\ x_0 = w}} \prod_{i     = 1}^{n} \text{Pr}(y_i \mid x_i) 
            + \left(\frac{1-p}{k-1}\right) \sum_{\substack{y \in \edits{\calk^n} \\ h(y) = z \\y_0 \neq w}} \sum_{\substack{x \in \mathcal{X} \\ x_0 = w}} \prod_{i = 1}^{n} \text{Pr}(y_i \mid x_i) \\
    =& \frac{1}{k^n} \sum_{\edits{z \in \mathcal{Z}}} \max_{w \in \mathcal{W}} 
        p  \sum_{\substack{y \in \edits{\calk^n} \\ h(y) = z \\y_0 =  w }} \sum_{i = 0}^{n-1} \binom{n-1}{i}p^{i} \left(\frac{1-p}{k-1}\right)^{n-1-i}
            + \left(\frac{1-p}{k-1}\right) \sum_{\substack{y \in \edits{\calk^n} \\ h(y) = z \\y_0 \neq  w}} \sum_{i = 0}^{n-1} \binom{n-1}{i}p^{i} \left(\frac{1-p}{k-1}\right)^{n-1-i}  \\ 
    =& \frac{1}{k^n} \sum_{\edits{z \in \mathcal{Z}}} \max_{w \in \mathcal{W}} 
        p  \sum_{\substack{y \in \edits{\calk^n} \\ h(y) = z \\y_0 =  w }} 1
            + \left(\frac{1-p}{k-1}\right) \sum_{\substack{y \in \edits{\calk^n} \\ h(y) = z \\y_0 \neq  w}} 1  \\ 
    =& \frac{1}{k^n} \sum_{\edits{z \in \mathcal{Z}}} \max_{\edits{\kappa}_i:n_i} \Bigg\{ 
        p  \sum_{\substack{y \in \edits{\calk^n} \\ h(y) = z \\y_0 =  \edits{\kappa}_0 }} 1
            + \left(\frac{1-p}{k-1}\right) \sum_{\substack{y \in \edits{\calk^n} \\ h(y) = z \\y_0 \neq \edits{\kappa}_0}} 1 , \\
        & \qquad p  \sum_{\substack{y \in \edits{\calk^n} \\ h(y) = z \\y_0 = \edits{\kappa}_1)}} 1
            + \left(\frac{1-p}{k-1}\right) \sum_{\substack{y \in \edits{\calk^n} \\ h(y) = z \\y_0 \neq \edits{\kappa}_1}} 1 , 
            \dots, \\
        & \qquad p  \sum_{\substack{y \in \edits{\calk^n} \\ h(y) = z \\y_0 = \edits{\kappa}_{k-1} }} 1
            + \left(\frac{1-p}{k-1}\right) \sum_{\substack{y \in \edits{\calk^n} \\ h(y) = z \\y_0 \neq \edits{\kappa}_{k-1}}} 1 \Bigg\} \\ 
    =& \frac{1}{k^n} \sum_{\edits{z \in \mathcal{Z}}} \max_{k_i:n_i} \Bigg\{ 
        p  {n-1\choose n_{0}-1,n_1,\ldots,n_{k{-}1}}
            + \left(\frac{1-p}{k-1}\right) \left({n\choose n_{0},n_1,\ldots,n_{k-1}} - {n-1\choose n_{0}-1,n_1,\ldots,n_{k{-}1}} \right) , \dots , \Bigg\} \\ 
    =& \frac{1}{k^n} \sum_{\edits{z \in \mathcal{Z}}} \max_{k_i:n_i} \Bigg\{ 
        p \frac{n_0}{n} {n\choose n_{0},n_1,\ldots,n_{k{-}1}}
            + \left(\frac{1-p}{k-1}\right) \frac{n - n_0}{n} {n\choose n_{0},n_1,\ldots,n_{k{-}1}} , \dots , \Bigg\} \\
    =& \frac{1}{k^n} \sum_{\edits{z \in \mathcal{Z}}}  {n\choose n_{0},n_1,\ldots,n_{k{-}1}} \max_{k_i:n_i} \Bigg\{ \frac{p \cdot n_0 + \NF{1-p}{k-1}(n - n_0)}{n} , \dots , \Bigg\} \\
\intertext{Let $n^*$ represent $\max \{n_0, n_1, \dots, n_{k-1}\}$.}
    =& \frac{1}{k^n} \sum_{\substack{ n_0, n_1, \dots, n_{k-1}}}  {n\choose n_{0},n_1,\ldots,n_{k{-}1}}\frac{p \cdot n^* + \NF{1-p}{k-1}(n - n^*)}{n}
\end{align*}
\end{proof}


\NSgenkApprox*
\begin{proof}
Let $n_j$ indicate $n_{\kappa_{j}}(x)$ such that $n_{\kappa_{0}} + n_{\kappa_{\edits{1}}} + \dots + n_{\kappa_{k-1}} = n$, per Definition~\ref{def:datasets}.
Recall Equation~\ref{equ:target_NSk}:
\begin{align*}
    \SV \hyperDist{\Conc{NS}} 
    =& \frac{1}{k^n} \sum_{n_0, n_{\edits{1}}, \dots, n_{k-1}} \binom{n}{n_1, n_2, \dots, n_k} 
    \times\frac{(n^{*})p + (n-n^{*})\NF{(1-p)}{(k-1)}}{n} \text{ where } n^{*} = \max(n_1, n_2, \dots, n_k)
    \intertext{For readability, let us denote the multinomial coefficient as $m$. We can rewrite the formula as follows.}
    =& \frac{1}{k^n} \sum_{n_0, n_{\edits{1}}, \dots, n_{k-1}} m \frac{(n^{*})p + (n-n^{*})\NF{(1-p)}{(k-1)}}{n}\\
    =& \frac{1}{k^n} \sum_{n_0, n_{\edits{1}}, \dots, n_{k-1}} m \left(  \frac{(n^{*})p}{n} + \frac{(n-n^{*})(1-p)}{n(k-1)} \right)\\
    =& \frac{1}{k^n}  \sum_{n_0, n_{\edits{1}}, \dots, n_{k-1}} m \frac{(n^{*})p}{n} + \frac{1}{k^n} \sum_{n_0, n_{\edits{1}}, \dots, n_{k-1}} m \frac{(n-n^{*})(1-p)}{n(k-1)}\\
    =& p\frac{1}{k^n}  \sum_{n_0, n_{\edits{1}}, \dots, n_{k-1}} m \frac{(n^{*})}{n} + \frac{(1-p)}{(k-1)}\frac{1}{k^n} \sum_{n_0, n_{\edits{1}}, \dots, n_{k-1}} m \frac{(n-n^{*})}{n}
    \intertext{Note the left half of the sum is $p \times \SV \hyperDist{\Conc{S}}$. }
    =& p\frac{1}{k^n}  \sum_{n_0, n_{\edits{1}}, \dots, n_{k-1}} m \frac{(n^{*})}{n} + \frac{(1-p)}{(k-1)}\frac{1}{k^n} \sum_{n_0, n_{\edits{1}}, \dots, n_{k-1}} m \left(1-\frac{(n^{*})}{n}\right)\\
    =& p\frac{1}{k^n}  \sum_{n_0, n_{\edits{1}}, \dots, n_{k-1}} m \frac{(n^{*})}{n} + \frac{(1-p)}{(k-1)}\frac{1}{k^n} \left(\sum_{n_0, n_{\edits{1}}, \dots, n_{k-1}} m - \sum_{n_0, n_{\edits{1}}, \dots, n_{k-1}} m \frac{(n^{*})}{n}\right)
    \intertext{The sum of multinomials with $k$ parts that add to $n$ is $k^n$.}
    =& p\frac{1}{k^n}  \sum_{n_0, n_{\edits{1}}, \dots, n_{k-1}} m \frac{(n^{*})}{n} + \frac{(1-p)}{(k-1)}\frac{1}{k^n} \left(k^n - \sum_{n_0, n_{\edits{1}}, \dots, n_{k-1}} m \frac{(n^{*})}{n}\right)\\
    =& p\frac{1}{k^n}  \sum_{n_0, n_{\edits{1}}, \dots, n_{k-1}} m \frac{(n^{*})}{n} + \frac{(1-p)}{(k-1)} \left(1 - \frac{1}{k^n}\sum_{n_0, n_{\edits{1}}, \dots, n_{k-1}} m \frac{(n^{*})}{n}\right)
    \intertext{Let us denote $\SV \hyperDist{\Conc{S}}$ as $S$.}
    =& p\times S + \frac{(1-p)}{(k-1)} \left(1 - S\right)\\
    =& \frac{1-p}{k-1} + S\left(\frac{kp-1}{k-1}\right)
    \intertext{Replace $S$ with the approximation for $\SV \hyperDist{\Conc{S}}$.}
    =& \frac{1-p}{k-1} + \left(\frac{1}{k} + \Theta \left(\sqrt{\frac{\ln k}{kn}}\right)\right)\left(\frac{kp-1}{k-1}\right)\\
    =& \frac{1-p}{k-1} + \frac{kp-1}{k(k-1)} + \Theta \left(\sqrt{\frac{\ln k}{kn}}\right) \left(\frac{kp-1}{k-1}\right)\\
    =& \frac{1}{k} + \Theta \left(\sqrt{\frac{\ln k}{kn}}\right) \left(\frac{kp-1}{k-1}\right)
\end{align*}
\end{proof}

\subsubsection{Proofs of Appendix~\ref{sec:model-reduced-krr}}
\label{sec:proofs-reduced-krr}

\nonequivalentN*

\begin{proof}
This follows from the fact that $\Conc{N}$ and 
$\Conc{N}^{r}$ do not even have the same input set and are, hence, incomparable.
\end{proof}

\commutativityReduced*

\begin{proof}
For simplicity, we provide a proof for the case $\calk$ has $k=2$ values;
the case when $k\geq2$ can be obtained as a direct generalization.

First, let us find a generic entry $(\Conc{N}\Conc{S}^{r})_{x,z}$, for 
a dataset $x \in \calk^{n}$ and a histogram $z$ on $\calk$:
\begin{align}
\label{eq:reduced00}
     (\Conc{N}\Conc{S}^{r})_{x,z} 
    =&\,\, \sum_{y \in \calk^{n}} \Conc{N}_{x,y} \Conc{S}^{r}_{y,z} & \text{(matrix multiplication)} \nonumber \\
    =&\,\, \sum_{y \in \calk^{n}: h(y) = z} \Conc{N}_{x,y} & \text{(Def.~\ref{def:shuffling-channel-reduced}: $\Conc{S}^{r}$)}
\end{align}

Now, let us find a generic entry $(\Conc{S}^{r}\Conc{N}^{r})_{x,z}$, for 
a dataset $x \in \calk^{n}$ and a histogram $z$ on $\calk$:
\begin{align}
    (\Conc{S}^{r}\Conc{N}^{r})_{x,z}
    =&\,\, \sum_{w} \Conc{S}^{r}_{x,w} \Conc{N}^{r}_{w,z} &  \text{(matrix multiplication)} \nonumber \\ 
    =&\,\, \Conc{N}^{r}_{h(x),z}  & \text{(Def.~\ref{def:shuffling-channel-reduced}: $\Conc{S}^{r}$)} \nonumber \\
    =&\,\,  \sum_{\substack{x' \in \calk^n: h(x')=h(x) \\ y \in \calk^{n}: h(y)=z}} \frac{1}{\#h(x)} \Conc{N}_{x',y} & \quad \text{(Def.~\ref{def:krr-reduced}: $\Conc{N}^{r}$)} \nonumber \\
    =&\,\,   \sum_{x' \in \calk^n: h(x')=h(x)} \frac{1}{\#h(x)}
    \sum_{y \in \calk^n: h(y)=\edits{z}}
    \Conc{N}_{x',y} & \quad \text{(rearranging)} \nonumber \\
    =&\,\, \sum_{y \in \calk^n: h(y)=\edits{z}}
    \Conc{N}_{x,y} \sum_{x' \in \calk^n: h(x')=h(x)} \frac{1}{\#h(x)}
    & \quad \text{(see below)} \label{eq:reduce01a} \\
    =&\,\, \sum_{y \in \calk^n: h(y)=\edits{z}}
    \Conc{N}_{x,y} & \quad \text{($\textstyle \sum_{x' \in \calk^n: h(x')=h(x)} \edits{\nicefrac{1}{\#h(x)}} = 1$)} \label{eq:reduced01}
\end{align}
Notice that ~\eqref{eq:reduce01a} follows from Lemma~\ref{lemma:aux}, because $\sum_{y \in \calk^n: h(y)=h(z)} \Conc{N}_{x',y}$ depends 
only on $h(x')$ and $h(y)$, 
and since $h(x')=h(x)$, this summation has the same value 
if we replace $x'$ with $x$, obtaining $\sum_{y \in \calk^n: h(y)=h(z)} \Conc{N}_{x,y}$.




To conclude the proof, notice that \eqref{eq:reduced00} and \eqref{eq:reduced01} are equal.
\end{proof}

\equivalenceCompositionsB*

\begin{proof}
Notice that by Prop.~\ref{theorem:commutativity-full}
$\Conc{NS} \equiv \Conc{SN}$, by what we just showed above $\Conc{N}\Conc{S} \equiv \Conc{N}\Conc{S}^{r}$, and by Prop.~\ref{theorem:commutativity-reduced} $\Conc{N}\Conc{S}^{r} = \Conc{S}^{r}\Conc{N}^{r}$. 
Hence, by transitivity, $\Conc{S}\Conc{N} \equiv \Conc{S}^{r}\Conc{N}^{r}$.
\end{proof}

\subsubsection{Proofs of Appendix~\ref{sec:brown}}
\label{sec:proofs-brown}
\equivBrown*

\begin{proof}
We can rewrite our formula with $\lambda^{*}$ on the left.
\begin{align}
    \sum_{\lambda_{n, k}} \lambda^{*} \binom{n}{\lambda} \binom{k}{\bar{\lambda} \quad k - \ell} \shrug& \sum_{|\lambda| = n} \lambda^{*} \frac{n!}{\lambda! \bar{\lambda}!} \ell! \binom{k}{\ell} \nonumber \\
    \sum_{\lambda_{n, k}} \lambda^{*} \frac{n!}{\lambda!} \frac{k!}{\bar{\lambda}!  (k-\ell)!}  \shrug& \sum_{|\lambda| = n} \lambda^{*} \frac{n!}{\lambda! \bar{\lambda}!} \ell! \frac{k!}{\ell! (k-\ell)!} \nonumber  \\
    \sum_{\lambda_{n, k}} \lambda^{*} \frac{n!}{\lambda!} \frac{k!}{\bar{\lambda}!  (k-\ell)!}  \shrug& \sum_{|\lambda| = n} \lambda^{*} \frac{n!}{\lambda! \bar{\lambda}!} \frac{k!}{(k-\ell)!}\nonumber  \\
    \sum_{\lambda_{n, k}} \lambda^{*} \frac{n! k!}{\lambda! \bar{\lambda}! (k-\ell)!} =& \sum_{|\lambda| = n} \lambda^{*} \frac{n! k!}{\lambda! \bar{\lambda}! (k-\ell)!}
\end{align}
\end{proof}

\end{document}